\documentclass[aip, jcp, preprint,showpacs,superscriptaddress,groupedaddress]{revtex4-1}

\usepackage{media9}

\usepackage[vcentermath]{youngtab}
\usepackage{young}
\usepackage{amsmath}
\usepackage{amssymb}
\usepackage{braket}
\usepackage{bm}
\usepackage{tikz}
\usetikzlibrary{decorations.pathmorphing}

\usepackage{subcaption}
\usepackage{colortbl}

\newtheorem{theorem}{Theorem}[section]
\newtheorem{lemma}[theorem]{Lemma}

\begin{document}

\title{Richardson-Gaudin states of non-zero seniority I: matrix elements}
\author{Paul A. Johnson}
 \email{paul.johnson@chm.ulaval.ca}
 \affiliation{D\'{e}partement de chimie, Universit\'{e} Laval, Qu\'{e}bec, Qu\'{e}bec, G1V 0A6, Canada}

\date{\today}

\begin{abstract}
Seniority-zero wavefunctions describe bond-breaking processes qualitatively. As eigenvectors of a model Hamiltonian, Richardson-Gaudin states provide a clear physical picture and allow for systematic improvement via standard single reference approaches. Until now, this treatment has been done in the seniority-zero channel. In this manuscript, the corresponding states with higher seniorities are identified, and their couplings through the Coulomb Hamiltonian are computed. In \emph{every} case, the couplings between the states are computed from the cofactors of their effective overlap matrix. Proof of principle calculations demonstrate that a single reference configuration interaction is comparable with seniority-based configuration interaction computations at a substantially reduced cost. The next manuscript in this series will identify the corresponding Slater-Condon rules and make the computations feasible.
\end{abstract}

\maketitle

\section{Introduction}
Weakly correlated electrons are perfectly understood in terms of Slater determinants. The physical behaviour is described by one Slater determinant upon which one can add cheap approximate corrections as in Kohn-Sham density functional theory or hierarchical systematic corrections as in coupled cluster (CC) theory.\cite{helgaker_book} Commercial and open-source software packages provide both approaches to the end-user. This is not the case for strongly correlated systems, loosely defined as those for which Slater determinants are a poor basis. Many Slater determinants are necessary which makes the physical picture unclear. Standard methods such as the complete active space self-consistent field (CASSCF)\cite{roos:1980,siegbahn:1980,siegbahn:1981,roos:1987} or selected configuration interactions (CI)\cite{huron:1973,sharma:2017,holmes:2017,li:2018,yao:2021} aim to pick the correct Slater determinants efficiently, which is often very difficult. Larger systems have been treated with approximate CASSCF solvers\cite{olsen:1988,malmqvist:1990,fleig:2001,ma:2011,manni:2013,thomas:2015,manni:2016,schriber:2016,levine:2020} and the density matrix renormalization group (DMRG).\cite{chan:2002,ghosh:2008,yanai:2009,wouters:2014,sun:2017,ma:2017} 

In ref. \citenum{bytautas:2011} and \citenum{bytautas:2015}, the authors demonstrated that strongly correlated chemical systems are well described by partitioning the Hilbert space based on the number of unpaired electrons, the \emph{seniority}. A CI based on seniority converges quickly even for the nitrogen molecule. The catch was that each seniority channel, computed in Slater determinants, scales faster than exponentially. By targeting the seniority-zero channel it was quickly discovered that products of closed shell pairs of electrons, in particular the antisymmetric product of 1 reference orbital geminals (AP1roG)\cite{limacher:2013,limacher:2014a,limacher:2014b,boguslawski:2014a,boguslawski:2014b,boguslawski:2014c,tecmer:2014} or equivalently pair coupled cluster doubles (pCCD),\cite{stein:2014,henderson:2014a,henderson:2014b} was indistinguishable from seniority-zero CI (historically doubly occupied configuration interaction\cite{weinhold:1967a,weinhold:1967b,veillard:1967,clementi:1967,cook:1975}) with mean-field cost, beating the known antisymmetric product of strongly orthogonal geminals (APSG),\cite{kutzelnigg:1964,surjan_book,surjan:2012} generalized valence-bond / perfect-pairing (GVB-PP),\cite{hurley:1953,hunt:1972,hay:1972,goddard:1973} and the antisymmetrized geminal power (AGP).\cite{coleman:1965,ortiz:1981,sarma:1989,rowe:1991,chen:1995,coleman:1997,rowe:2001} These are all \emph{geminal} wavefunctions, which originated very early in quantum chemistry.\cite{fock:1950,mcweeny:1959,mcweeny:1960,mcweeny:1963,silver:1969,silver:1970a,silver:1970b,silver:1970c} Much work has been devoted to pCCD\cite{boguslawski:2015,boguslawski:2016a,boguslawski:2016b,boguslawski:2017,boguslawski:2019,boguslawski:2021,marie:2021,kossoski:2021,baran:2021} and AGP since.\cite{neuscamman:2012,neuscamman:2013,henderson:2019,khamoshi:2019,henderson:2020,khamoshi:2020,dutta:2020,khamoshi:2021,dutta:2021} 

Unfortunately there is no obvious way to add higher seniorities into the picture, though methods have been discussed. The power of Slater determinants is that they are a basis for the Hilbert space so that corrections can be added systematically. This is not straightforward for seniority-zero wavefunction ans\"{a}tze. The space does not have a basis of pCCD vectors for example, but one can demand exact treatment of quadruples or hextuples.\cite{parkhill:2009,parkhill:2010,lehtola:2016,lehtola:2018} The products of pairs can include open-shell configurations at a substantially increased (unfeasible) cost, but this will still miss effects of weak correlation.\cite{cassam:2006,cassam:2010,cassam:2023,johnson:2024c} Adding higher seniorities with the random-phase approximation on top of seniority-zero CI is difficult.\cite{vu:2019} As weakly correlated systems are well treated with excitation-based CI and strongly correlated systems are well treated with seniority-based CI, an interesting approach is the hierarchical CI (hCI) that classifies Slater determinants by a single parameter balancing both effects.\cite{kossoski:2022} In the end, weakly correlated systems require more excitations, and strongly correlated systems require more seniorities making this balance difficult. 

The eigenvectors of the reduced Bardeen-Cooper-Schrieffer (BCS)\cite{bardeen:1957a,bardeen:1957b,schrieffer_book} Hamiltonian, the Richardson\cite{richardson:1963,richardson:1964,richardson:1965}-Gaudin\cite{gaudin:1976,gaudin_book} (RG) states, are a basis for the Hilbert space built from weakly interacting pairs of electrons. The idea is that systems that are strongly correlated in terms of electrons are weakly correlated in terms of RG pairs. This is not simply wishful thinking. Seniority-zero CI is dominated by one RG state\cite{johnson:2020,fecteau:2022} while second order Epstein\cite{epstein:1926}-Nesbet\cite{nesbet:1955} perturbation theory (ENPT) accounts for the rest.\cite{johnson:2024b} This comes at a cost. RG states are built from solutions of non-linear equations which were difficult to solve for a long time,\cite{rombouts:2004,guan:2012,pogosov:2012,debaerdemacker:2012,claeys:2015} though through a change of variables the procedure is now cheap and robust.\cite{faribault:2011,elaraby:2012} Density matrix elements were also difficult to compute,\cite{sklyanin:1999,amico:2002,faribault:2008,faribault:2010,gorohovsky:2011,fecteau:2020,johnson:2021,moisset:2022a} but are now simply produced from a single linear algebra operation.\cite{faribault:2022,johnson:2024b} Analogues to the Slater-Condon rules follow from counting the number of near-zero singular values in the overlap matrix, which is the fundamental basis for Wick's theorem.\cite{chen:2023} Up to now only seniority-zero RG states have been considered. In this manuscript the matrix elements for RG states of higher seniorities, in particular up to and including seniority-four, are constructed and reduced to sums of cofactors of the effective overlap matrix. The development follows the same lines as the seniority-zero case\cite{faribault:2022} but is substantially more tedious. There are also \emph{many} more types of element to compute. A complete list is presented for a spin-preserving two-body operator, such as the Coulomb Hamiltonian. The Wigner-Eckart theorem ensures that spin-dependent operators will have the same results decorated with Clebsch-Gordan coefficients. Starting at seniority-four, the issue of linear dependence causes the final expressions to be much less clean. However, they are all computable from the same primitive elements. The development of the matrix elements is lengthy and difficult enough that it is presented on its own. The next manuscript will develop the corresponding Slater-Condon rules so that CI or ENPT are feasible.

This manuscript is organized as follows. Section \ref{sec:prelim} briefly introduces the relevant Lie algebras, here su(2) and sp(N), the reduced BCS Hamiltonian, RG states and their norms. Section \ref{sec:s0} presents the matrix elements for seniority-zero RG states in a short synopsis of ref. \citenum{faribault:2022}. All the couplings involving seniorities up to two are computed in Section \ref{sec:s2} while those for seniorities up to four are computed in Section \ref{sec:s4}. Feasibility and completeness are discussed in Section \ref{sec:dis}, before some proof of principle calculations are presented. As a reference, the results of spin-coupling (bypassing iterative constructions) is included as Appendix \ref{sec:spin_coupling}. Finally, alternative expressions for the matrix elements in terms of other variables are summarized in Appendix \ref{sec:rap_cf}. These are only to be used as intermediate checks. Numerically they are \emph{disastrous}.

\section{Preliminaries} \label{sec:prelim}
This section briefly outlines the Lie algebras for pairs of electrons, as well as RG states and their corresponding norms. For a more complete description, see ref. \citenum{johnson:2024a}.
\subsection{Electron pairs}
\subsubsection{su(2)}
From a collection of second-quantized operators 
\begin{align}
	[a^{\dagger}_{p\sigma},a_{q\tau}]_+ = \delta_{pq} \delta_{\sigma\tau}
\end{align}
which create/remove electrons in individual spin-orbitals, one can construct objects that create/remove pairs of electrons. Closed shell pairs of electrons are built with the Lie algebra su(2): in a given spatial orbital $p$ there are three operators
\begin{align} \label{eq:pair_su2}
	S^+_p = a^{\dagger}_{p\uparrow}a^{\dagger}_{p\downarrow}, 
	\quad S^-_p = a_{p\downarrow}a_{p\uparrow},
	\quad S^z_p = \frac{1}{2}\left( a^{\dagger}_{p\uparrow}a_{p\uparrow} + a^{\dagger}_{p\downarrow}a_{p\downarrow} -1 \right),
\end{align}
with the structure
\begin{align}
	[S^+_p, S^-_q] &= 2 \delta_{pq} S^z_p \\
	[S^z_p, S^{\pm}_q] &= \pm \delta_{pq} S^{\pm}_p.
\end{align}
Here $S^+_p$ creates a pair of electrons in the spatial orbital $p$ and $S^-_p$ removes a pair from $p$. $S^z_p$ effectively counts the number of pairs in the spatial orbital: acting on a full spatial orbital returns a value of $+\frac{1}{2}$ while acting on an empty spatial orbital returns a value of $-\frac{1}{2}$. Rather than $S^z_p$ it is convenient to employ the number operator
\begin{align}
	\hat{n}_p &= 2S^z_p + 1 \\
	&= a^{\dagger}_{p\uparrow} a_{p\uparrow} + a^{\dagger}_{p\downarrow} a_{p\downarrow}.
\end{align}

Restricted Slater determinants (RSD), in which all electrons occur in up/down spin partners, are constructed with these objects. In particular, the RSD with the $M$ spatial orbitals $i_1,\dots i_M = \{i\}$ occupied is explicitly
\begin{align} \label{eq:rsd}
	\ket{\{i\}} = S^+_{i_1} S^+_{i_2} \dots S^+_{i_M} \ket{\theta} 
\end{align}
where $\ket{\theta}$ is the physical vacuum. RSD are always closed-shell singlets of non-interacting electrons. Unrestricted Slater determinants (USD), e.g.
\begin{align} \label{eq:usd}
	\ket{\{i\},a\sigma b\tau} = S^+_{i_1} S^+_{i_2} \dots S^+_{i_M} a^{\dagger}_{a\sigma} a^{\dagger}_{b\tau} \ket{\theta} 
\end{align}
allow for single occupation of individual orbitals by adding unpaired electrons on top of an RSD. A USD is understood as having a ``paired'' part, a string of $S^+_i$, and an ``unpaired'' part, a string of $a^{\dagger}_{a\sigma}$. The same structure is present in RG states: the paired indices are \emph{unblocked} and are labelled $i,j,k,l$ while the unpaired indices are \emph{blocked} and labelled $a,b,c,d$. A general index, blocked or unblocked, will be labelled $p,q,r,s$. The term blocked indices stems from the fact that they do not participate in the pairing. For example, with $a^{\dagger}_{b\sigma}$, the su(2) operators for spatial orbital $b$ all yield zero:
\begin{align}
	S^+_b a^{\dagger}_{b\sigma} \ket{\theta} = S^-_b a^{\dagger}_{b\sigma} \ket{\theta} = S^z_b a^{\dagger}_{b\sigma} \ket{\theta} = 0.
\end{align}

Notice that USDs are not spin eigenfunctions, but they can be made into configuration state functions (CSF) by coupling the blocked indices only. In the present contribution the focus is singlets, which are constructed from pairs of electrons in different spatial orbitals, so-called open-shell singlets. 

\subsubsection{sp(N)}
Open-shell singlet pairs can be created from the objects
\begin{align} \label{eq:spn_creator}
	A^+_{pq} &= \frac{1}{\sqrt{2}} 
	\left(
	a^{\dagger}_{p\uparrow} a^{\dagger}_{q\downarrow} - a^{\dagger}_{p\downarrow} a^{\dagger}_{q\uparrow}
	\right)
\end{align}
and removed by
\begin{align} \label{eq:spn_ann}
	A^-_{pq} &= \frac{1}{\sqrt{2}} \left(
	a_{q\downarrow}a_{p\uparrow} - a_{q\uparrow}a_{p\downarrow}
	\right).
\end{align}
The use of these objects necessarily introduces a third
\begin{align}
	A^0_{pq} = a^{\dagger}_{p\uparrow} a_{q\uparrow} + a^{\dagger}_{p\downarrow} a_{q\downarrow}
\end{align}
and indeed a \emph{fourth} since $A^0_{qp} \neq A^0_{pq}$. These last two objects are singlet excitation operators. On their own, they close the Lie algebra u(N),\cite{helgaker_book} but when taken together with $A^+_{pq}$ and $A^-_{pq}$ the Lie algebra is sp(N). In particular, for each choice $p<q$ there are 4 distinct operators $A^+_{pq}, A^-_{pq},A^0_{pq},A^0_{qp}$. With $N$ spatial orbitals there are $\binom{N}{2}$ such choices and thus $4\binom{N}{2}$ operators. Each spatial orbital also contributes 3 operators $S^+_{p},S^-_{p},S^z_{p}$, giving a total of $N(2N+1)$ operators, the dimension of the Lie algebra sp(N). In ref. \citenum{johnson:2024c}, the operators were normalized to make the sp(N) structure constants as symmetric as possible. While not incorrect, such a choice is inconvenient when dealing with representations. In this contribution, the required structure is
\begin{align}
	[A^0_{pq},A^0_{rs}] &= \delta_{qr} A^0_{ps} - \delta_{ps} A^0_{rq} \label{eq:spn_a0} \\
	[A^0_{pq},A^+_{rs}] &= \delta_{qr} A^+_{ps} + \delta_{qs} A^+_{pr} \\
	[A^0_{pq},S^+_r]    &= \sqrt{2} \delta_{qr} A^+_{pq}
\end{align}
while the rest of the sp(N) commutators are not even pertinent enough to mention. The choice of normalization of the open-shell pair creators is made so that
\begin{align}
	\ket{(pq)} =  A^+_{pq}  \ket{\theta}
\end{align}
is normalized. In this convention, the open-shell pair creators and annihilators are adjoints of one another, as are the singlet excitation operators
\begin{align}
	\left(A^+_{pq}\right)^{\dagger} &= A^-_{pq} \\
	\left(A^0_{pq}\right)^{\dagger} &= A^0_{qp}.
\end{align}
Strictly speaking, the open-shell pair creators \eqref{eq:spn_creator} are defined for $p<q$ but as swapping the indices in the RHS causes no change $A^+_{pq} = A^+_{qp}$, it is convenient to lift this restriction on the indices. Diagonal elements are also well-defined
\begin{align}
	A^{\pm}_{pp} &= \sqrt{2} S^{\pm}_p \\
	A^0_{pp} &= \hat{n}_p.
\end{align}
Finally, there are consequences of the Pauli principle that do not follow directly from the Lie algebra structure. Pairs of electrons cannot be created more than once
\begin{align}
	S^+_p S^+_p = 0,
\end{align}
nor can they be created in a set of levels partially occupied by an open-shell singlet
\begin{align}
	S^+_p A^+_{pq} = S^+_q A^+_{pq} = 0.
\end{align}
On the other hand, open-shell singlet creators can act twice, with the result
\begin{align}
	A^+_{pq} A^+_{pq} = - S^+_p S^+_q,
\end{align}
or if only one of the indices is shared
\begin{align}
	A^+_{pq} A^+_{pr} = -\frac{1}{\sqrt{2}} S^+_p A^+_{qr}.
\end{align}
Finally, for a collection of four distinct indices it is easily verified that
\begin{align}
	A^+_{pq} A^+_{rs} + A^+_{pr} A^+_{qs} + A^+_{ps}A^+_{qr} = 0.
\end{align}
This last property is what makes open-shell singlets tedious: while there are three ways to create two open-shell singlets across four spatial orbitals only two of them are linearly independent.

A set of indices $\{a,b,c,d\}$ will be said to be in \emph{natural order}, provided $a<b<c<d$. It is convenient to adopt the shorthand
\begin{align}
	A^+_{ab}A^+_{cd}\ket{\theta} = \ket{(ab)(cd)}
\end{align}
and label the three possible states with an integer $\omega$
\begin{align}
	\ket{(ab)(cd)} &\rightarrow \omega = 1 \nonumber \\
	\ket{(ac)(bd)} &\rightarrow \omega = 2 \nonumber \\
	\ket{(ad)(bc)} &\rightarrow \omega = 3.	
\end{align}
Notice that while there are $4!=24$ possible orderings of the indices, only three are distinct since $A^+_{ab} = A^+_{ba}$ and $A^+_{ab}A^+_{cd} = A^+_{cd}A^+_{ab}$. For convenience, the 24 possible states, grouped by $\omega$ are summarized in Table \ref{tab:abcd}. 
\begin{table}[ht!] % [h] for here, you can change this option according to your needs
	\centering % Center the table
	\begin{tabular}{c|c|c} % Specify the number of columns and alignment\
		$\omega=1$ & $\omega=2$ & $\omega=3$ \\
		\hline % Draw another horizontal line
		(ab)(cd) & (ac)(bd) & (ad)(bc) \\ 
		(ba)(cd) & (ca)(bd) & (da)(bc) \\
		(ab)(dc) & (ac)(db) & (ad)(cb) \\
		(ba)(dc) & (ca)(db) & (da)(cb) \\
		(cd)(ab) & (bd)(ac) & (bc)(ad) \\
		(cd)(ba) & (bd)(ca) & (bc)(da) \\
		(dc)(ab) & (db)(ac) & (cb)(ad) \\
		(dc)(ba) & (db)(ca) & (cb)(da) \\
	\end{tabular}
	\caption{Equivalent orders} % Table caption
	\label{tab:abcd} 
\end{table}

For a set of indices $\{a,b,c,d\}$ in natural order, building four-electron singlets with Clebsch-Gordan coupling leads to the choice
\begin{align}
	\ket{\varphi^{(1)}_{abcd}} &= A^+_{ab} A^+_{cd} \ket{\theta} \\
	\ket{\varphi^{(2)}_{acbd}} &= \frac{1}{\sqrt{3}} \left( A^+_{ac}A^+_{bd} - A^+_{ad}A^+_{bc} \right) \ket{\theta}
\end{align}
which are easily verified to be orthonormal. Seniority-four singlet CSFs are RSDs acting on these two ``vacuums''
\begin{align}
	\ket{\{i\},\varphi^{(\mu)}_{abcd}} = S^+_{i_1} S^+_{i_2} \dots S^+_{i_M} \ket{\varphi^{\mu}_{abcd}}
\end{align}
so the two states $\ket{\varphi^{(1)}_{abcd}}$ and $\ket{\varphi^{(2)}_{abcd}}$ will be referred to as the seniority-four vacuums for the set $\{a,b,c,d\}$ in natural order. The situation becomes much worse as the number of open-shell singlet pairs grows and choosing an orthogonal basis seems to be incredibly difficult. A complete solution is possible in terms of diagrams and Young tableaux, though as the present contribution requires only seniorities two and four, this construction is summarized in Appendix \ref{sec:spin_coupling}. Other choices could be made, via e.g. L\"{o}wdin orthogonalization, but there is no perfect choice. All will suffer the difficulties of section \ref{sec:comb_tedious} and the present construction is the easiest to generalize. 

\subsection{RG states}
RG states are the eigenvectors of the reduced BCS Hamiltonian
\begin{align} \label{eq:hbcs}
	\hat{H}_{BCS} = \frac{1}{2} \sum^N_{p=1} \varepsilon_p \hat{n}_p - \frac{g}{2} \sum^N_{p,q=1} S^+_p S^-_q.
\end{align}
With pair creators
\begin{align} \label{eq:gaudin_pair}
	S^+(v) = \sum^N_{p=1} \frac{S^+_p}{v - \varepsilon_p},
\end{align}
defined in terms of complex numbers $v$, called \emph{rapidities}, the seniority-zero RG states for $M$ pairs of electrons are
\begin{align} \label{eq:RG_state_0}
	\ket{ \{v\}^M} = S^+(v_1) S^+(v_2) \dots S^+(v_M) \ket{\theta}.
\end{align}
For the states \eqref{eq:RG_state_0} to be eigenvectors of \eqref{eq:hbcs}, the rapidities must satisfy Richardson's equations
\begin{align} \label{eq:rich_0}
	\frac{2}{g} + \sum^N_{p=1} \frac{1}{v_{\alpha} - \varepsilon_p} + \sum^M_{\beta (\neq \alpha)=1} \frac{2}{v_{\beta} - v_{\alpha}} = 0, \quad \forall \alpha =1,\dots,M.
\end{align}
Rapidities will be indexed with lowercase greek indices. It is known that for $M$ pairs there are $\binom{N}{M}$ distinct solutions of Richardson's equations.\cite{richardson:1965} A set of $\{v\}$ that solves Richardson's equations and the corresponding RG state $\ket{ \{v\}^M}$ are referred to as \emph{on-shell} while a set of arbitrary $\{u\}$ and the corresponding RG state $\ket{\{u\}^M}$ are \emph{off-shell}. The BCS eigenvalue of an on-shell seniority-zero RG state is the sum of its rapidities
\begin{align}
	E^M_{BCS} = \sum^M_{\alpha} v_{\alpha}.
\end{align}
Solving Richardson's equations numerically for rapidities is not the focus of this manuscript. It is possible, but difficult near critical points at which rapidities want to coincide with particular single-particle energies $\{\varepsilon\}$.\cite{debaerdemacker:2012,claeys:2015} 

Rapidities are convenient when studying analytic properties of RG states. Numerical applications necessitate the use of eigenvalue-based variables (EBV)
\begin{align} \label{eq:ebv_0}
	V_p = \sum^M_{\alpha = 1} \frac{g}{\varepsilon_p - v_{\alpha}}.
\end{align}
Richardson's equations for the rapidities are satisfied provided the EBV satisfy the equations
\begin{align} \label{eq:ebv_eq_0}
	V^2_p - 2 V_p - g \sum^N_{q (\neq p)=1} \frac{V_q - V_p}{\varepsilon_q - \varepsilon_p} = 0, \quad \forall p=1,\dots,M
\end{align}
in addition to the normalization
\begin{align}
	\sum^N_{p=1} V_p = 2 M.
\end{align}
These equations are much easier to solve numerically in a robust manner.\cite{faribault:2011,elaraby:2012,fecteau:2022} In short, the EBV are evolved from known solutions at $g=0$, where the EBV equations \eqref{eq:ebv_eq_0} decouple
\begin{align}
	V_p (V_p - 2) = 0,
\end{align}
whose corresponding solutions are $M$ EBV equal to 2 and $N-M$ EBV equal to zero. The interaction in the reduced BCS Hamiltonian disappears and the RG states are RSDs, defined by which spatial orbitals are doubly-occupied ($V_p=2$) and which are empty ($V_p=0$). This information is cleanly summarized as an ordered list of 1s (full) and 0s (empty) called a \emph{bitstring}. The key property is that the EBV evolve uniquely from $g=0$ so that RG states, \emph{at any $g$}, can be labelled unambiguously based on their $g=0$ RSD representative. The ground state of the reduced BCS model is \emph{always} the state labelled by $M$ 1s followed by $N-M$ 0s, and the highest excited state is \emph{always} the state labelled by $N-M$ 0s followed by $M$ 1s. The other states have crossings at intermediate $g$, but they are always strict crossings and not avoided crossings.\cite{yuz:2002,yuz:2003,yuz:2005} 

RG states of higher seniorities are defined in essentially the same way, the only difference being that the vacuum will include a set of blocked levels that do not participate in the pairing effects. In particular, a singlet RG state of $2M$ electrons with seniority two
\begin{align}
	\ket{ \{v\}^{M-1}, (ab) } = S^+(v_1) S^+(v_2) \dots S^+(v_{M-1}) A^+_{ab} \ket{\theta}
\end{align}
is an eigenvector of \eqref{eq:hbcs} provided the $M-1$ rapidities $\{v\}$ satisfy the corresponding set of Richardson's equations
\begin{align} \label{eq:rich_2}
	\frac{2}{g} + \sum^N_{p=1 (\neq a,b)} \frac{1}{v_{\alpha} - \varepsilon_p} + \sum^{M-1}_{\beta (\neq \alpha)=1} \frac{2}{v_{\beta} - v_{\alpha}} = 0, 
	\quad \forall \alpha = 1,\dots,M-1.
\end{align}
There are $\binom{N}{2}$ choices for blocking two levels, and for each there are $\binom{N-2}{M-1}$ solutions of \eqref{eq:rich_2} leading to $\binom{N}{2}\binom{N-2}{M-1}$ seniority-two singlet RG states with $2M$ electrons. Note: the blocked levels do not appear in Richardson's equations \eqref{eq:rich_2} but \emph{do} appear in the pair creators \eqref{eq:gaudin_pair}. The corresponding eigenvalue is
\begin{align}
	E^{M-1,ab}_{BCS} = \frac{1}{2} (\varepsilon_a + \varepsilon_b) + \sum^{M-1}_{\alpha=1} v_{\alpha}.
\end{align}
EBV are defined in the exact same way, except that those corresponding to blocked levels are not defined. In the computation of matrix elements, it will turn out to be convenient to interpret them as being equal to zero. The corresponding set of non-linear equations are, 
\begin{align} \label{eq:ebv_2}
	V^2_p - 2 V_p - g \sum^N_{q (\neq p,a,b)=1} \frac{V_q - V_p}{\varepsilon_q - \varepsilon_p} &= 0, \qquad \forall p=1,\dots,N (\neq a,b) \\
		\sum^N_{p (\neq a,b)=1} V_p &= 2 (M-1).
\end{align}
With a label x for blocked sites, the RG states are again labelled as bitstrings, e.g. 11x0x100, which would be the bitstring 110100 with levels 3 and 5 blocked. The EBV equations are solved by evolving from the initial RSD with $V_1 = V_2 = V_6 = 2$ and $V_4 = V_7 = V_8 = 0$, omitting contributions from $\varepsilon_3$ and $\varepsilon_5$. The seniority-two RG states are thus no more complicated than the seniority-zero RG states. 

Seniority-four RG states are essentially the same, the only difference being the two vacuums: for $\mu=1,2$, the states
\begin{align}
	\ket{\{v\}^{M-2},\varphi^{(\mu)}_{abcd}} = S^+(v_1) S^+(v_2) \dots S^+(v_{M-2}) \ket{\varphi^{(\mu)}_{abcd}}
\end{align}
are eigenvectors of \eqref{eq:hbcs} provided the $M-2$ rapidities are solutions of Richardson's equations
\begin{align} \label{eq:rich_4}
	\frac{2}{g} + \sum^N_{p=1 (\neq a,b,c,d)} \frac{1}{v_{\alpha} - \varepsilon_p} + \sum^{M-2}_{\beta (\neq \alpha)=1} \frac{2}{v_{\beta} - v_{\alpha}} = 0
	\quad \forall \alpha = 1,\dots,M-2.
\end{align}
There are $\binom{N}{4}$ choices for blocking four levels with two linearly independent singlets for each choice, and $\binom{N-4}{M-2}$ solutions of \eqref{eq:rich_4} giving $2\binom{N}{4}\binom{N-4}{M-2}$ seniority-four singlet RG states of $2M$ electrons. The corresponding eigenvalues are
\begin{align}
	E^{M-2,abcd}_{BCS} = \frac{1}{2} (\varepsilon_a + \varepsilon_b + \varepsilon_c + \varepsilon_d) + \sum^{M-2}_{\alpha=1} v_{\alpha}.
\end{align}
Notice that the two states $\ket{\{v\},\varphi^{(1)}_{abcd}}$ and $\ket{\{v\},\varphi^{(2)}_{abcd}}$ are built with the \emph{same} set of rapidities $\{v\}$. Further, the blocked levels could be coupled to a triplet or a quintuplet without changing the structure. The corresponding EBV satisfy the nonlinear equations
\begin{align} \label{eq:ebv_4}
	V^2_p - 2 V_p - g \sum^N_{q (\neq p,a,b,c,d)=1} \frac{V_q - V_p}{\varepsilon_q - \varepsilon_p} &= 0, \qquad \forall p=1,\dots,N (\neq a,b,c,d) \\
	\sum^N_{p (\neq a,b,c,d)=1} V_p &= 2 (M-2).
\end{align}
Seniority-four RG states are labelled in the same way as seniority-two RG states, keeping in mind that each bitstring corresponds to two states. RG states of any seniority can thus be easily deduced and constructed numerically.

\subsection{Norms and scalar products}
Scalar products and correlation functions in terms of rapidities are known, but lead to numerical problems. They are thus relegated to appendix \ref{sec:rap_cf}. For $\ket{\{v\}^M}$ an on-shell RG state and $\ket{\{u\}^M}$ arbitrary, the scalar product is a single $N\times N$ determinant in terms of the corresponding EBV\cite{faribault:2012,claeys:2017b}
\begin{align} \label{eq:ebv_scalar}
	\braket{ \{v\}^M | \{u\}^M } = \eta (0) \det J
\end{align}
where, in terms of the seniority $\Omega$ 
\begin{align}
	\eta (\Omega) = (-1)^{N-M-\frac{\Omega}{2}} \left( \frac{1}{2} \right)^{N-2M} g^{\Omega-2M}
\end{align}
and
\begin{align} \label{eq:matsuno}
	J_{pq} =
	\begin{cases}
		U_p + V_p - 2 + \sum^N_{r (\neq p)=1} \frac{g}{\varepsilon_r - \varepsilon_p}, \quad & p=q, \\
		\frac{g}{\varepsilon_p - \varepsilon_q}, \quad & p\neq q.
	\end{cases} 
\end{align}
$J$ will be referred to as the effective overlap matrix. When the two states are the same, this matrix becomes the Jacobian of the EBV equations and is emphasized $\bar{J}$. When the two RG states are on-shell but distinct, $J$ becomes singular.

For seniority-two RG states, the scalar product between $\ket{\{v\}^{M-1},(ab)}$ an on-shell RG state and $\ket{\{u\}^{M-1},(ab)}$ arbitrary depends on an $(N-2) \times (N-2)$ determinant 
\begin{align} \label{eq:ebv_scalar_2}
	\braket{ \{v\}^{M-1},(ab) | \{u\}^{M-1},(ab) } = \eta (2) \det J(ab) \braket{(ab)|(ab)},
\end{align} 
where $J(ab)$ is \eqref{eq:matsuno} \emph{without} any contributions from levels $a$ and $b$: there are no rows and columns for $a$ and $b$, and the diagonal elements receive no contributions from $a$ and $b$. Rather than deal with an $(N-2)\times (N-2)$ matrix, it will be convenient to extend it to $N\times N$ with unit diagonal elements $J(ab)_{aa} = J(ab)_{bb}=1$
\begin{align} \label{eq:matsuno_2}
	J(ab)_{pq} =
	\begin{cases}
		1, \quad & p=q\; (=a,b), \\
		U_p + V_p - 2 + \sum^N_{r (\neq p,a,b)=1} \frac{g}{\varepsilon_r - \varepsilon_p}, \quad & p=q\; (\neq a,b), \\
		\frac{g}{\varepsilon_p - \varepsilon_q}, \quad & p\neq q \; (\neq a,b).
	\end{cases} 
\end{align}
Notice that the scalar product \eqref{eq:ebv_scalar_2} is only non-zero if the blocked levels in the bra and the ket are identical. Finally, the norm 
\begin{align}
	\braket{ \{v\}^{M-1},(ab) | \{v\}^{M-1},(ab) } = \eta(2) \det \bar{J}(ab)
\end{align}
is proportional to the Jacobian of the EBV equations with blocked levels $a$ and $b$.

Seniority-four RG states are the same, with a minor complication arising from the two linearly independent seniority-four singlets. For $\{v\}$ on-shell and $\{u\}$ arbitrary, the scalar product is
\begin{align}
	\braket{ \{v\}^{M-2},\varphi^{(\mu)}_{abcd} | \{u\}^{M-2},\varphi^{(\nu)}_{abcd} } &= 
	\eta (4) \det J(abcd) \braket{ \varphi^{(\mu)}_{abcd} | \varphi^{(\nu)}_{abcd}}
\end{align}
where
\begin{align} \label{eq:matsuno_4}
	J(abcd)_{pq} =
	\begin{cases}
		1, \quad & p=q\; (=a,b,c,d), \\
		U_p + V_p - 2 + \sum^N_{r (\neq p,a,b,c,d)=1} \frac{g}{\varepsilon_r - \varepsilon_i}, \quad & p=q\; (\neq a,b,c,d), \\
		\frac{g}{\varepsilon_p - \varepsilon_q}, \quad & p\neq q \; (\neq a,b,c,d).
	\end{cases} 
\end{align}
The norms of the two seniority-four states are identical and the scalar product between the two vacuums $\braket{ \varphi^{(1)}_{abcd} | \varphi^{(2)}_{abcd}}$ is necessarily zero.

The remainder of the manuscript will report the matrix elements of a spin conserving 2-body operator, in particular the Coulomb Hamiltonian which has an explicit expression in terms of the singlet excitation operators $A^0_{pq}$
\begin{align} \label{eq:HC}
	\hat{H}_C = \sum_{pq} h_{pq} A^0_{pq} + \frac{1}{2} \sum_{pqrs} V_{pqrs} \left( A^0_{pq}A^0_{rs} - \delta_{qr} A^0_{ps} \right) + V_{NN}.
\end{align}
The one- and two-electron integrals
\begin{align}
	h_{pq} &= \int d\textbf{x}\; \phi^*_{p} (\textbf{x})
	\left( -\frac{\nabla}{2} - \sum_{I} \frac{Z_I}{\vert \textbf{x} - \textbf{R}_I \vert} \right)  \phi_q(\textbf{x}) \\
	V_{pqrs} &= \int d\textbf{x}_1 d\textbf{x}_2\;
	\frac{\phi^*(\textbf{x}_1) \phi_q(\textbf{x}_1) \phi^*_r (\textbf{x}_2) \phi_s (\textbf{x}_2)}
	{\vert \textbf{x}_1 - \textbf{x}_2\vert}
\end{align}
are computed in a basis of orbitals $\{\phi\}$. While in this contribution it will be assumed that the orbitals are restricted, an extension to unrestricted and generalized sets of orbitals is possible. \emph{Chemists'} notation is employed for the 2-electron integrals.

For two states $\ket{\Phi}$ and $\ket{\Phi'}$, the matrix element
\begin{align}
	\braket{\Phi | \hat{H}_c | \Phi'} = \sum_{pq} h_{pq} \gamma^{\Phi\Phi'}_{pq}
	+ \frac{1}{2} \sum_{pqrs} V_{pqrs} \Gamma^{\Phi\Phi'}_{pqrs}
	+ V_{NN} \braket{\Phi | \Phi'}
\end{align}
is computed from the 1- and 2-body density matrix (DM) elements
\begin{subequations}
	\begin{align}
		\gamma^{\Phi \Phi'}_{pq} &:= \braket{\Phi | A^0_{pq} | \Phi'} \\
		\Gamma^{\Phi \Phi'}_{pqrs} &:= \braket{\Phi | (A^0_{pq}A^0_{rs} - \delta_{qr}A^0_{ps}) | \Phi'}.
	\end{align}
\end{subequations}
Notice that \eqref{eq:spn_a0} forces $\Gamma^{\Phi \Phi'}_{pqrs} = \Gamma^{\Phi \Phi'}_{rspq}$. These are reduced density matrix (RDM) elements when $\ket{\Phi}$ and $\ket{\Phi'}$ are the same and transition density matrix (TDM) elements otherwise. 

The DM elements for seniority-zero RG states will now be quickly reviewed in the present conventions before passing to seniorities two and four.

\section{Seniority-zero} \label{sec:s0}
For two M-pair RG states $\ket{\{u\}^M}$ and $\ket{\{v\}^M}$, the only non-zero 1-body DM elements are diagonal. This can be shown in a tedious manner by moving $A^0_{kk}$ to the right until it acts on the vacuum, giving a sum
\begin{align}
	\gamma^{VU}_{kk} = \sum^M_{\alpha =1} \frac{\braket{\{v\}^M | S^+_k | \{u\}^M_{\alpha} }}{u_{\alpha}-\varepsilon_k}
\end{align}
where $\{u\}_{\alpha}$ is the set of rapidities $\{u\}$ without $u_{\alpha}$. The reduced scalar products $\braket{\{v\}^M | S^+_k | \{u\}^M_{\alpha} }$ are referred to as \emph{form factors}. 

Likewise, there are only a few non-zero two-electron DM elements. First notice that the \emph{diagonal} element is just the 1-DM element
\begin{align}
	\Gamma^{VU}_{kkkk} = \gamma^{VU}_{kk},
\end{align}
which can be shown in the same manner, keeping in mind that the same pair of electrons $S^+_i$ cannot be created twice. The \emph{direct} elements for $k\neq l$ are
\begin{align}
	\Gamma^{VU}_{kkll} = 4 \sum^M_{\alpha = 1}\sum^M_{\beta (\neq \alpha)=1}
	\frac{\braket{ \{v\}^M | S^+_k S^+_l | \{u\}^M_{\alpha,\beta} }}
	{(u_{\alpha} - \varepsilon_k)(u_{\beta} - \varepsilon_l)}
\end{align}
while the \emph{exchange} elements are
\begin{align}
	\Gamma^{VU}_{kllk} = - \frac{1}{2} \Gamma^{VU}_{kkll}.
\end{align}
As these elements are scalar multiples of one another, no distinction was made in previous papers. In non-zero seniorities it will become a theme that exchange DM elements involving blocked levels do not behave intuitively. Finally, there are \emph{pair-transfer} elements for $k\neq l$
\begin{align}
	\Gamma^{VU}_{klkl} = 
	 2 \sum^M_{\alpha=1} \frac{\braket{\{v\}^M | S^+_k | \{u\}^M_{\alpha} }}{u_{\alpha} - \varepsilon_l}
	-2 \sum^M_{\alpha=1} \sum^M_{\beta (\neq \alpha)=1}
	\frac{\braket{ \{v\}^M | S^+_k S^+_l | \{u\}^M_{\alpha,\beta} }}
	{(u_{\alpha} - \varepsilon_l)(u_{\beta} - \varepsilon_l)}
\end{align}
which can be seen to match previous conventions since $A^0_{kl}A^0_{kl} = 2 S^+_k S^-_l$.

With the rapidities $\{u\}$ and $\{v\}$ the form factors can be computed as residues of the scalar product $\braket{\{v\}^M | \{u\}^M}$. However, rapidities behave \emph{very poorly} for numerical purposes, with separate expressions required for RDM and TDM elements. While rapidity-based expressions were computed for this contribution, they are relegated to appendix \ref{sec:rap_cf}.

Expressions for the DM elements are now understood directly in terms of the EBV which applies equally to RDM and TDM elements, differing only in the manner the primitive summands are computed.\cite{faribault:2022,johnson:2024b} However, as RG states do not have a succinct representation in terms of EBV, the intermediate steps involve rapidities. The development for the 1-DM elements will now be summarized, and the interested reader is directed to appendix B of ref. \citenum{faribault:2022} for the details of the 2-DM elements.

Manipulation of determinants of the matrix $J$ \eqref{eq:matsuno} are simplified with two practical lemmas (see appendix A of ref. \citenum{faribault:2022} for proofs):
\begin{lemma} \label{lem:rank1}
	For $z$ an arbitrary complex number distinct from $\{\varepsilon\}$, the diagonal rank-N update
	\begin{align}
		\Lambda(z) = 
		\begin{pmatrix}
			\frac{g}{\varepsilon_1 - z} & 0 & \dots & 0 \\
			0 & \frac{g}{\varepsilon_2 - z} & \dots & 0 \\
			\vdots & \vdots & \ddots & \vdots \\
			0 & 0 & \dots & \frac{g}{\varepsilon_N - z}
		\end{pmatrix}
	\end{align}
	to the determinant of $J$ is equivalent to the rank-1 update
	\begin{align}
		\det (J - \Lambda(z)) = \det (J - \textbf{x}(z) \textbf{1}^T)
	\end{align}
	in terms of the vectors
	\begin{align}
		\textbf{x}(z)^T &=
		\begin{pmatrix}
			\frac{g}{\varepsilon_1 - z} & \frac{g}{\varepsilon_2 - z} & \dots & \frac{g}{\varepsilon_N - z}
		\end{pmatrix} \\
		\textbf{1}^T &=
		\begin{pmatrix}
			1 & 1 & \dots & 1
		\end{pmatrix}.
	\end{align}
\end{lemma}
\begin{lemma} \label{lem:sum_rank_1}
	For $J$, an $N\times N$ invertible matrix, the sum of $M$ rank-1 updates of arbitrary vectors $\textbf{x}_{\alpha}$ and a specific vector $\textbf{y}^T$ is the single determinant
	\begin{align}
		\sum^M_{\alpha=1} \lambda_{\alpha} \det (J - \textbf{x}_{\alpha} \textbf{y}^T) = 
		\det 
		\begin{pmatrix}
			\sum^M_{\alpha = 1} \lambda_{\alpha} & \textbf{y}^T \\
			\sum^M_{\alpha = 1} \lambda_{\alpha} \textbf{x}_{\alpha} & J
		\end{pmatrix},
	\end{align}
	 where $\lambda_{\alpha}$ are scalars.
\end{lemma}

Since the local pair creators $S^+_k$ are the residues of the RG pair creators at the simple poles
\begin{align}
	S^+_k = \lim_{u \rightarrow \varepsilon_k} (u - \varepsilon_k) S^+(u),
\end{align}
the form factors are evaluated as the residues of the scalar product \eqref{eq:ebv_scalar}
\begin{align}
	\braket{\{v\}^M | S^+_k | \{u\}^M_{\alpha} } = \lim_{u_{\alpha} \rightarrow \varepsilon_k} 
	(u_{\alpha} - \varepsilon_k) \braket{\{v\}^M | \{u\}^M}. 
\end{align}
Notice that the simple pole $u_{\alpha} \rightarrow \varepsilon_k$ occurs in the $k$th diagonal element of $J$
\begin{align}
	\lim_{u_{\alpha} \rightarrow \varepsilon_k} (u_{\alpha} - \varepsilon_k) J_{kk} = -g
\end{align}
while the remaining diagonal elements ($p\neq k$) are modified
\begin{align}
	\lim_{u_{\alpha} \rightarrow \varepsilon_k} J_{pp} = J_{pp} 
	+ \frac{g}{\varepsilon_p - \varepsilon_k}
	- \frac{g}{\varepsilon_p - u_{\alpha}}.
\end{align}
The resulting determinant has a single non-zero element in the $k$th row and $k$th column, giving
\begin{align}
	\braket{\{v\}^M | S^+_k | \{u\}^M_{\alpha} } = - g \eta(0) 
	\det \left(J^{k,k} + \Lambda^{k,k}(\varepsilon_k) - \Lambda^{k,k} (u_{\alpha}) \right)
\end{align}
with $J^{k,k}$ the matrix $J$ without the $k$th row and $k$th column etc. Now lemma \ref{lem:rank1} can be used to reduce the update $\Lambda^{k,k}(u_{\alpha})$ to the rank-1 update $\textbf{x}^k(u_{\alpha}) \textbf{1}^T$, so that
\begin{align}
	\gamma^{VU}_{kk} = - 2 g \eta (0) \sum^M_{\alpha=1}
	\frac{\det \left( J^{k,k} + \Lambda^{k,k}(\varepsilon_k) - \textbf{x}^k(u_{\alpha}) \textbf{1}^T \right)}
	{u_{\alpha} - \varepsilon_k}.
\end{align}
Lemma \ref{lem:sum_rank_1} reduces this sum to the single determinant
\begin{align}
	\gamma^{VU}_{kk} = 2 \eta(0) \det
	\begin{pmatrix}
		\sum_{\alpha} \frac{g}{\varepsilon_k - u_{\alpha}} & \textbf{1}^T \\
		\sum_{\alpha} \frac{g \textbf{x}^k(u_{\alpha})}{\varepsilon_k - u_{\alpha}} & J^{k,k} + d^{k,k}(\varepsilon_k)
	\end{pmatrix}.
\end{align}
The (1,1) element of this determinant is $U_k$, while the other elements of the first column are
\begin{align}
	\sum_{\alpha} \frac{g}{(\varepsilon_p - u_{\alpha})} \frac{g}{(\varepsilon_k - u_{\alpha})}
	= -g \frac{U_k - U_p}{\varepsilon_k - \varepsilon_p}.
\end{align}
The ``damage'' to $J$ will be repaired with row operations. Add $\frac{g}{\varepsilon_k - \varepsilon_p}$ times the first row to each of $p$th rows to
\begin{enumerate}
	\item remove the factor of $g\frac{U_k}{\varepsilon_k - \varepsilon_p}$ from the first column
	\item remove the update $\Lambda^{k,k}(\varepsilon_k)$ to the diagonal elements of $J^{k,k}$
	\item scale the off-diagonal elements of $J^{k,k}$
	\begin{align}
		\frac{g}{\varepsilon_p - \varepsilon_q} + \frac{g}{\varepsilon_k - \varepsilon_p} =
		\frac{g}{(\varepsilon_p - \varepsilon_q)} \frac{(\varepsilon_k - \varepsilon_q)}{(\varepsilon_k - \varepsilon_p)}.
	\end{align}
\end{enumerate}
Remove the factor of $\frac{g}{\varepsilon_k - \varepsilon_p}$ from each of the $p$ rows (except the first) and the factor $\frac{\varepsilon_k - \varepsilon_q}{g}$ from each of the $j$ columns (except the first). These factors cancel one another. Finally, reorder the rows and columns such that (1,1) element is moved to the ($k,k$) position. No sign is introduced as this permutation involves the same number of row and column interchanges. The result 
\begin{align}
	\gamma^{VU}_{kk} = 2 \eta(0) \det J (k\rightarrow \textbf{U})
\end{align}
involves the original matrix $J$ whose $k$th column has been replaced by the vector of EBV $\textbf{U}$, which can be written as 
\begin{align}
	\gamma^{VU}_{kk} = 2 \eta(0) \mathcal{D}^{(0)}_{kk} (J)
\end{align}
in terms of a summation
\begin{align} \label{eq:jsum_d0_kk}
	\mathcal{D}^{(0)}_{kk} (J) = \sum^N_{p=1} U_p [J]^{p,k}
\end{align}
in $J$'s first cofactors
\begin{align}
	[J]^{p,q} = (-1)^{p+q} \det J^{p,q}.
\end{align}
$\mathcal{D}^{(0)}_{kk} (J)$ will be referred to as a \emph{J-sum}, and will reappear in seniority-conserving couplings in other seniority channels. The superscript refers to the change in seniority for the $J$-sum. The same approach, with \emph{many} tedious intermediate steps, follows for the elements of the 2-DM. \emph{Second} cofactors of the matrix $J$ 
\begin{align}
	[J]^{pq,rs} = (-1)^{p+q+r+s+h(p,q)+h(r,s)} \det J^{pq,rs}
\end{align}
are required. To be unambiguous, their definition involves the factor $h(p,q)$
\begin{align}
	h(p,q) = 
	\begin{cases}
		1 & p > q \\
		0 & p < q.
	\end{cases}
\end{align}
If only second cofactors are required, then a Heaviside function can replace $h(p,q)$, but this does not generalize directly. The correct observation is that cofactors are antisymmetric in their row or column indices. Third and fourth cofactors will be required to couple RG states with different seniorities, which can be defined in exactly the same manner. Third cofactors
\begin{align}
	[J]^{p_1 p_2 p_3,q_1 q_2 q_3} = (-1)^{p_1 + p_2 + p_3 + q_1 + q_2 + q_3 + h(p_1,p_2,p_3) + h(q_1,q_2,q_3)} \det J^{p_1 p_2 p_3,q_1 q_2 q_3}
\end{align}
are $N-3 \times N-3$ determinants with an appropriate sign, and fourth cofactors are
\begin{align}
	[J]^{p_1 p_2 p_3 p_4,q_1 q_2 q_3 q_4} = (-1)^{p_1 + p_2 + p_3 + p_4 + q_1 + q_2 + q_3 + q_4 + h(p_1,p_2,p_3,p_4) + h(q_1,q_2,q_3,q_4)} 
	\det J^{p_1 p_2 p_3 p_4,q_1 q_2 q_3 q_4}
\end{align}
are $N-4 \times N-4$ determinants with an appropriate sign. Given a set of indices in natural order $p_1 < \dots < p_k$, the indicator function $h(p_1,\dots,p_k)$ returns the value zero. If the indices are not in natural order, then $h(p_1,\dots,p_k)$ returns zero (one) if the indices can be ordered with an even (odd) permutation. 

The direct 2-DM elements are
\begin{align}
	\Gamma^{VU}_{kkll} = 4 \eta(0) \mathcal{D}^{(0)}_{kkll} (J)
\end{align}
in terms of the $J$-sum
\begin{align} \label{eq:clean_d2D_sum}
	\mathcal{D}^{(0)}_{kkll} (J) = 
	\sum^N_{p =1} \sum^N_{q =i+1} \frac{\mathfrak{p}^{kl}_{pq}}{\mathfrak{d}^{kl}_{pq}}
	K_{pq}[J]^{pq,kl}.
\end{align}
These elements require the factors 
\begin{align} \label{eq:kmat}
	K_{pq} = U_p U_q + g \frac{U_p - U_q}{\varepsilon_p - \varepsilon_q}
	= \det \begin{pmatrix}
		U_p + \frac{g}{\varepsilon_q - \varepsilon_p} & \frac{g}{\varepsilon_p - \varepsilon_q} \\
		\frac{g}{\varepsilon_q - \varepsilon_p} & U_q + \frac{g}{\varepsilon_p - \varepsilon_q}
	\end{pmatrix}
\end{align}
along with Cauchy determinants
\begin{align}
	\mathfrak{d}^{pq}_{rs} = \begin{vmatrix}
		\frac{1}{\varepsilon_p - \varepsilon_r} & \frac{1}{\varepsilon_p - \varepsilon_s} \\
		\frac{1}{\varepsilon_q - \varepsilon_r} & \frac{1}{\varepsilon_q - \varepsilon_s}
	\end{vmatrix}
	= \frac{(\varepsilon_p - \varepsilon_q)(\varepsilon_s - \varepsilon_r)}
	{(\varepsilon_p - \varepsilon_r)(\varepsilon_q - \varepsilon_r)(\varepsilon_p - \varepsilon_s)(\varepsilon_q - \varepsilon_s)}
\end{align}
and permanents
\begin{align}
	\mathfrak{p}^{pq}_{rs} = \begin{vmatrix}
		\frac{1}{\varepsilon_p - \varepsilon_r} & \frac{1}{\varepsilon_p - \varepsilon_s} \\
		\frac{1}{\varepsilon_q - \varepsilon_r} & \frac{1}{\varepsilon_q - \varepsilon_s}
	\end{vmatrix}^+
	= \frac{1}{(\varepsilon_p - \varepsilon_r)(\varepsilon_q - \varepsilon_s)} 
	+ \frac{1}{(\varepsilon_p - \varepsilon_s)(\varepsilon_q - \varepsilon_r)}.
\end{align}
The factors $K_{pq}$ must be computed for each state, but the Cauchy determinants and permanents can be computed and stored once. They are necessary for \emph{many} of the matrix elements between RG states. It is understood that
\begin{align}
	\frac{\mathfrak{p}^{pq}_{pq}}{\mathfrak{d}^{pq}_{pq}} =
	\frac{\mathfrak{p}^{ps}_{pq}}{\mathfrak{d}^{ps}_{pq}} =
	\frac{\mathfrak{p}^{rq}_{pq}}{\mathfrak{d}^{rq}_{pq}} =
	- \frac{\mathfrak{p}^{rp}_{pq}}{\mathfrak{d}^{rp}_{pq}} =
	- \frac{\mathfrak{p}^{qs}_{pq}}{\mathfrak{d}^{qs}_{pq}} = 1.
\end{align}
Equation \eqref{eq:clean_d2D_sum} is very clean and suggestive for higher ranks of correlation functions: the $z$th-order diagonal-correlation 
\begin{align}
	\Gamma^{VU}_{k_1 k_1 k_2 k_2 \dots k_z k_z} &= \braket{\{v\}^M | A^0_{k_1 k_1} A^0_{k_2 k_2} \dots A^0_{k_z k_z} |\{u\}^M} \\
	&= 2^z \eta (0) \sum_{p_1 < p_2 < \dots p_z} \frac{\mathfrak{p}^{k_1 k_2 \dots k_z}_{p_1 p_2 \dots p_z}}{\mathfrak{d}^{k_1 k_2 \dots k_z}_{p_1 p_2 \dots p_z}} K_{p_1 p_2 \dots p_z} [J]^{p_1 p_2 \dots p_z,k_1 k_2 \dots k_z}
\end{align}
should be directly expressible in terms of $z \times z$ Cauchy permanents and determinants, a $z \times z$ extension of \eqref{eq:kmat}, and $z$th cofactors of the matrix $J$. 

The pair-transfer elements are
\begin{align}
	\Gamma^{VU}_{klkl} = 2 \eta(0) \mathcal{P}^{(0)}_{klkl} (J)
\end{align}
in terms of the $J$-sum
\begin{align} \label{eq:clean_d2P_sum}
	\mathcal{P}^{(0)}_{klkl} (J) &= U_l \left(1 + \frac{\varepsilon_k - \varepsilon_l}{g} (U_l - J_{ll}) \right)[J]^{l,k}
	+ \sum^N_{p(\neq k,l)} \frac{\varepsilon_p - \varepsilon_k}{\varepsilon_p - \varepsilon_l} ([J]^{p,k} - 2 K_{pl} [J]^{pl,kl}) \nonumber \\
	&-2 \sum^N_{p < q (\neq k,l)} \frac{(\varepsilon_k - \varepsilon_p)(\varepsilon_k - \varepsilon_q)}
	{(\varepsilon_k - \varepsilon_l)(\varepsilon_q - \varepsilon_p)} K_{pq} [J]^{pq,kl}
\end{align}
Unfortunately, this expression does not lend itself to conjecture.

These expressions are equally valid for RDM and TDM elements, the only difference being in how $J$'s cofactors are evaluated. When the states are the same $J$ is invertible and the elements of the inverse \emph{are} the scaled cofactors. When the states are different, $J$ is necessarily singular, but the singular value decomposition still allows for an efficient computation that bypasses the direct construction.\cite{chen:2023,johnson:2024b}

With the seniority-zero DM elements understood, the procedure will be repeated for RG states of seniorities zero, two and four. In every case the matrix elements are computable from $J$-sums.

\section{Seniority-two} \label{sec:s2}
Considering states with seniorities zero and two leads to coupling between seniority-zero and seniority-two, as well as coupling between seniority-two states. 
\subsection{0 - 2 coupling}
The simplest case is the coupling of seniority-zero RG states with seniority-two RG states. The same approach will be taken, omitting the intermediate summation details, highlighting only the necessary modifications as they arise.

First the 1-electron elements, the only non-zero ones being
\begin{align}
	\gamma^{V,Uab}_{ab} &= \braket{ \{v\}^M | A^0_{ab} | \{u\}^{M-1},(ab) } \\
	\gamma^{V,Uab}_{ba} &= \braket{ \{v\}^M | A^0_{ba} | \{u\}^{M-1},(ab) }
\end{align}	
as all other elements will yield scalar products between states with distinct seniorities or distinct blocked levels. With
\begin{align}
	[A^0_{ab}, S^+(u)]   &= \frac{\sqrt{2} A^+_{ab}}{u - \varepsilon_b} \\
	[A^0_{ab}, A^+_{ab}] &= \sqrt{2} S^+_a \\
	A^+_{ab} A^+_{ab}    &= - S^+_a S^+_b,
\end{align}
$A^0_{ab}$ can be moved to the right until it destroys the vacuum $\ket{\theta}$ yielding
\begin{align} \label{eq:s02_1_ff}
	\gamma^{V,Uab}_{ab} = \sqrt{2} \braket{ \{v\}^M | S^+_a | \{u\}^{M-1}}
	- \sqrt{2} \sum^{M-1}_{\alpha = 1} \frac{ \braket{ \{v\}^M | S^+_a S^+_b | \{u\}^{M-1}_{\alpha} } }
		{u_{\alpha} - \varepsilon_b},
\end{align}
a sum involving only form factors of seniority-zero RG states. The scalar product could be computed in the opposite manner, leading to a sum of form factors of seniority-two RG states. The end result should be the same, and the present development is easier. The two contributions in \eqref{eq:s02_1_ff} represent two distinct physical processes. In the first, $A^0_{ab}$ acts on the open-shell singlet to make the closed-shell pair $S^+_a$. The second process is more complicated: the RG pair $S^+(u_{\alpha})$ is broken into the open-shell pair $A^+_{ab}$ which couples to the already present open-shell singlet to yield the closed shell $-S^+_a S^+_b$. If CSFs were used instead of RG states, this second process would not occur at all. This will be a common theme.

The set $\{v\}$ is on-shell in seniority-zero, but the set $\{u\}$ is \emph{not}. However, the determinant expression for the scalar product \eqref{eq:ebv_scalar} requires only one of the RG states to be on-shell and can thus still be used. In particular, by adding an aritifical rapidity $w$, the EBV scalar product \eqref{eq:ebv_scalar} can be used directly
\begin{align}
	\braket{ \{v\}^M | \{u\}^{M-1} \cup w} = \eta(0) \det (J + \Lambda(w)),
\end{align}
with only a relabelling of the diagonal elements. Since the EBV $\{U\}$ now involve only $M-1$ rapidities, the diagonal elements are
\begin{align}
	(J + \Lambda(w))_{pp} = V_p + U_p + \frac{g}{\varepsilon_p - w} - \frac{2}{g} + \sum^N_{q (\neq p) = 1} \frac{g}{\varepsilon_q - \varepsilon_p}.
\end{align}
As this determinant is for $\{v\}$ on-shell, it is $N \times N$ involving all the single particle levels, including those that are blocked. This means that the undefined EBV $U_a$ and $U_b$ appear but will vanish in the final expressions. It will be convenient to understand them as being equal to zero.

The required form factors are again evaluated as residues of the scalar product. Simple poles appear only in the diagonal elements, with residues
\begin{align}
	\lim_{w \rightarrow \varepsilon_a} (w - \varepsilon_a) (J + \Lambda(w))_{aa} = - g
\end{align}
while the other diagonal elements are modified
\begin{align}
	\lim_{w \rightarrow \varepsilon_a} (J + \Lambda(w))_{pp} = J_{pp} + \frac{g}{\varepsilon_p - \varepsilon_a}.
\end{align}
The form factors are thus
\begin{align}
	\braket{ \{v\}^M | S^+_a | \{u\}^{M-1} } &= - g \eta(0) \det (J^{a,a} + \Lambda^{a,a} (\varepsilon_a)) \\
	\braket{ \{v\}^M | S^+_a S^+_b | \{u\}^{M-1}_{\alpha} } &=
	g^2 \eta(0) \det (J^{ab,ab} + \Lambda^{ab,ab} (\varepsilon_a) + \Lambda^{ab,ab} (\varepsilon_b) 
	- \Lambda^{ab,ab} (u_{\alpha}) ).		
\end{align}	

The transition elements are
\begin{align}
	\gamma^{V,Uab}_{ab} = - g \sqrt{2} \eta(0) \mathcal{D}^{(2)}_{ab} (J)
\end{align}
which follows a similar development as for the seniority-zero 1-electron elements. Using lemmas \ref{lem:rank1} and \ref{lem:sum_rank_1}, and repairing the damage in exactly the same way, the required $J$-sum is
\begin{align}
	\mathcal{D}^{(2)}_{ab} (J) = \det (J^{a,a} + \Lambda^{a,a} (\varepsilon_a)) 
	- \det 
	\begin{pmatrix}
		U_b & \frac{g}{\varepsilon_b - \bm{\varepsilon} } \\
		\textbf{U} & J^{ab,ab} + \Lambda^{ab,ab} (\varepsilon_a)
	\end{pmatrix}
\end{align}
where the first row in the second determinant involves all the single particle energies except $\varepsilon_a$ and $\varepsilon_b$, while the vector $\textbf{U}$ is all the EBV except $U_a$ and $U_b$. This result may be simplified: in the first determinant, $U_b$ appears in the $b$th diagonal element, while it only appears in the top left element of the second determinant. In both cases, the weight is $J^{ab,ab} + \Lambda^{ab,ab}(\varepsilon_a)$ occuring with opposite signs, erasing any dependence on $U_b$. Such a cancellation will occur for all the matrix elements in this manuscript. As $U_b$ gives no contribution, any choice can be made for its value, and it is convenient to choose $U_b=0$: the final expressions will involve only cofactors of the matrix $J$. $U_a$ does not appear at all as only the submatrix $J^{a,a}$ is involved. The same arguments applied to $\mathcal{D}^{(2)}_{ba} (J)$ allows the choice $U_a=0$ to be made. 

To repair the damage $\Lambda^{a,a}(\varepsilon_a)$, the first determinant can be lifted to an $N\times N$ determinant with a first row of $N$ 1s, and the remainder of the first column of zeros. Likewise, the second determinant should be extended with a row of 1s \emph{except} with a 0 in the second column, and a first column of zeros. $\frac{g}{\varepsilon_a - \varepsilon_p}$ times the first row is added to each of the other rows, and the rows and columns are again scaled. Finally, the first determinant is expanded along the first column, and the second determinant is expanded on the first two columns to yield
\begin{align} \label{eq:jsum_d2_ab}
	\mathcal{D}^{(2)}_{ab} (J) = \sum^N_{p=1} [J]^{p,a} + \sum^N_{p = 1} 
	\frac{\varepsilon_a - \varepsilon_p}{\varepsilon_a - \varepsilon_b} U_p 
	\sum^N_{q = 1}[J]^{pq,ab}
\end{align}
a sum involving only cofactors of the matrix $J$, it being understood that $U_a = U_b =0$.

2-electron couplings are possible that involve both the blocked levels and 1 repeated index. If the repeated index is one of the blocked levels, the non-zero elements are
\begin{align}
	\Gamma^{V,Uab}_{babb} &= \Gamma^{V,Uab}_{bbba} = \gamma^{V,Uab}_{ba} \\
	\Gamma^{V,Uab}_{aaab} &= \Gamma^{V,Uab}_{abaa} = \gamma^{V,Uab}_{ab},
\end{align}
while $\Gamma^{V,Uab}_{bbab} = \Gamma^{V,Uab}_{abbb} = \Gamma^{V,Uab}_{aaba} = \Gamma^{V,Uab}_{baaa} = 0$. There are three types of elements when the repeated index is not one of the blocked levels. First, there are direct elements, which expanded in form factors become
\begin{align}
	\Gamma^{V,Uab}_{kkab} &= 2 \sqrt{2} \sum^{M-1}_{\alpha=1} 
	\frac{\braket{ \{v\}^M | S^+_k S^+_a | \{u\}^{M-1}_{\alpha} }}{u_{\alpha} - \varepsilon_k}
	- 2 \sqrt{2} \sum_{\alpha < \beta} \mathfrak{p}^{\alpha \beta}_{kb}
	\braket{ \{v\}^M | S^+_k S^+_a S^+_b | \{u\}^{M-1}_{\alpha,\beta} }.
\end{align}
Here, a Cauchy permanent with greek indices refers to
\begin{align}
	\mathfrak{p}^{\alpha \beta}_{rs} = \begin{vmatrix}
		\frac{1}{u_{\alpha} - \varepsilon_r} & \frac{1}{u_{\alpha} - \varepsilon_s} \\
		\frac{1}{u_{\beta}  - \varepsilon_r} & \frac{1}{u_{\beta}  - \varepsilon_s}
	\end{vmatrix}^+
	= \frac{1}{(u_{\alpha} - \varepsilon_r)(u_{\beta} - \varepsilon_s)} 
	+ \frac{1}{(u_{\alpha} - \varepsilon_s)(u_{\beta} - \varepsilon_r)}.
\end{align}
An expression in terms of EBV is obtainable following the approach for the direct elements in appendix B of ref. \citenum{faribault:2022}, then extending the resulting determinants to $N \times N$ as above. This procedure is \emph{incredibly tedious} but more or less straightforward following the procedures outlined in ref. \citenum{faribault:2022}. This will be the case for the remainder of the $J$-sums in this manuscript so that only the final results will be reported. Here, 
\begin{align}
	\Gamma^{V,Uab}_{kkab} = - g 2 \sqrt{2} \eta(0) \mathcal{D}^{(2)}_{kkab} (J) 
\end{align}
with
\begin{align} 
	\mathcal{D}^{(2)}_{kkab} (J) &= \sum^N_{p (\neq a,b) = 1} 
	\frac{\varepsilon_a - \varepsilon_p}{\varepsilon_a - \varepsilon_k} U_p
	\left(
	\sum^N_{l = 1} [J]^{pq,ka} + \frac{g}{\varepsilon_b - \varepsilon_p} \sum^N_{q = 1}
	[J]^{pqb,kab}
	\right) \nonumber \\
	&+ \sum^N_{p (\neq k,a,b) = 1} \frac{\varepsilon_a - \varepsilon_p}{\varepsilon_a - \varepsilon_b} K_{kp}
	\sum^N_{q =1} [J]^{kpq,kab} \nonumber \\
	&+ \sum_{p<q (\neq k,a,b)} 
	\frac{(\varepsilon_a - \varepsilon_p)(\varepsilon_a - \varepsilon_q)}
	     {(\varepsilon_a - \varepsilon_k)(\varepsilon_a - \varepsilon_b)}
	\frac{\mathfrak{p}^{kb}_{pq}}{\mathfrak{d}^{kb}_{pq}} K_{pq}
	\sum^N_{r=1} [J]^{pqr,kab}, \label{eq:direct_sum_02}
	\end{align}
an expression involving only the cofactors of the common matrix $J$. Once again, the EBV for the blocked levels $U_a$ and $U_b$ do not appear in the final expression, so that when interpreted $U_a=U_b=0$, the $J$-sum can be cleaned up substantially:
\begin{align} \label{eq:clean_direct_sum_02}
	\mathcal{D}^{(2)}_{kkab} &= \sum^N_{p=1} \frac{(\varepsilon_a - \varepsilon_p)}{(\varepsilon_a - \varepsilon_k)} U_p \sum^N_{q=1} [J]^{pq,ka}
	+ \sum_{p<q} \frac{(\varepsilon_a - \varepsilon_p)(\varepsilon_a - \varepsilon_q)}
	{(\varepsilon_a - \varepsilon_k)(\varepsilon_a - \varepsilon_b)}
	\frac{\mathfrak{p}^{kb}_{pq}}{\mathfrak{d}^{kb}_{pq}} K_{pq} \sum^N_{r=1} [J]^{pqr,kab}.
\end{align}
As it is understood that $U_b=0$,
\begin{align}
	K_{pb} = \frac{g U_p}{\varepsilon_p - \varepsilon_b}.
\end{align}
The corresponding \emph{exchange} elements for the unblocked levels are simply
\begin{align}
	\Gamma^{V,Uab}_{kbak} = - \frac{1}{2} \Gamma^{V,Uab}_{kkab}.
\end{align}
The other direct element $\Gamma^{V,Uab}_{kkba}$ is obtained by exchanging the roles of $a$ and $b$ in equation \eqref{eq:clean_direct_sum_02}.

The second type of elements are \emph{pair-forming} 
\begin{align}
	\Gamma^{V,Uab}_{kakb} &= \sqrt{2} \braket{ \{v\}^M | S^+_k | \{u\}^{M-1} } + \sqrt{2} \sum_{\alpha < \beta} \mathfrak{p}^{\alpha \beta}_{ab} 
	\braket{ \{v\}^M | S^+_k S^+_a S^+_b | \{u\}^{M-1}_{\alpha,\beta} } \nonumber \\
	&- \sqrt{2} \sum^M_{\alpha = 1} \frac{\braket{ \{v\}^M | S^+_k S^+_a | \{u\}^{M-1}_{\alpha} }}{u_{\alpha} - \varepsilon_a}
	 - \sqrt{2} \sum^M_{\alpha = 1} \frac{\braket{ \{v\}^M | S^+_k S^+_b | \{u\}^{M-1}_{\alpha} }}{u_{\alpha} - \varepsilon_b},
\end{align}
so-called since
\begin{align}
	A^0_{ka} A^0_{kb} = \sqrt{2} S^+_k A^-_{ab}.
\end{align}
The result of the cofactor summation gives
\begin{align}
	\Gamma^{V,Uab}_{kakb} = - g \sqrt{2} \eta(0) \mathcal{P}^{(2)}_{kakb} (J)
\end{align}
with
\begin{align}
	\mathcal{P}^{(2)}_{kakb} (J) &= \sum^N_{p=1} [J]^{p,k} \nonumber \\
	&+ \sum^N_{p (\neq k,a,b)=1} \frac{\varepsilon_k - \varepsilon_p}{\varepsilon_k - \varepsilon_a} U_p
	\left(
	\sum^N_{q (\neq p)=1} [J]^{pq,ka} + \frac{g}{\varepsilon_b - \varepsilon_p}\sum^N_{q (\neq p,b)=1} [J]^{pqb,kab}
	\right) \nonumber \\
	&+ \sum^N_{p (\neq k,a,b)=1} \frac{\varepsilon_k - \varepsilon_p}{\varepsilon_k - \varepsilon_b} U_p
	\left(
	\sum^N_{q (\neq p)=1} [J]^{pq,kb} + \frac{g}{\varepsilon_a - \varepsilon_p}\sum^N_{q (\neq p,a)=1} [J]^{paq,kab}
	\right) \nonumber \\
	&+ \sum_{p<q (\neq k,a,b)} 
	\frac{(\varepsilon_k - \varepsilon_p)(\varepsilon_k - \varepsilon_q)}
	     {(\varepsilon_k - \varepsilon_a)(\varepsilon_k - \varepsilon_b)}
	\frac{\mathfrak{p}^{ab}_{pq}}{\mathfrak{d}^{ab}_{pq}} K_{pq}
	\sum^N_{r (\neq p,q)=1} [J]^{pqr,kab}. 
\end{align}
Again, by interpreting $U_a = U_b = 0$, the $J$-sum can be cleaned up to give
\begin{align} \label{eq:jsum_p2_kakb}
	\mathcal{P}^{(2)}_{kakb} (J) &= \sum^N_{p=1} [J]^{p,k} 
	+ \sum^N_{p=1} (\varepsilon_k - \varepsilon_p) U_p \sum^N_{q=1}
	\left(
	\frac{[J]^{pq,ka}}{(\varepsilon_k - \varepsilon_a)} + \frac{[J]^{pq,kb}}{(\varepsilon_k - \varepsilon_b)}
	\right) \nonumber \\
	&+ \sum_{p<q} \frac{(\varepsilon_k - \varepsilon_p)(\varepsilon_k - \varepsilon_q)}
	{(\varepsilon_k - \varepsilon_a)(\varepsilon_k - \varepsilon_b)}
	\frac{\mathfrak{p}^{ab}_{pq}}{\mathfrak{d}^{ab}_{pq}} K_{pq} \sum^N_{r=1} [J]^{pqr,kab}
\end{align}
where the summations over the indices $p,q,r$ are complete.

Finally there are \emph{pair-breaking} elements
\begin{align}
	\Gamma^{V,Uab}_{akbk} &= - \sqrt{2} \sum^{M-1}_{\alpha=1} 
	\frac{\braket{ \{v\} | S^+_a S^+_b | \{u\}^{M-1}_{\alpha} } }{u_{\alpha} - \varepsilon_k}
	+ \sqrt{2} \sum_{\alpha < \beta} \mathfrak{p}^{\alpha \beta}_{kk}
	\braket{ \{v\}^M | S^+_k S^+_a S^+_b | \{u\}^{M-1}_{\alpha,\beta} },
\end{align}
since
\begin{align}
	A^0_{ak} A^0_{bk} = \sqrt{2} A^+_{ab} S^-_k.
\end{align}
These elements only occur from breaking at least one RG pair, which is forbidden for CSFs. The form factor summation follows the same lines as that of the pair-transfer elements in appendix B of ref. \citenum{faribault:2022}, with the similar modfications as above. The result is
\begin{align}
	\Gamma^{V,Uab}_{akbk} = - g \sqrt{2} \eta(0) \mathcal{P}^{(2)}_{akbk} (J)
\end{align}
with
\begin{align} 
	\mathcal{P}^{(2)}_{akbk} (J) &= 
	\frac{\varepsilon_a - \varepsilon_k}{\varepsilon_a - \varepsilon_b} U_k 
	\left(
	1 + \frac{\varepsilon_b - \varepsilon_k}{g} U_k
	\right)
	\sum^N_{p =1} [J]^{kp,ab} \nonumber \\
	&+ \sum^N_{p (\neq k)=1} \left(
	  \frac{\varepsilon_b - \varepsilon_k}{\varepsilon_p - \varepsilon_k}U_k 
	+ \frac{\varepsilon_b - \varepsilon_p}{\varepsilon_k - \varepsilon_p}U_p 
	\right)
	\frac{\varepsilon_a - \varepsilon_p}{\varepsilon_a - \varepsilon_b} \sum^N_{q =1} [J]^{pq,ab} \nonumber \\
	& + \frac{2}{(\varepsilon_a - \varepsilon_k)(\varepsilon_b - \varepsilon_k)}
	\sum_{p<q (\neq a,b)} \frac{K_{pq}}{\mathfrak{d}^{ab}_{pq}} \sum^N_{r =1} [J]^{pqr,kab}. \label{eq:jsum_p2_akbk}
\end{align}
Here, a $J$-sum is encountered that cannot be written as clean complete summations. The $p=k$ term in the second row is indeterminate, while the restriction on the double summation in the last row could be relaxed as $\frac{1}{\mathfrak{d}^{ab}_{ia}}=\frac{1}{\mathfrak{d}^{ab}_{ib}}=0$, but this is dangerous if we compute and store the complete list of $\mathfrak{d}^{pq}_{rs}$. This expression does not appear to be symmetric in $a$ and $b$, though numerically it is.

For each pair of states $\ket{\{v\}^M}$ and $\ket{\{u\}^{M-1},(ab)}$, the required information for the matrix elements involves the single particle energies $\{\varepsilon\}$, the pairing strength $g$, the EBV for each state and the cofactors of the matrix $J$. 

\subsection{2 - 2 coupling}
Seniority-two RG states can couple with each other in many different ways, depending on how many blocked levels are shared between the two states.

\subsubsection{2 shared blocked levels}
If both states have the same blocked levels, i.e. the states are $\ket{\{v\}^{M-1},(ab)}$ and $\ket{\{u\}^{M-1},(ab)}$, the couplings are the same as those between seniority-zero RG states with minor modifications. The scalar product between the two states is
\begin{align}
	\braket{\{v\}^{M-1},(ab) | \{u\}^{M-1},(ab)} = \eta(2) \det (J(ab))
\end{align}  
where $J(ab)$ is the $N \times N$ matrix \eqref{eq:matsuno_2}. Again, the single particle energies $\varepsilon_a$ and $\varepsilon_b$ do not contribute to any of the diagonal elements. DM elements for unblocked indices follow directly
\begin{align}
	\gamma^{Vab,Uab}_{kk} &= 2 \eta(2) \mathcal{D}^{(0)}_{kk} (J(ab))     \\
	\Gamma^{Vab,Uab}_{kkll} &= 4 \eta(2) \mathcal{D}^{(0)}_{kkll} (J(ab)) \\
	\Gamma^{Vab,Uab}_{klkl} &= 2 \eta(2) \mathcal{P}^{(0)}_{klkl} (J(ab))
\end{align}
in terms of the $J$-sums \eqref{eq:jsum_d0_kk}, \eqref{eq:clean_d2D_sum} and \eqref{eq:clean_d2P_sum} evaluated with $J(ab)$ rather than $J$. For $J(ab)$, the only non-zero cofactors involving $a$ and $b$ are $[J]^{a,a}$, $[J]^{b,b}$ and $[J]^{ab,ab}$, which do not appear in these $J$-sums. Exchange elements in the unblocked levels remain the same
\begin{align}
	\Gamma^{Vab,Uab}_{kllk} = - \frac{1}{2} \Gamma^{Vab,Uab}_{kkll}.
\end{align}
It remains to specify the elements involving blocked levels. First, by a direct evaluation one can verify that
\begin{align}
	\gamma^{Vab,Uab}_{aa} = \gamma^{Vab,Uab}_{bb} = \Gamma^{Vab,Uab}_{aabb} = \Gamma^{Vab,Uab}_{abba} = 
	\braket{ \{v\}^{M-1} | \{u\}^{M-1} },
\end{align}
where the scalar product on the right is zero if the two states are distinct. Otherwise, the scalar product becomes the norm and these elements are 1. Notice that the direct $\Gamma^{Vab,Uab}_{aabb}$ and exchange $\Gamma^{Vab,Uab}_{abba}$ elements are equal. A common theme is that \emph{exchange elements involving blocked levels are not intuitive}. The diagonal and pair-transfer elements in the blocked levels
\begin{align}
	\Gamma^{Vab,Uab}_{aaaa} = \Gamma^{Vab,Uab}_{bbbb} = \Gamma^{Vab,Uab}_{abab} = \Gamma^{Vab,Uab}_{baba} =
	\Gamma^{Vab,Uab}_{akak} = \Gamma^{Vab,Uab}_{kaka} = 0
\end{align}
all vanish, as one cannot remove nor create a pair of electrons in a singly-occupied spatial orbital. Finally, there are direct and exchange elements involving one blocked level
\begin{align}
	\Gamma^{Vab,Uab}_{aakk} = -2 \Gamma^{Vab,Uab}_{akka} = \gamma^{Vab,Uab}_{kk}.
\end{align}

\subsubsection{1 shared blocked level}
For two seniority-two RG states that share one blocked level, $\ket{\{v\}^{M-1},(ab)}$ and $\ket{ \{u\}^{M-1},(ac)}$, there are 1- and 2-electron DM elements. The 1-electron elements involve the indices of the unshared blocked levels. The \emph{forward} scattering element
\begin{align}
	\gamma^{Vab,Uac}_{bc} &= \braket{ \{v\}^{M-1}, (ab) | \{u\}^{M-1}, (ab) } - \sum^{M-1}_{\alpha=1} 
	\frac{\braket{ \{v\}^{M-1}, (ab) | S^+_c | \{u\}^{M-1}_{\alpha},(ab) }}{u_{\alpha} - \varepsilon_c} \label{eq:22_f} \\
	&= \eta(2) \mathcal{F}^{(0)}_{bc} (J(ab))	
\end{align}
is the one expected for CSFs. The two contributions in \eqref{eq:22_f} represent two different physical processes. In the first, the electron in the blocked level $c$ is transferred to blocked level $b$, hence the name forward scattering. In the second process, an RG pair is broken and replaced with an open-shell singlet pair $A^+_{bc}$, which couples to $A^+_{ac}$ to yield a closed-shell singlet $S^+_c$ and an open-shell singlet $A^+_{ab}$. The form factor summation follows along the same lines
\begin{align} \label{eq:jsum_f0_bc}
	\mathcal{F}^{(0)}_{bc} (J(ab))	 = \det(J(ab)) - \sum^N_{p =1} U_p [J(ab)]^{p,c}.
\end{align}
Notice again that the undefined $U_c$ gives no contribution: both terms produce $U_c [J(ab)]^{c,c}$ but with opposite signs, so it is convenient to set $U_c=0$. The $p=a$ and $p=b$ terms give no contribution since the cofactors $[J(ab)]^{a,c}=[J(ab)]^{b,c}=0$ vanish. A final common theme is that forward scattering elements are reducible, in this case with the $J$-sum \eqref{eq:jsum_d0_kk}
\begin{align}
	\mathcal{F}^{(0)}_{bc} (J(ab)) = \det(J(ab)) - \mathcal{D}^{(0)}_{cc} (J(ab)).
\end{align}
It is not difficult to verify that the direct and exchange elements involving $a$ indexed twice are
\begin{align}
	\Gamma^{Vab,Uac}_{aabc} = \Gamma^{Vab,Uac}_{acba} = \gamma^{Vab,Uac}_{bc},
\end{align}
while the corresponding elements with $b$ or $c$ listed twice are zero. 

The \emph{backward} scattering element
\begin{align}
	\gamma^{Vab,Uac}_{cb} &= - \sum^{M-1}_{\alpha=1} \frac{\braket{ \{v\}^{M-1},(ab) | S^+_c | \{u\}^{M-1},(ab) }}
	{u_{\alpha} - \varepsilon_b} \\
	&= \eta (2) \mathcal{B}^{(0)}_{cb} (J(ab))	
\end{align}
is usually much smaller numerically as it can only occur through breaking an RG pair to create an open-shell singlet, like the second contribution in \eqref{eq:22_f}. This element vanishes for CSFs. Again, the form factor summation presents no difficulty
\begin{align}
	\mathcal{B}^{(0)}_{cb} (J(ab))	 = - \sum^N_{p (\neq b)=1}
	\left(
	\frac{\varepsilon_b - \varepsilon_c}{\varepsilon_b - \varepsilon_p} U_b +
	\frac{\varepsilon_c - \varepsilon_p}{\varepsilon_b - \varepsilon_p} U_p
	\right)
	[J(ab)]^{p,c}, \label{eq:jsum_b0_cb}
\end{align}
and again there is no dependance on $U_c$. The $p=b$ term is indeterminate, and should be interpreted as zero, but it is safer to exclude it from the summation. The backward scattering elements are irreducible. In the blocked levels, there are direct elements
\begin{align}
	\Gamma^{Vab,Uac}_{aacb} = \Gamma^{Vab,Uac}_{bbcb} = \Gamma^{Vab,Uac}_{cccb} = \gamma^{Vab,Uac}_{cb}.
\end{align}
There are also exchange elements
\begin{align}
	\Gamma^{Vab,Uac}_{abca} = - 2 \gamma^{Vab,Uac}_{cb}
\end{align}
which again behave in a counter-intuitive manner.

There are distinct direct forward $\Gamma^{Vab,Uac}_{kkbc}$ and direct backward $\Gamma^{Vab,Uac}_{kkcb}$ elements, with corresponding exchange elements
\begin{align}
	\Gamma^{Vab,Uac}_{kcbk} &= - \frac{1}{2} \Gamma^{Vab,Uac}_{kkbc} \\
	\Gamma^{Vab,Uac}_{kbck} &= - \frac{1}{2} \Gamma^{Vab,Uac}_{kkcb}.
\end{align}
The direct forward elements are
\begin{align}
	\Gamma^{Vab,Uac}_{kkbc} &= 2 \sum^{M-1}_{\alpha=1} \frac{\braket{ \{v\}^{M-1},(ab) | S^+_k | \{u\}^{M-1}_{\alpha},(ab)}}{u_{\alpha} - \varepsilon_k} \nonumber \\
	&-2 \sum_{\alpha < \beta} \mathfrak{p}^{\alpha \beta}_{kc} 
	\braket{ \{v\}^{M-1},(ab) | S^+_k S^+_c | \{u\}^{M-1}_{\alpha,\beta} ,(ab) } \\
	&= 2 \eta(2) \mathcal{F}^{(0)}_{kkbc} (J(ab)).
\end{align}
The form factor summation follows the same steps as the direct elements in appendix B of ref. \citenum{faribault:2022}. The dependance on $U_c$ once again vanishes, though in a more elaborate manner. The final result
\begin{align}
	\mathcal{F}^{(0)}_{kkbc} (J(ab)) &= \sum^N_{p (\neq a,b,c)} U_p 
	\left(
	[J(ab)]^{p,k} - \frac{g}{\varepsilon_p - \varepsilon_c} [J(ab)]^{pc,kc}
	\right) \nonumber \\
	&- \sum^N_{p (\neq k,a,b,c)=1} K_{kp} [J(ab)]^{kp,kc} - \sum_{p<q (\neq k,a,b,c)} 
	\frac{\mathfrak{p}^{kc}_{pq}}{\mathfrak{d}^{kc}_{pq}} K_{pq} [J(ab)]^{pq,kc} \label{eq:jsum_f0_kkbc} \\
	&= \mathcal{D}^{(0)}_{kk} (J(ab)) - \mathcal{D}^{(0)}_{kkcc} (J(ab)) 
\end{align}
is reducible in terms of the $J$-sums \eqref{eq:jsum_d0_kk} and \eqref{eq:clean_d2D_sum}. 

The direct backward elements 
\begin{align}
	\Gamma^{Vab,Uac}_{kkcb} &= -2 \sum_{\alpha < \beta} \mathfrak{p}^{\alpha \beta}_{kb}
	\braket{ \{v\}^{M-1},(ab) | S^+_k S^+_c | \{u\}^{M-1}_{\alpha,\beta},(ab) } \\
	&= -2 \eta(2) \mathcal{B}^{(0)}_{kkcb} (J(ab))
\end{align}
have no dependance on $U_c$ at all
\begin{align}
	\mathcal{B}^{(0)}_{kkcb} (J(ab)) &=
	\sum^N_{p (\neq k,b)=1} \left(
	\frac{\varepsilon_c - \varepsilon_p}{\varepsilon_b - \varepsilon_p}K_{kp} +
	\frac{\varepsilon_c - \varepsilon_b}{\varepsilon_p - \varepsilon_b}K_{kb}
	\right) [J(ab)]^{kp,kc} \nonumber \\
	&+ \sum^N_{p (\neq k,b,c)=1} \frac{1}{\varepsilon_k - \varepsilon_p} \frac{K_{pb}}{\mathfrak{d}^{kc}_{pb}}
	\sum^N_{q (\neq b)=1} \frac{[J(ab)]^{pq,kc}}{\varepsilon_q - \varepsilon_b}
	+ \sum_{p<q (\neq k,b,c)} \frac{\mathfrak{p}^{kb}_{pq}}{\mathfrak{d}^{kc}_{pq}} K_{pq} [J(ab)]^{pq,kc}. \label{eq:jsum_b0_kkcb}
\end{align}
While the summation on the first line can be absorbed into the summations on the second line, it is not productive as in each case the result would be an indeterminate form that reduces to the written expression. Thus, again, the backward scattering element is irreducible and much less clean than the corresponding forward scattering element.

The pair-breaking elements
\begin{align}
	\Gamma^{Vab,Uac}_{bkck} &= - \sum^{M-1}_{\alpha = 1} \frac{\braket{ \{v\}^{M-1},(ab) | S^+_c | \{u\}^{M-1}_{\alpha},(ab) }}{u_{\alpha} - \varepsilon_k} \nonumber \\
	&+ 2 \sum_{\alpha < \beta} 
	\frac{\braket{ \{v\}^{M-1},(ab) | S^+_k S^+_c | \{u\}^{M-1}_{\alpha,\beta},(ab) }}
	{(u_{\alpha}-\varepsilon_k)(u_{\beta}-\varepsilon_i)} \\
	&= \eta(2) \mathcal{P}^{(0)}_{ckbk} (J(ab))
\end{align}
are computed from the $J$-sum
\begin{align}
	\mathcal{P}^{(0)}_{ckbk} (J(ab)) &= - U_k \left(
	1 + \frac{\varepsilon_c - \varepsilon_k}{g} (U_i - J(ab)_{kk})
	\right) [J(ab)]^{i,c} \nonumber \\
	&- \sum^N_{p (\neq k,a,b,c)=1} \frac{\varepsilon_c - \varepsilon_p}{\varepsilon_k - \varepsilon_p}
	\left(
	U_p [J(ab)]^{p,c} - 2 K_{kp} [J(ab)]^{pk,ck}
	\right) \nonumber \\
	&+ 2 \sum_{p<q (\neq k,a,b,c)}
	\frac{(\varepsilon_c - \varepsilon_p)(\varepsilon_c - \varepsilon_q)}
	{(\varepsilon_c - \varepsilon_k)(\varepsilon_q - \varepsilon_p)} K_{pq} [J(ab)]^{pq,ck} \label{eq:jsum_p0_ckbk} \\
	&= \mathcal{P}^{(0)}_{ckck}(J(ab)).
\end{align}
This $J$-sum is reducible to \eqref{eq:clean_d2P_sum} since all the cofactors with $a$ and $b$ that appear in the summations will vanish. 

Finally, the pair-forming elements
\begin{align}
	\Gamma^{Vab,Uac}_{kbkc} &= - \sum^{M-1}_{\alpha=1} \frac{\braket{ \{v\}^{M-1},(ab) | S^+_k | \{u\}^{M-1}_{\alpha},(ab) }}{u_{\alpha}-\varepsilon_b} \nonumber \\
	&+ \sum_{\alpha < \beta} \mathfrak{p}^{\alpha \beta}_{bc} \braket{ \{v\}^{M-1},(ab) | S^+_k S^+_c | \{u\}^{M-1}_{\alpha},(ab)} \\
	&= \eta(2) \mathcal{P}^{(0)}_{kbkc} (J(ab))
\end{align}
with $J$-sum
\begin{align}
	\mathcal{P}^{(0)}_{kbkc} (J(ab)) &= 
	- \sum^N_{p (\neq b,c)=1}
	\left(
	\frac{\varepsilon_b - \varepsilon_k}{\varepsilon_b - \varepsilon_p} U_b +
	\frac{\varepsilon_k - \varepsilon_p}{\varepsilon_b - \varepsilon_p} U_p
	\right)
	\left(
	[J(ab)]^{p,k} - \frac{g}{\varepsilon_p - \varepsilon_c} [J(ab)]^{pc,kc}
	\right) \nonumber \\
	&- \sum^N_{p (\neq k,b,c)=1} \frac{1}{\varepsilon_p - \varepsilon_c} \frac{K_{pb}}{\mathfrak{d}^{kc}_{pb}}
	\sum^N_{q (\neq b)=1} \frac{[J(ab)]^{pq,kc}}{\varepsilon_q - \varepsilon_b} 
	+ \sum_{p<q (\neq k,b,c)} \frac{\mathfrak{p}^{bc}_{pq}}{\mathfrak{d}^{kc}_{pq}} K_{pq} [J(ab)]^{pq,kc}, \label{eq:jsum_p0_kbkc}
\end{align}
are irreducible. This expression is clearly asymmetric in terms of the indices $b$ and $c$ as $U_c=0$, but $U_b \neq 0$. To be coherent, the blocked index listed first (here $b$) should refer to level that is blocked in the left state but unblocked in the right state.

Again, in all cases all that is required is $\{\varepsilon\}$, the EBV for each state, and cofactors of the matrix $J(ab)$.

\subsubsection{No shared blocked levels}
Finally, if there are no common blocked levels between the two seniority-two RG states $\ket{\{v\}^{M-1},(ab)}$ and $\ket{ \{u\}^{M-1},(cd)}$, there are three types of 2-electron transition elements. In the first type, the electrons in the blocked levels both scatter forward
\begin{align}
	\Gamma^{Vab,Ucd}_{acbd} &= \braket{ \{v\}^{M-1},(ab) | \{u\}^{M-1},(ab) } \nonumber \\
	&- \sum^{M-1}_{\alpha = 1} \frac{ \braket{ \{v\}^{M-1},(ab) | S^+_c | \{u\}^{M-1}_{\alpha},(ab) } }{u_{\alpha} - \varepsilon_c}
	 - \sum^{M-1}_{\alpha = 1} \frac{ \braket{ \{v\}^{M-1},(ab) | S^+_d | \{u\}^{M-1}_{\alpha},(ab) } }{u_{\alpha} - \varepsilon_d} \nonumber \\
	&+ \sum_{\alpha < \beta} \mathfrak{p}^{\alpha \beta}_{cd} \braket{ \{v\}^{M-1},(ab) | S^+_c S^+_d | \{u\}^{M-1}_{\alpha,\beta},(ab) } \\
	&= \eta(2) \mathcal{F}^{(0)}_{acbd} (J(ab))
\end{align}
with a $J$-sum
\begin{align}
	\mathcal{F}^{(0)}_{acbd} (J(ab)) &= \det (J(ab)) 
	- \sum^N_{p (\neq a,b,c,d)=1} U_p ([J(ab)]^{p,c} + [J(ab)]^{p,d}) \nonumber \\
	&+ \sum^N_{p (\neq a,b,c,d)=1} U_p 
	\left(
	\frac{g}{\varepsilon_p - \varepsilon_c}[J(ab)]^{cp,cd} + \frac{g}{\varepsilon_p - \varepsilon_d}[J(ab)]^{pd,cd}
	\right) \nonumber \\
	&+ \sum_{p<q (\neq a,b,c,d)} \frac{\mathfrak{p}^{cd}_{pq}}{\mathfrak{d}^{cd}_{pq}} K_{pq} [J(ab)]^{pq,cd} \label{eq:jsum_f0_acbd} \\
	&= \det (J(ab)) - \mathcal{D}^{(0)}_{cc} (J(ab)) - \mathcal{D}^{(0)}_{dd} (J(ab)) + \mathcal{D}^{(0)}_{ccdd} (J(ab))
\end{align}
that is reducible to \eqref{eq:jsum_d0_kk} and \eqref{eq:clean_d2D_sum} since the $a$ and $b$ cofactors will vanish, $U_c=U_d=0$ and $K_{pc} = \frac{g U_p}{\varepsilon_p - \varepsilon_c}$ etc. The scattering targets do not matter, i.e. these elements are symmetric with exchange
\begin{align}
	\Gamma^{Vab,Ucd}_{adbc} &= \Gamma^{Vab,Ucd}_{acbd}.
\end{align}

The second type of element occurs in ``direct/exchange'' pairs
\begin{align}
	\Gamma^{Vab,Ucd}_{abcd} &= - 2 \Gamma^{Vab,Ucd}_{adcb} \\
	\Gamma^{Vab,Ucd}_{bacd} &= - 2 \Gamma^{Vab,Ucd}_{bdca} \\
	\Gamma^{Vab,Ucd}_{abdc} &= - 2 \Gamma^{Vab,Ucd}_{acdb} \\
	\Gamma^{Vab,Ucd}_{badc} &= - 2 \Gamma^{Vab,Ucd}_{bcda}	
\end{align}
which have distinct interpretations. In the ``direct'' elements, e.g.
\begin{align}
	\Gamma^{Vab,Ucd}_{abcd} &= 2 \sum^{M-1}_{\alpha = 1} \frac{\braket{ \{v\}^{M-1},(ab) | S^+_c | \{u\}^{M-1}_{\alpha},(ab) }}{u_{\alpha} - \varepsilon_b} \nonumber \\
	&- 2 \sum_{\alpha<\beta} \mathfrak{p}^{\alpha \beta}_{bd} \braket{ \{v\}^{M-1},(ab) | S^+_c S^+_d | \{u\}^{M-1}_{\alpha,\beta},(ab) } \\
	&= 2 \eta(2) \mathcal{X}^{(0)}_{abcd} (J(ab))
\end{align}
the open-shell pair $A^+_{cd}$ condenses to $S^+_c$ while an RG pair is broken to give the open-shell pair $A^+_{ab}$. In the corresponding ``exchange'' element, $\Gamma^{Vab,Ucd}_{adcb}$, there is a forward scattering $d\rightarrow a$ and a backward scattering $c \leftarrow b$. The $J$-sum
\begin{align}
	\mathcal{X}^{(0)}_{abcd} (J(ab)) &= U_b
	\left(
	[J(ab)]^{c,c} + \frac{\varepsilon_c-\varepsilon_b}{\varepsilon_d - \varepsilon_b}[J(ab)]^{d,c} + \frac{g}{\varepsilon_d - \varepsilon_b}[J(ab)]^{cd,cd}
	\right) \nonumber \\
	&+\sum^N_{p (\neq a,b,c,d)=1} 
	\frac{\varepsilon_c - \varepsilon_b}{\varepsilon_p - \varepsilon_b} U_b
	\left(
	[J(ab)]^{p,c} + \frac{g}{\varepsilon_d - \varepsilon_b}[J(ab)]^{pd,cd} 
	\right) \nonumber \\
	&+\sum^N_{p (\neq a,b,c,d)=1} 
	\frac{\varepsilon_c - \varepsilon_p}{\varepsilon_b - \varepsilon_k} U_p
	\left(
	[J(ab)]^{p,c} + \frac{g}{\varepsilon_d - \varepsilon_p}[J(ab)]^{pd,cd} 
	\right) \nonumber \\
	&+ \sum^N_{p (\neq a,b,c,d)=1} \frac{1}{\varepsilon_d - \varepsilon_p} \frac{K_{pb}}{\mathfrak{d}^{cd}_{pb}} \sum^N_{q (\neq p,a,b)=1} \frac{[J(ab)]^{pq,cd}}{\varepsilon_b - \varepsilon_q} \nonumber \\
	&- \sum_{p<q (\neq a,b,c,d)} \frac{\mathfrak{p}^{bd}_{pq}}{\mathfrak{d}^{cd}_{pq}} K_{pq} [J(ab)]^{pq,cd} \label{eq:jsum_x0_abcd} \\
	&= - \mathcal{B}^{(0)}_{cb} (J(ab)) - \mathcal{B}^{(0)}_{ddcb} (J(ab))
\end{align}
is again reducible, but to the backward scattering $J$-sums \eqref{eq:jsum_b0_cb} and \eqref{eq:jsum_b0_kkcb}.

In the last type, both electrons scatter backwards
\begin{align}
	\Gamma^{Vab,Ucd}_{cadb} &= \sum_{\alpha < \beta} \mathfrak{p}^{\alpha \beta}_{ab} \braket{ \{v\},(ab) | S^+_c S^+_d | \{u\}^{M-1}_{\alpha,\beta},(ab)} \\
	&= \eta(2) \mathcal{B}^{(0)}_{cadb} (J(ab))
\end{align}
with the irreducible $J$-sum
\begin{align}
	\mathcal{B}^{(0)}_{cadb} (J(ab)) &= K_{ab} [J(ab)]^{cd,cd} \nonumber \\
	&- \sum^N_{p (\neq a,b,c,d)=1} \frac{1}{\varepsilon_a - \varepsilon_p} \frac{K_{pb}}{\mathfrak{d}^{cd}_{pb}} \sum^N_{q (\neq a,b)=1} \frac{[J(ab)]^{pq,cd}}{\varepsilon_b - \varepsilon_q} \nonumber \\
	&- \sum^N_{p (\neq a,b,c,d)=1} \frac{1}{\varepsilon_b - \varepsilon_p} \frac{K_{pa}}{\mathfrak{d}^{cd}_{pa}} \sum^N_{q (\neq a,b)=1} \frac{[J(ab)]^{pq,cd}}{\varepsilon_a - \varepsilon_q} \nonumber \\
	&+ \frac{K_{ab}}{\mathfrak{d}^{cd}_{ab}} \sum^N_{p (\neq a,b,c,d)=1} ( \mathfrak{d}^{pc}_{ab} [J(ab)]^{pc,cd} + \mathfrak{d}^{pd}_{ab}[J(ab)]^{pd,cd} ) \nonumber \\
	&+ \sum_{p<q (\neq a,b,c,d)}
	\left(
	\frac{\mathfrak{p}^{ab}_{pq}}{\mathfrak{d}^{cd}_{pq}} K_{pq} + \frac{\mathfrak{d}^{pq}_{ab}}{\mathfrak{d}^{cd}_{ab}} K_{ab}.
	\right) [J(ab)]^{pq,cd}. \label{eq:jsum_b0_cadb}
\end{align}
These elements are symmetric with respect to exchange
\begin{align}
	\Gamma^{Vab,Ucd}_{cbda} &= \Gamma^{Vab,Ucd}_{cadb}
\end{align}
like the double forward scattering elements. 

This completes the list of possible couplings up to seniority-two. 
\section{Seniority-four} \label{sec:s4}
Seniority-four states themselves present no additional difficulty. Almost all of the required $J$-sums have already been computed. The complication is that the behaviour of the two seniority-four vacuums is different, and hence most of the results must be tabulated.

\subsection{0 - 4 coupling}
There is a single type of element to compute, and once again by moving the $A^0_{pq}$ operators to the right, the resulting expressions depend only upon seniority-zero matrix elements
\begin{align}
	\Gamma^{V,Uabcd(1)}_{abcd{}} &= 2 \braket{\{v\}^M | S^+_a S^+_c | \{u\}^{M-2} } \nonumber \\
	&- 2 \sum^{M-2}_{\alpha=1} \frac{1}{u_{\alpha} - \varepsilon_b} \braket{\{v\}^M | S^+_a S^+_b S^+_c | \{u\}^{M-2}_{\alpha} } \nonumber \\
	&- 2 \sum^{M-2}_{\alpha=1} \frac{1}{u_{\alpha} - \varepsilon_d} \braket{\{v\}^M | S^+_a S^+_c S^+_d | \{u\}^{M-2}_{\alpha} } \nonumber \\
	&+ 2 \sum_{\alpha < \beta} \mathfrak{p}^{\alpha \beta}_{bd} 
	\braket{\{v\}^M | S^+_a S^+_b S^+_c S^+_d | \{u\}^{M-2}_{\alpha,\beta}} \\
	&= 2 \eta(0) g^2 \mathcal{D}^{(4)}_{abcd}.
\end{align}
Again, the summation of the form factors is lengthy and tedious, but follows the same established patterns with no particular difficulty. Similar to the case for seniority-zero coupling to seniority-two, the determinants must be extended. Here, this must be done twice, resulting in fourth-cofactors of the common matrix $J$ \eqref{eq:matsuno}. The irreducible $J$-sum is
\begin{align}
	\mathcal{D}^{(4)}_{abcd} (J) &= \frac{\mathcal{W}^{ac}}{(\varepsilon_a - \varepsilon_c)}  \nonumber \\
	&- \sum^N_{p (\neq a,b,c,d)=1} 
	\frac{(\varepsilon_a - \varepsilon_p)(\varepsilon_c - \varepsilon_p)}
	     {(\varepsilon_a - \varepsilon_b)(\varepsilon_c - \varepsilon_b)} \frac{U_p}{(\varepsilon_a - \varepsilon_c)}
	     \left(
	     \mathcal{W}^{acb}_p - \frac{g}{\varepsilon_p - \varepsilon_d}\mathcal{W}^{acbd}_{pd}
	     \right) \nonumber \\
	&- \sum^N_{p (\neq a,b,c,d)=1} 
	\frac{(\varepsilon_a - \varepsilon_p)(\varepsilon_c - \varepsilon_p)}
		{(\varepsilon_a - \varepsilon_d)(\varepsilon_c - \varepsilon_d)} \frac{U_p}{(\varepsilon_a - \varepsilon_c)}
		\left(
		\mathcal{W}^{acd}_p - \frac{g}{\varepsilon_p - \varepsilon_b}\mathcal{W}^{acbd}_{bp}
		\right) \nonumber \\
	&+ \sum_{p<q (\neq a,b,c,d)} 
	\frac{\mathfrak{d}^{ac}_{bd}\mathfrak{p}^{bd}_{pq}}
	     {\mathfrak{d}^{ac}_{pq}\mathfrak{d}^{bd}_{pq}}
	     \frac{(\varepsilon_p - \varepsilon_q)}{(\varepsilon_b - \varepsilon_d)} \frac{K_{pq}}{(\varepsilon_a - \varepsilon_c)} \mathcal{W}^{acbd}_{pq} \label{eq:jsum_d4_abcd}
\end{align}
in terms of the intermediates
\begin{align}
	\mathcal{W}^{ac} &= \sum_{r<s} (\varepsilon_r - \varepsilon_s) [J]^{rs,ac} \\
	\mathcal{W}^{acb}_{p} &= \sum_{r<s} (\varepsilon_r - \varepsilon_s) [J]^{rsp,acb} \\
	\mathcal{W}^{acd}_{p} &= \sum_{r<s} (\varepsilon_r - \varepsilon_s) [J]^{rsp,acd} \\
	\mathcal{W}^{acbd}_{pq} &= \sum_{r<s} (\varepsilon_r - \varepsilon_s) [J]^{rspq,acbd}.
\end{align}
For numerical purposes it is safest to restrict the summations in the $J$-sum so that they do not include any of the four blocked levels. Noticing that the intermediates $\mathcal{W}$ are antisymmetric in the upper indices, they may be computed once for each combination of indices. There are 6 unique reduced summations to compute, but 12 unique DM elements for each seniority-four vacuum. These are summarized in Table \ref{tab:s04_coupling}.
\begin{table}[ht!] % [h] for here, you can change this option according to your needs
	\centering % Center the table
	\begin{tabular}{c|c|c || c|c|c || c|c|c} % Specify the number of columns and alignment\
		& $\bra{\theta} \ket{\varphi^{(1)}_{abcd}}$ & $\bra{\theta} \ket{\varphi^{(2)}_{abcd}}$ &
		& $\bra{\theta} \ket{\varphi^{(1)}_{abcd}}$ & $\bra{\theta} \ket{\varphi^{(2)}_{abcd}}$ & 
		& $\bra{\theta} \ket{\varphi^{(1)}_{abcd}}$ & $\bra{\theta} \ket{\varphi^{(2)}_{abcd}}$\\
		\hline % Draw another horizontal line
		$\Gamma_{abcd}$ & $2 \mathcal{D}^{(4)}_{abcd}$ & 0 &
		$\Gamma_{acdb}$ & $- \mathcal{D}^{(4)}_{abdc}$ & $ \sqrt{3} \mathcal{D}^{(4)}_{abdc}$ &
		$\Gamma_{adcb}$ & $- \mathcal{D}^{(4)}_{abcd}$ & $-\sqrt{3} \mathcal{D}^{(4)}_{abcd}$ \\
		
		$\Gamma_{abdc}$ & $2\mathcal{D}^{(4)}_{abdc}$ & 0 &
		$\Gamma_{bdca}$ & $-\mathcal{D}^{(4)}_{bacd}$ & $ \sqrt{3}$ $\mathcal{D}^{(4)}_{bacd}$ &
		$\Gamma_{bcda}$ & $-\mathcal{D}^{(4)}_{badc}$ & $-\sqrt{3}$ $\mathcal{D}^{(4)}_{badc}$ \\
		
		$\Gamma_{bacd}$ & $2\mathcal{D}^{(4)}_{bacd}$ & 0 &	
		$\Gamma_{acbd}$ & $-\mathcal{D}^{(4)}_{acbd}$ & $ \sqrt{3}$ $\mathcal{D}^{(4)}_{acbd}$ &
		$\Gamma_{adbc}$ & $-\mathcal{D}^{(4)}_{acbd}$ & $-\sqrt{3}$ $\mathcal{D}^{(4)}_{acbd}$ \\
			
		$\Gamma_{badc}$ & $2\mathcal{D}^{(4)}_{badc}$ & 0 &		
		$\Gamma_{cadb}$ & $-\mathcal{D}^{(4)}_{cadb}$ & $ \sqrt{3}$ $\mathcal{D}^{(4)}_{cadb}$ &
		$\Gamma_{cbda}$ & $-\mathcal{D}^{(4)}_{cadb}$ & $-\sqrt{3}$ $\mathcal{D}^{(4)}_{cadb}$ \\
	\end{tabular}
	\caption{TDM elements between between seniority-zero and seniority-four RG states. Each element necessarily includes a factor of $\eta(0)g^2$ and the $J$-sums \eqref{eq:jsum_d4_abcd} are all computed from the common matrix $J$ \eqref{eq:matsuno}.} % Table caption
	\label{tab:s04_coupling} 
\end{table}
The columns of Table \ref{tab:s04_coupling} are separated based on the grouping of indices. The three possible groupings behave differently as the choice of seniority-four vacuums is asymmetric. 

\subsection{2 - 4 coupling}
Non-zero couplings occur between seniority-two and seniority-four states that share two common blocked levels. These elements are essentially the same as those between seniority-zero and seniority-two, with some modifications of the prefactors. The bigger problem is the difference between $\ket{\varphi^{(1)}_{abcd}}$ and $\ket{\varphi^{(2)}_{abcd}}$. 

\subsubsection{2 shared blocked levels}
\paragraph{ab pattern}
There are different results depending on which two blocked levels are shared. First, suppose $a$ and $b$ are the common blocked levels. The results are the same if the shared levels are $c$ and $d$, up to the interchange of indices $(a,b) \leftrightarrow (c,d)$. The 1-body matrix elements involving the first seniority-four state are
\begin{align}
	\gamma^{Vab,Uabcd(1)}_{cd} &= - \sqrt{2} g \eta(2) \mathcal{D}^{(2)}_{cd} (J(ab)) \\
	\gamma^{Vab,Uabcd(1)}_{dc} &= - \sqrt{2} g \eta(2) \mathcal{D}^{(2)}_{dc} (J(ab)),
\end{align}
where the $J$-sum \eqref{eq:jsum_d2_ab} has already been computed for seniority-zero coupling to seniority-two. No modifications to the expressions are necessary as $J(ab)$ is understood as \eqref{eq:matsuno_2}, the $N \times N$ matrix with 1s in the $(a,a)$ and $(b,b)$ positions and zeros elsewhere in the $a$ and $b$ rows and columns. As before, the additional cofactors included in the summation are strictly zero. 

For the first seniority-four vacuum there are direct 
\begin{align}
	\Gamma^{Vab,Uabcd(1)}_{aacd} = \Gamma^{Vab,Uabcd(1)}_{bbcd} = \Gamma^{Vab,Uabcd(1)}_{cccd} &= \gamma^{Vab,Uabcd(1)}_{cd} \\
	\Gamma^{Vab,Uabcd(1)}_{aadc} = \Gamma^{Vab,Uabcd(1)}_{bbdc} = \Gamma^{Vab,Uabcd(1)}_{dddc} &= \gamma^{Vab,Uabcd(1)}_{dc}
\end{align}
and exchange elements
\begin{align}
	\Gamma^{Vab,Uabcd(1)}_{adca} = \Gamma^{Vab,Uabcd(1)}_{bdcb} &= - \frac{1}{2} \gamma^{Vab,Uabcd(1)}_{cd} \\
	\Gamma^{Vab,Uabcd(1)}_{acda} = \Gamma^{Vab,Uabcd(1)}_{bcdb} &= - \frac{1}{2} \gamma^{Vab,Uabcd(1)}_{dc}
\end{align}
involving the blocked levels, in addition to the 2-electron elements computed from $J$-sums \eqref{eq:clean_direct_sum_02}, \eqref{eq:jsum_p2_kakb} and \eqref{eq:jsum_p2_akbk}
\begin{align}
	\Gamma^{Vab,Uabcd(1)}_{kkcd} &= - 2 \sqrt{2} g \eta (2) \mathcal{D}^{(2)}_{kkcd} (J(ab)) \\
	\Gamma^{Vab,Uabcd(1)}_{kkdc} &= - 2 \sqrt{2} g \eta (2) \mathcal{D}^{(2)}_{kkdc} (J(ab)) \\
	\Gamma^{Vab,Uabcd(1)}_{kckd} &= -   \sqrt{2} g \eta (2) \mathcal{P}^{(2)}_{kckd} (J(ab)) \\
	\Gamma^{Vab,Uabcd(1)}_{ckdk} &= -   \sqrt{2} g \eta (2) \mathcal{P}^{(2)}_{ckdk} (J(ab))
\end{align}
along with the corresponding
\begin{align}
	\Gamma^{Vab,Uabcd(1)}_{kdck} &= -\frac{1}{2} \Gamma^{Vab,Uabcd(1)}_{kkcd} \\
	\Gamma^{Vab,Uabcd(1)}_{kcdk} &= -\frac{1}{2} \Gamma^{Vab,Uabcd(1)}_{kkdc}.
\end{align}
In this case, the second seniority-four vacuum only has non-zero exchange elements within blocked levels
\begin{align}
	\Gamma^{Vab,Uabcd(2)}_{adca} = - \Gamma^{Vab,Uabcd(2)}_{bdcb} &= - g \eta (2) \sqrt{\frac{3}{2}} \mathcal{D}^{(2)}_{cd} (J(ab)) \\
	\Gamma^{Vab,Uabcd(2)}_{acda} = - \Gamma^{Vab,Uabcd(2)}_{bcdb} &= \;\;\; g \eta (2) \sqrt{\frac{3}{2}} \mathcal{D}^{(2)}_{dc} (J(ab)),   
\end{align}
computed from the $J$-sum \eqref{eq:jsum_d2_ab}. There are no other couplings with the second seniority-four RG vacuum if the shared levels are $a$ and $b$.

\paragraph{ac and ad patterns}
If the two shared indices are $a$ and $c$ (equivalently $b$ and $d$) the elements are different, but are computed from the same $J$-sums as shown in Table \ref{tab:s24_2shared_ac}. The results are the same for the first seniority-four vacuum for shared levels $a$ and $d$, while for the second seniority-four vacuum, there is an additional sign. Thus we take $\rho = 1$ for shared levels $a$ and $c$, and $\rho = -1$ for shared levels $a$ and $d$. The results are otherwise identical up to an interchange of the relevant indices. 

\begin{table}[ht!] % [h] for here, you can change this option according to your needs
	\centering % Center the table
	\begin{tabular}{c|c|c} % Specify the number of columns and alignment\
		& $\braket{(ac) || \varphi^{(1)}_{abcd}}$ & $\braket{(ac) || \varphi^{(2)}_{abcd}}$ \\
		\hline % Draw another horizontal line
		$\gamma_{bd}$ & $\frac{1}{\sqrt{2}} g \eta (2) \mathcal{D}^{(2)}_{bd}$ & $- \rho \sqrt{\frac{3}{2}} g \eta (2) \mathcal{D}^{(2)}_{bd}$ \\
		$\Gamma_{adba}$ & $-2 \gamma_{bd}$ & 0 \\
		$\Gamma_{cdbd}$ & $   \gamma_{bd}$ & $- \gamma_{bd}$ \\
		\hline
		$\gamma_{db}$ & $\frac{1}{\sqrt{2}} g \eta (2) \mathcal{D}^{(2)}_{db}$ & $- \rho \sqrt{\frac{3}{2}} g \eta (2) \mathcal{D}^{(2)}_{db}$ \\
		$\Gamma_{cbdc}$ & $-2 \gamma_{db}$ & 0 \\
		$\Gamma_{abda}$ & $   \gamma_{db}$ & $- \gamma_{db}$ \\
		\hline
		$\Gamma_{kkbd}$ & $\sqrt{2} g \eta(2) \mathcal{D}^{(2)}_{kkbd}$ & $-2\rho \sqrt{\frac{3}{2}} g \eta(2) \mathcal{D}^{(2)}_{kkbd}$ \\
		$\Gamma_{kkdb}$ & $\sqrt{2} g \eta(2) \mathcal{D}^{(2)}_{kkdb}$ & $-2\rho \sqrt{\frac{3}{2}} g \eta(2) \mathcal{D}^{(2)}_{kkdb}$ \\
		$\Gamma_{kbkd}$ & $\frac{1}{\sqrt{2}} g \eta(2) \mathcal{P}^{(2)}_{kbkd}$ & $- \rho \sqrt{\frac{3}{2}} g \eta(2) \mathcal{P}^{(2)}_{kbkd}$ \\
		$\Gamma_{bkdk}$ & $\frac{1}{\sqrt{2}} g \eta(2) \mathcal{P}^{(2)}_{bkdk}$ & $- \rho \sqrt{\frac{3}{2}} g \eta(2) \mathcal{P}^{(2)}_{bkdk}$
	\end{tabular}
	\caption{Couplings of seniority-two and seniority-four states that share two blocked levels in an $ac$ $(\rho=1)$ pattern. For an $ad$ pattern, take $\rho=-1$ and swap $c$ with $d$. $J$-sums \eqref{eq:jsum_d2_ab}, \eqref{eq:clean_direct_sum_02}, \eqref{eq:jsum_p2_kakb} and \eqref{eq:jsum_p2_akbk} are computed with the common matrix $J(ac)$ \eqref{eq:matsuno_2}.} % Table caption
	\label{tab:s24_2shared_ac} 
\end{table}

Direct 2-body elements for both vacuums ($\mu=1,2$) follow the same pattern
\begin{align}
	\Gamma^{Vac,Uabcd(\mu)}_{aabd} = \Gamma^{Vac,Uabcd(\mu)}_{bbbd} = \Gamma^{Vac,Uabcd(\mu)}_{ccbd} &= \gamma^{Vac,Uabcd(\mu)}_{bd} \\
	\Gamma^{Vac,Uabcd(\mu)}_{aadb} = \Gamma^{Vac,Uabcd(\mu)}_{ccdb} = \Gamma^{Vac,Uabcd(\mu)}_{dddb} &= \gamma^{Vac,Uabcd(\mu)}_{db},
\end{align}
and so do exchange elements
\begin{align}
	\Gamma^{Vac,Uabcd(\mu)}_{kdbk} &= - \frac{1}{2} \Gamma^{Vac,Uabcd(\mu)}_{kkbd} \\
	\Gamma^{Vac,Uabcd(\mu)}_{kbdk} &= - \frac{1}{2} \Gamma^{Vac,Uabcd(\mu)}_{kkdb}.
\end{align}

\subsubsection{1 shared blocked level}
Coupling between seniority-two and seniority-four states with one common blocked level are also possible: two blocked levels can form a seniority-zero pair, while a third blocked index can be changed. Evidently this process was not possible when coupling seniority-zero with seniority-two, but this will occur for all other combinations of seniorities differing by two. 

Assume that the shared index is $a$, i.e. the seniority-two state is $\ket{ \{v\}^{M-1}, (ae) }$ while the seniority-four states have blocked levels $a,b,c,d$. The other cases are accessible by swapping indices from the same set of matrix elements with one subtlety. The labelling of seniority-four singlets has been chosen such that $a<b<c<d$, but these indices may be swapped freely in $\ket{\varphi^{(1)}_{abcd}}$ since $A^+_{ab} = A^+_{ba}$
\begin{align}
	\ket{\varphi^{(1)}_{abcd}} =
	\ket{\varphi^{(1)}_{bacd}} =
	\ket{\varphi^{(1)}_{abdc}} =
	\ket{\varphi^{(1)}_{badc}} =
	\ket{\varphi^{(1)}_{cdab}} =
	\ket{\varphi^{(1)}_{cdba}} =
	\ket{\varphi^{(1)}_{dcab}} =
	\ket{\varphi^{(1)}_{dcba}}
\end{align}
there the 8 symmetries one would expect. However,
\begin{align}
	\ket{\varphi^{(2)}_{abcd}} =
	- \ket{\varphi^{(2)}_{bacd}} =
	- \ket{\varphi^{(2)}_{abdc}} =
	\ket{\varphi^{(2)}_{badc}} =
	\ket{\varphi^{(2)}_{cdab}} =
	- \ket{\varphi^{(2)}_{cdba}} =
	- \ket{\varphi^{(2)}_{dcab}} =
	\ket{\varphi^{(2)}_{dcba}}
\end{align}
there are fewer symmetries for the second state. This can be accounted for either by introducing a sign to the corresponding matrix elements, or by swapping another pair of indices: if $a>b$, then either introduce an extra sign to all the elements for $\ket{\varphi^{(2)}_{abcd}}$ or swap the labels $c$ and $d$. 

In the first type of element, one of the electrons scatters forward, while the other two condense to a pair,
\begin{align}
	\Gamma^{Vae,Uabcd(1)}_{ebcd} &= \sqrt{2} \braket{\{v\}^{M-1},(ae) | S^+_c | \{u\}^{M-2},(ae)} \nonumber \\
	&- \sqrt{2} \sum^{M-2}_{\alpha=1} \frac{1}{u_{\alpha} - \varepsilon_b} 
	   \braket{\{v\}^{M-1},(ae) | S^+_b S^+_c | \{u\}^{M-2}_{\alpha},(ae)} \nonumber \\
	&- \sqrt{2} \sum^{M-2}_{\alpha=1} \frac{1}{u_{\alpha} - \varepsilon_d}
	   \braket{\{v\}^{M-1},(ae) | S^+_c S^+_d | \{u\}^{M-2}_{\alpha},(ae)} \nonumber \\
	&+ \sqrt{2} \sum_{\alpha < \beta} \mathfrak{p}^{\alpha\beta}_{bd} 
	   \braket{ \{v\}^{M-1},(ae) | S^+_b S^+_c S^+_d | \{u\}^{M-2}_{\alpha,\beta},(ae)} \\
	&= - \sqrt{2} g \eta(2) \mathcal{F}^{(2)}_{ebcd} (J(ae))
\end{align}
with the $J$-sum
\begin{align}
	\mathcal{F}^{(2)}_{ebcd} (J(ae)) &= \sum^N_{p =1} [J(ae)]^{p,c} \nonumber \\
	&+ \sum^N_{p (\neq b,c,d) =1} \frac{(\varepsilon_c - \varepsilon_p)}{(\varepsilon_c - \varepsilon_b)} U_p
	\sum^N_{q =1}
	\left(
	[J(ae)]^{pq,cb} - \frac{g}{\varepsilon_p - \varepsilon_d} [J(ae)]^{pqd,cbd}
	\right) \nonumber \\
	&+ \sum^N_{p (\neq b,c,d) =1} \frac{(\varepsilon_c - \varepsilon_p)}{(\varepsilon_c - \varepsilon_d)} U_p
	\sum^N_{q =1}
	\left(
	[J(ae)]^{pq,cd} - \frac{g}{\varepsilon_p - \varepsilon_b} [J(ae)]^{pqb,cdb}
	\right) \nonumber \\
	&+ \sum_{p<q (\neq b,c,d)} 
	\frac{(\varepsilon_b - \varepsilon_p)(\varepsilon_b - \varepsilon_q)}
	     {(\varepsilon_b - \varepsilon_d)(\varepsilon_b - \varepsilon_c)}
	\frac{\mathfrak{p}^{bd}_{pq}}{\mathfrak{d}^{cd}_{pq}} K_{pq}
	\sum^N_{r = 1} [J(ae)]^{pqr,bcd}. \label{eq:jsum_f2_ebcd}
\end{align}
This is the only forward scattering $J$-sum that does not appear to be reducible. The corresponding exchange element is
\begin{align}
	\Gamma^{Vae,Uabcd(1)}_{cbed} = - \frac{1}{2} \Gamma^{Vae,Uabcd(1)}_{ebcd},
\end{align}
which counts a forward scattering $d \rightarrow e$ along with a singlet excitation $A^+_{ab}\rightarrow A^+_{ac}$ which couples to the already present $A^+_{cd}$ to yield $A^+_{ae}$. There are three possible choices of indices, and as they are not all summarized intuitively they are presented in Table \ref{tab:s24_coupling}.
\begin{table}[ht!] % [h] for here, you can change this option according to your needs
	\centering % Center the table
	\begin{tabular}{c|c|c} % Specify the number of columns and alignment\
		& $\braket{(ae) || \varphi^{(1)}_{abcd}}$ & $\braket{(ae) || \varphi^{(2)}_{abcd}}$ \\
		\hline % Draw another horizontal line
		$\Gamma_{ebcd}$ & $ - \sqrt{2} g \eta(2) \mathcal{F}^{(2)}_{ebcd} (J(ae))$ & 0 \\ 
		$\Gamma_{bced}$ & $ \;\;\; \frac{1}{\sqrt{2}} g \eta(2) \mathcal{F}^{(2)}_{ebcd} (J(ae))$ &
		       $ \;\;\;\sqrt{\frac{3}{2}} g \eta(2) \mathcal{F}^{(2)}_{ebcd} (J(ae))$ \\
		
		$\Gamma_{ebdc}$ & $ - \sqrt{2} g \eta(2) \mathcal{F}^{(2)}_{ebdc} (J(ae))$ & 0 \\
		$\Gamma_{dbec}$ & $ \;\;\; \frac{1}{\sqrt{2}} g \eta(2) \mathcal{F}^{(2)}_{ebdc} (J(ae))$ & 
		       $ - \sqrt{\frac{3}{2}} g \eta(2) \mathcal{F}^{(2)}_{ebdc} (J(ae))$ \\
		
		$\Gamma_{ecbd}$ & $  \;\;\; \frac{1}{\sqrt{2}} g \eta(2) \mathcal{F}^{(2)}_{ecbd} (J(ae))$ &
		       $  \;\;\; \sqrt{\frac{3}{2}} g \eta(2) \mathcal{F}^{(2)}_{ecbd} (J(ae))$ \\
		$\Gamma_{bced}$ & $  \;\;\; \frac{1}{\sqrt{2}} g \eta(2) \mathcal{F}^{(2)}_{ecbd} (J(ae))$ & 
		       $ - \sqrt{\frac{3}{2}} g \eta(2) \mathcal{F}^{(2)}_{ecbd} (J(ae))$ \\
	\end{tabular}
	\caption{Forward scattering elements between seniority-two and seniority-four with one shared blocked level. The $J$-sum \eqref{eq:jsum_f2_ebcd} is computed with the matrix $J(ae)$ \eqref{eq:matsuno_2}.} % Table caption
	\label{tab:s24_coupling} 
\end{table}

In the second type of element one electron scatters backwards while the others condense to a pair
\begin{align}
	\Gamma^{Vae,Uabcd(1)}_{bedc} &= - \sqrt{2} \sum^{M-2}_{\alpha = 1} \frac{1}{u_{\alpha} - \varepsilon_e}
	\braket{ \{v\}^{M-1},(ae) | S^+_b S^+_d | \{u\}^{M-2}_{\alpha}, (ae) } \nonumber \\
	&+   \sqrt{2} \sum_{\alpha < \beta} \mathfrak{p}^{\alpha\beta}_{ce} 
	     \braket{ \{v\}^{M-1},(ae) | S^+_b S^+_c S^+_d | \{u\}^{M-2}_{\alpha,\beta},(ae)  } \\
	&= - \sqrt{2} g \eta(2) \mathcal{B}^{(2)}_{bedc} (J(ae))
\end{align}
with $J$-sum
\begin{align}
	\mathcal{B}^{(2)}_{bedc} (J(ae)) &= 
	\sum^N_{p (\neq e)=1} \frac{(\varepsilon_b - \varepsilon_p)}{(\varepsilon_b - \varepsilon_d)}
	\frac{(\varepsilon_e - \varepsilon_d)}{(\varepsilon_e - \varepsilon_p)} U_e 
	\sum^N_{q =1}
	\left(
	[J(ae)]^{pq,bd} 
	- \frac{g}{\varepsilon_e - \varepsilon_c}
	 [J(ae)]^{pqc,bdc} 
	\right) \nonumber \\
	&+ \sum^N_{p (\neq c,e)=1} \frac{(\varepsilon_b - \varepsilon_p)}{(\varepsilon_b - \varepsilon_d)}
	\frac{(\varepsilon_d - \varepsilon_p)}{(\varepsilon_e - \varepsilon_p)} U_p
	\sum^N_{q =1}
	\left(
	[J(ae)]^{pq,bd} 
	- \frac{g}{\varepsilon_p - \varepsilon_c}
	 [J(ae)]^{pqc,bdc} 
	\right) \nonumber \\
	&+ \frac{(\varepsilon_e - \varepsilon_c)}{(\varepsilon_b - \varepsilon_c)(\varepsilon_d - \varepsilon_c)}
	\sum^N_{p (\neq b,c,d,e)=1} \frac{K_{pe}}{\mathfrak{d}^{bd}_{pe}}
	\sum_{q<r (\neq e)} \frac{(\varepsilon_q - \varepsilon_r)}{(\varepsilon_q - \varepsilon_e)(\varepsilon_r - \varepsilon_e)}[J(ae)]^{pqr,bdc}
	\nonumber \\
	&+ \sum_{p<q (\neq b,c,d,e)} 
	\frac{(\varepsilon_c - \varepsilon_p)(\varepsilon_c - \varepsilon_q)}
	     {(\varepsilon_c - \varepsilon_b)(\varepsilon_c - \varepsilon_d)}
	\frac{\mathfrak{p}^{ce}_{pq}}{\mathfrak{d}^{bd}_{pq}} K_{pq} \sum^N_{r =1} [J(ae)]^{pqr,bdc}. \label{eq:jsum_b2_bedc}
\end{align}
Again, there are three distinct choices which are summarized in Table \ref{tab:s24_coupling_2}.
\begin{table}[ht!] % [h] for here, you can change this option according to your needs
	\centering % Center the table
	\begin{tabular}{c|c|c} % Specify the number of columns and alignment\
		& $\braket{(ae) || \varphi^{(1)}_{abcd}}$ & $\braket{(ae) || \varphi^{(2)}_{abcd}}$ \\
		\hline % Draw another horizontal line
		$\Gamma_{bedc}$ & $ - \sqrt{2} g \eta(2) \mathcal{B}^{(2)}_{bedc} (J(ae))$ & 0 \\ 
		$\Gamma_{bcde}$ & $ \;\;\; \frac{1}{\sqrt{2}} g \eta(2) \mathcal{B}^{(2)}_{bedc} (J(ae))$ &
		$ \;\;\;\sqrt{\frac{3}{2}} g \eta(2) \mathcal{B}^{(2)}_{bedc} (J(ae))$ \\
		
		$\Gamma_{becd}$ & $ - \sqrt{2} g \eta(2) \mathcal{B}^{(2)}_{becd} (J(ae))$ & 0 \\
		$\Gamma_{bdce}$ & $ \;\;\; \frac{1}{\sqrt{2}} g \eta(2) \mathcal{B}^{(2)}_{becd} (J(ae))$ & 
		$ - \sqrt{\frac{3}{2}} g \eta(2) \mathcal{B}^{(2)}_{becd} (J(ae))$ \\
		
		$\Gamma_{cbde}$ & $  \;\;\; \frac{1}{\sqrt{2}} g \eta(2) \mathcal{B}^{(2)}_{cbde} (J(ae))$ &
		$  \;\;\; \sqrt{\frac{3}{2}} g \eta(2) \mathcal{B}^{(2)}_{cbde} (J(ae))$ \\
		$\Gamma_{cedb}$ & $  \;\;\; \frac{1}{\sqrt{2}} g \eta(2) \mathcal{B}^{(2)}_{cbde} (J(ae))$ & 
		$ - \sqrt{\frac{3}{2}} g \eta(2) \mathcal{B}^{(2)}_{cbde} (J(ae))$ \\
	\end{tabular}
	\caption{Backward scattering elements between seniority-two and seniority-four with one shared blocked level. The $J$-sum \eqref{eq:jsum_b2_bedc} is computed with the matrix $J(ae)$ \eqref{eq:matsuno_2}.} % Table caption
	\label{tab:s24_coupling_2} 
\end{table}

\subsection{4 - 4 coupling}
Seniority-four states couple to other seniority-four states in the same manner as seniority-two states couple to other seniority-two states: the blocked levels may be identical, differ by one, or differ by two.
\subsubsection{4 shared blocked levels}
As expected, the couplings between seniority-four states that share 4 common blocked levels are the usual seniority-zero type elements. Further complications are possible as couplings between $\ket{\varphi^{(1)}_{abcd}}$ and $\ket{\varphi^{(2)}_{abcd}}$ could occur. Thankfully, the only non-zero couplings between the two vacuums are the exchange elements
\begin{align}
	  \Gamma^{Vabcd(1),Uabcd(2)}_{adda} =  
	  \Gamma^{Vabcd(1),Uabcd(2)}_{bccb} &=   \;\;\;\frac{\sqrt{3}}{2} \braket{\{v\}^{M-2} | \{u\}^{M-2}} \\
	  \Gamma^{Vabcd(1),Uabcd(2)}_{acca} =
	  \Gamma^{Vabcd(1),Uabcd(2)}_{bddb} &= - \frac{\sqrt{3}}{2} \braket{\{v\}^{M-2} | \{u\}^{M-2}}.
\end{align}
Again, the rapidities $\{u\}$ and $\{v\}$ are both on-shell for the same set of Richardson's equations. Thus, the scalar product $\braket{\{v\}^{M-2}|\{u\}^{M-2}}$ is zero if they are distinct, and the norm of the state if they are the same. There are no other couplings between the two vacuums.

For \emph{both} vacuums the 1-body elements
\begin{align}
	\gamma^{Vabcd(\mu),Uabcd(\mu)}_{aa} = \gamma^{Vabcd(\mu),Uabcd(\mu)}_{bb} = \gamma^{Vabcd(\mu),Uabcd(\mu)}_{cc} = \gamma^{Vabcd(\mu),Uabcd(\mu)}_{dd} =
	\braket{\{v\}^{M-2} | \{u\}^{M-2}}
\end{align}
become 1 when normalized, and the same is true for the direct elements in the blocked levels $\Gamma^{Vabcd(\mu),Uabcd(\mu)}_{aabb}$, $\Gamma^{Vabcd(\mu),Uabcd(\mu)}_{aacc}$, $\Gamma^{Vabcd(\mu),Uabcd(\mu)}_{aadd}$, $\Gamma^{Vabcd(\mu),Uabcd(\mu)}_{bbcc}$, $\Gamma^{Vabcd(\mu),Uabcd(\mu)}_{bbdd}$ and $\Gamma^{Vabcd(\mu),Uabcd(\mu)}_{ccdd}$. Exchange elements in the blocked levels are however distinct for each vacuum: 
\begin{align}
	\Gamma^{Vabcd(1),Uabcd(1)}_{abba} = \Gamma^{Vabcd(1),Uabcd(1)}_{cddc} &= \;\;\;\braket{\{v\}^{M-2} | \{u\}^{M-2}} \\
	\Gamma^{Vabcd(2),Uabcd(2)}_{abba} = \Gamma^{Vabcd(2),Uabcd(2)}_{cddc} &= - \braket{\{v\}^{M-2} | \{u\}^{M-2}}
\end{align}
while
\begin{align}
	\Gamma^{Vabcd(1)}_{acca} = \Gamma^{Vabcd(1)}_{adda} = \Gamma^{Vabcd(1)}_{bccb} = \Gamma^{Vabcd(1)}_{bddb} 
	&= - \frac{1}{2} \braket{\{v\}^{M-2} | \{v\}^{M-2}} \\
	\Gamma^{Vabcd(2)}_{acca} = \Gamma^{Vabcd(2)}_{adda} = \Gamma^{Vabcd(2)}_{bccb} = \Gamma^{Vabcd(2)}_{bddb} 
	&= \;\;\;\frac{1}{2} \braket{\{v\}^{M-2} | \{v\}^{M-2}}.
\end{align}
The remainder of the non-zero elements are the same as above for each vacuum, with no coupling between them
\begin{align}
	\gamma^{Vabcd(\mu),Uabcd(\mu)}_{kk}   &= 2 \eta (4) \mathcal{D}^{(0)}_{kk}(J(abcd)) \\
	\Gamma^{Vabcd(\mu),Uabcd(\mu)}_{kkll} &= 4 \eta (4) \mathcal{D}^{(0)}_{kkll}(J(abcd)) \\
	\Gamma^{Vabcd(\mu),Uabcd(\mu)}_{klkl} &= 2 \eta (4) \mathcal{P}^{(0)}_{klkl}(J(abcd)),
\end{align}
in terms of the $J$-sums \eqref{eq:jsum_d0_kk}, \eqref{eq:clean_d2D_sum} and \eqref{eq:clean_d2P_sum} evaluated with the common matrix $J(abcd)$ \eqref{eq:matsuno_4}, with exchange elements
\begin{align}
	\Gamma^{Vabcd(\mu),Uabcd(\mu)}_{kllk} = - \frac{1}{2} \Gamma^{Vabcd(\mu),Uabcd(\mu)}_{kkll}
\end{align}
and for each of the blocked levels
\begin{align}
	\Gamma^{Vabcd(\mu),Uabcd(\mu)}_{aakk} = - 2 \Gamma^{Vabcd(\mu),Uabcd(\mu)}_{akka} = \gamma^{Vabcd(\mu),Uabcd(\mu)}_{kk}.
\end{align}

\subsubsection{Seniority-Four Excitations} \label{sec:comb_tedious}
The last two possibilities introduce a final difficulty. Unfortunately, singlet excitations upon seniority-four states \emph{do not respect} an orthonormal basis. For example, consider the singlet excitation $A^0_{a'a}$, acting on the two vacuums
\begin{align}
	A^0_{a'a} \ket{ \varphi^{(1)}_{abcd} } &= \ket{(a'b)(cd)} \\
	A^0_{a'a} \ket{ \varphi^{(2)}_{abcd} } &= \frac{1}{\sqrt{3}} \left( \ket{(a'c)(bd)} - \ket{(a'd)(bc)} \right).
\end{align}
There is no reason the states on the right-hand side (RHS) should be the chosen orthonormal basis for the set $\{a',b,c,d\}$ in natural order. This would happen no matter the choice of basis, and thus leads to a tedious development with group theory. 

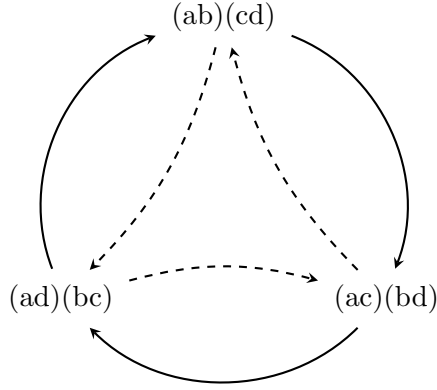
\begin{figure} 
	\begin{tikzpicture}[>=stealth,thick]
		% Define vertices
		\node (A) at (90:2.5)  {(ab)(cd)};
		\node (B) at (210:2.5) {(ad)(bc)};
		\node (C) at (330:2.5) {(ac)(bd)};
		
		% Inner cycle
		\draw[dashed, ->, bend left=15, shorten <=2pt, shorten >=2pt] (A) to (B);
		\draw[dashed, ->, bend left=15, shorten <=2pt, shorten >=2pt] (B) to (C);
		\draw[dashed, ->, bend left=15, shorten <=2pt, shorten >=2pt] (C) to (A);
		
		% Outer cycle
		\draw[->, bend left=45, shorten <=2pt, shorten >=2pt] (A) to (C);
		\draw[->, bend left=45, shorten <=2pt, shorten >=2pt] (C) to (B);
		\draw[->, bend left=45, shorten <=2pt, shorten >=2pt] (B) to (A);
	\end{tikzpicture} 
	\caption[]{Cyclic labelling of seniority-four states: for  $\omega[(ab)(cd)] = 1$ or  $\omega[(ab)(cd)] = 3$ \emph{clockwise} if $(a < b \;\text{and}\; c<d)$ or $(a > b \;\text{and}\; c>d)$, otherwise \emph{counterclockwise}. For $\omega[(ab)(cd)] = 2$ \emph{counterclockwise} if $(a < b \;\text{and}\; c<d)$ or $(a > b \;\text{and}\; c>d)$, otherwise \emph{clockwise}.}
	\label{fig:s4_vacs}
\end{figure}

Given an ordered set of indices $\{a,b,c,d\}$, \emph{not necessarily in natural order}, the first task is to determine the values of $\omega$ for the three vectors. In particular, define
\begin{align}
	\omega_1 [abcd] &= \omega[(ab)(cd)] \\
	\omega_2 [abcd] &= \omega[(ac)(bd)] \\
	\omega_3 [abcd] &= \omega[(ad)(bc)].
\end{align}
If $\ket{(ab)(cd)}$ is equivalent (see Table \ref{tab:abcd}) to the ordering in which the indices can be written in completely ascending order, then $\omega[(ab)(cd)] = 1$ etc. Identifying $\omega$ for each of the states is thus possible with a few discrete comparisons. In particular, if $(a < b \;\text{and}\; c<d)$ or $(a > b \;\text{and}\; c>d)$ then $\omega$ follows the \emph{clockwise} (solid) path in Figure \ref{fig:s4_vacs} when  $\omega[(ab)(cd)] = 1$ or  $\omega[(ab)(cd)] = 3$, and the \emph{counterclockwise} (dashed) path when  $\omega[(ab)(cd)] = 2$.
This is perhaps more clear in Table \ref{tab:abcd_labelling}, where the complete list of possibilities are written for $\{a,b,c,d\}=\{1,2,3,4\}$.
\begin{table}[ht!] 
	\centering % Center the table
	\begin{tabular}{c|c|c||c|c|c||c|c|c} % Specify the number of columns and alignment
		(ab)(cd)&(ac)(bd)&(ad)(bc)&(ab)(cd)&(ac)(bd)&(ad)(bc)&(ab)(cd)&(ac)(bd)&(ad)(bc)  \\ % Table header row
		\hline % Draw another horizontal line 
		(12)(34)\cellcolor{  red!25}&(13)(24)\cellcolor{ blue!25}&(14)(23)\cellcolor{green!25}
		&(13)(24)\cellcolor{ blue!25}&(12)(34)\cellcolor{  red!25}&(14)(32)\cellcolor{green!25}
		&(14)(23)\cellcolor{green!25}&(12)(43)\cellcolor{  red!25}&(13)(42)\cellcolor{ blue!25}  \\ % Table content row 1
		
		(21)(34)\cellcolor{  red!25}&(23)(14)\cellcolor{green!25}&(24)(13)\cellcolor{ blue!25}
		&(31)(24)\cellcolor{ blue!25}&(32)(14)\cellcolor{green!25}&(34)(12)\cellcolor{  red!25} 
		&(41)(23)\cellcolor{green!25}&(42)(13)\cellcolor{ blue!25}&(43)(12)\cellcolor{  red!25}  \\ % Table content row 2
		
		(12)(43)\cellcolor{  red!25}&(14)(23)\cellcolor{green!25}&(13)(24)\cellcolor{ blue!25}
		&(13)(42)\cellcolor{ blue!25}&(14)(32)\cellcolor{green!25}&(12)(34)\cellcolor{  red!25}
		&(14)(32)\cellcolor{green!25}&(13)(42)\cellcolor{ blue!25}&(12)(43)\cellcolor{  red!25}  \\ % Table content row 3
		
		(21)(43)\cellcolor{  red!25}&(24)(13)\cellcolor{ blue!25}&(23)(14)\cellcolor{green!25}
		&(31)(42)\cellcolor{ blue!25}&(34)(12)\cellcolor{  red!25}&(32)(14)\cellcolor{green!25}
		&(41)(32)\cellcolor{green!25}&(43)(12)\cellcolor{  red!25}&(42)(13)\cellcolor{ blue!25}  \\ % Table content row 4
		
		(34)(12)\cellcolor{  red!25}&(31)(42)\cellcolor{ blue!25}&(32)(41)\cellcolor{green!25} 
		&(24)(13)\cellcolor{ blue!25}&(21)(43)\cellcolor{  red!25}&(23)(41)\cellcolor{green!25}
		&(23)(14)\cellcolor{green!25}&(21)(34)\cellcolor{  red!25}&(24)(31)\cellcolor{ blue!25}  \\ % Table content row 5
		
		(34)(21)\cellcolor{  red!25}&(32)(41)\cellcolor{green!25}&(31)(42)\cellcolor{ blue!25}
		&(24)(31)\cellcolor{ blue!25}&(23)(41)\cellcolor{green!25}&(21)(43)\cellcolor{  red!25} 
		&(23)(41)\cellcolor{green!25}&(24)(31)\cellcolor{ blue!25}&(21)(34)\cellcolor{  red!25}  \\ % Table content row 6
		
		(43)(12)\cellcolor{  red!25}&(41)(32)\cellcolor{green!25}&(42)(31)\cellcolor{ blue!25}
		&(42)(13)\cellcolor{ blue!25}&(41)(23)\cellcolor{green!25}&(43)(21)\cellcolor{  red!25} 
		&(32)(14)\cellcolor{green!25}&(31)(24)\cellcolor{ blue!25}&(34)(21)\cellcolor{  red!25}  \\ % Table content row 7
		
		(43)(21)\cellcolor{  red!25}&(42)(31)\cellcolor{ blue!25}&(41)(32)\cellcolor{green!25}
		&(42)(31)\cellcolor{ blue!25}&(43)(21)\cellcolor{  red!25}&(41)(23)\cellcolor{green!25} 
		&(32)(41)\cellcolor{green!25}&(34)(21)\cellcolor{  red!25}&(31)(24)\cellcolor{ blue!25}  \\ % Table content row 8
	\end{tabular}
	\caption{Identification of seniority-four states: red ($\omega=1$), blue ($\omega=2)$ and green ($\omega=3$).} % Table caption
	\label{tab:abcd_labelling} % Label for referencing the table
\end{table}
$\omega_1[abcd]=1$ in the first three columns of Table \ref{tab:abcd_labelling}, while $\omega_1[abcd]=2$ in the three middle columns, and $\omega_1[abcd]=3$ in the last three columns. The corresponding values of $\omega_2[abcd]$ and $\omega_3[abcd]$ will differ, and give the rules summarized in Figure \ref{fig:s4_vacs}.

The 3 $\times$ 3 matrix 
\begin{align}
	\kappa_{ij} = \left(- \frac{1}{2} \right)^{ \vert \text{sgn} (i - \omega[j]) \vert },
\end{align}
has elements that represent the overlap between the seniority-four states in the natural order (rows) and the given order (columns). If the given order is the natural order, then
\begin{align}
	\kappa = \begin{pmatrix}
		1 & - \frac{1}{2} & - \frac{1}{2} \\
		- \frac{1}{2} & 1 & - \frac{1}{2} \\
		- \frac{1}{2} & - \frac{1}{2} & 1
	\end{pmatrix},
\end{align}
otherwise the columns of $\kappa$ are permuted. Unfortunately, all $3!=6$ permutations are possible. Density matrix elements coupling two distinct seniority-four states will depend on the elements of the \emph{vacuum overlap matrices}:
\begin{align}
	\Upsilon^{(1)} &= \begin{pmatrix}
		\kappa_{11} & \frac{1}{\sqrt{3}} (\kappa_{12} - \kappa_{13})  \\
		\frac{1}{\sqrt{3}} (\kappa_{21} - \kappa_{31})& \frac{1}{3} (\kappa_{22} - \kappa_{32} - \kappa_{23} + \kappa_{33})
	\end{pmatrix} \\
	\Upsilon^{(2)} &= \begin{pmatrix}
		\kappa_{12} & \frac{1}{\sqrt{3}} (\kappa_{11} - \kappa_{13}) \\
		\frac{1}{\sqrt{3}} (\kappa_{22} - \kappa_{32}) & \frac{1}{3} (\kappa_{21} - \kappa_{31} - \kappa_{23} + \kappa_{33})
	\end{pmatrix} \\
	\Upsilon^{(3)} &= \begin{pmatrix}
		\kappa_{13} & \frac{1}{\sqrt{3}} (\kappa_{11} - \kappa_{12}) \\
		\frac{1}{\sqrt{3}} (\kappa_{23} - \kappa_{33}) & \frac{1}{3} (\kappa_{21} - \kappa_{31} - \kappa_{22} + \kappa_{32})
	\end{pmatrix}.
\end{align}
Notice that this is not the best choice if, for some reason, one cared about the group theory of these objects. The correct choice is to have the second columns of $\Upsilon^{(3)}$ multiplied by a sign: in this way, going from $\Upsilon^{(1)}\rightarrow \Upsilon^{(2)}$ one exchanges the second indices of 1 with 2, and going from $\Upsilon^{(2)}\rightarrow \Upsilon^{(3)}$ one replaces the second indices of 2 with 3. But, that would introduce extra signs in some of the matrix elements to compute, leading to expressions that appeared asymetric and, likely, mistakes.

The vacuum overlap matrices can be computed for each possible $\kappa$, and the results are summarized in Table \ref{tab:kappa_matrices}.
\begin{table}[ht!] % [h] for here, you can change this option according to your needs
	\centering % Center the table
	\begin{tabular}{c|c|c|c} % Specify the number of columns and alignment\
		$\kappa$ & $\Upsilon^{(1)}$ & $\Upsilon^{(2)}$ & $\Upsilon^{(3)}$ \\
		\hline % Draw another horizontal line
		$\begin{pmatrix}
			1 & - \frac{1}{2} & - \frac{1}{2} \\
			- \frac{1}{2} & 1 & - \frac{1}{2} \\
			- \frac{1}{2} & - \frac{1}{2} & 1
		\end{pmatrix}$ 
		& $\begin{pmatrix}
			1 & 0 \\
			0 & 1
		\end{pmatrix}$
		& $\begin{pmatrix}
			- \frac{1}{2} & \frac{\sqrt{3}}{2} \\
			\frac{\sqrt{3}}{2} & \frac{1}{2}
		\end{pmatrix}$
		& $\begin{pmatrix}
			- \frac{1}{2} & - \frac{\sqrt{3}}{2} \\
			\frac{\sqrt{3}}{2} & - \frac{1}{2}
		\end{pmatrix}$ \\
		$\begin{pmatrix}
			1 & - \frac{1}{2} & - \frac{1}{2} \\
			- \frac{1}{2} & - \frac{1}{2} & 1\\
			- \frac{1}{2} & 1 & - \frac{1}{2}
		\end{pmatrix}$  
		& $\begin{pmatrix}
			1 & 0 \\
			0 & -1
		\end{pmatrix}$
		& $\begin{pmatrix}
			- \frac{1}{2} & - \frac{\sqrt{3}}{2} \\
			\frac{\sqrt{3}}{2} & - \frac{1}{2}
		\end{pmatrix}$
		& $\begin{pmatrix}
			- \frac{1}{2} & \frac{\sqrt{3}}{2} \\
			\frac{\sqrt{3}}{2} &   \frac{1}{2}
		\end{pmatrix}$ \\
		\hline
		$\begin{pmatrix}
			- \frac{1}{2} & 1 & - \frac{1}{2} \\
			1 & - \frac{1}{2} & - \frac{1}{2} \\
			- \frac{1}{2} & - \frac{1}{2} & 1
		\end{pmatrix}$
		& $\begin{pmatrix}
			- \frac{1}{2} & \frac{\sqrt{3}}{2} \\
			\frac{\sqrt{3}}{2} & \frac{1}{2}
		\end{pmatrix}$
		& $\begin{pmatrix}
			1 & 0 \\
			0 & 1
		\end{pmatrix}$
		& $\begin{pmatrix}
			- \frac{1}{2} & - \frac{\sqrt{3}}{2} \\
			- \frac{\sqrt{3}}{2} & \frac{1}{2}
		\end{pmatrix}$ \\
		$\begin{pmatrix}
			- \frac{1}{2} & 1 & - \frac{1}{2} \\
			- \frac{1}{2} & - \frac{1}{2} & 1 \\
			1 & - \frac{1}{2} & - \frac{1}{2}
		\end{pmatrix}$
		& $\begin{pmatrix}
			- \frac{1}{2} & - \frac{\sqrt{3}}{2} \\
			\frac{\sqrt{3}}{2} & - \frac{1}{2}
		\end{pmatrix}$
		& $\begin{pmatrix}
			1 & 0 \\
			0 & -1
		\end{pmatrix}$
		& $\begin{pmatrix}
			- \frac{1}{2} &  \frac{\sqrt{3}}{2} \\
			- \frac{\sqrt{3}}{2} & - \frac{1}{2}
		\end{pmatrix}$ \\
		\hline
		$\begin{pmatrix}
			- \frac{1}{2} & - \frac{1}{2} & 1 \\
			1 & - \frac{1}{2} & - \frac{1}{2} \\
			- \frac{1}{2} & 1 & - \frac{1}{2}
		\end{pmatrix}$
		& $\begin{pmatrix}
			- \frac{1}{2} &  \frac{\sqrt{3}}{2} \\
			- \frac{\sqrt{3}}{2} & - \frac{1}{2}
		\end{pmatrix}$
		& $\begin{pmatrix}
			- \frac{1}{2} & - \frac{\sqrt{3}}{2} \\
			- \frac{\sqrt{3}}{2} &   \frac{1}{2}
		\end{pmatrix}$
		& $\begin{pmatrix}
			1 & 0 \\
			0 & 1
		\end{pmatrix}$ \\
		$\begin{pmatrix}
			- \frac{1}{2} & - \frac{1}{2} & 1 \\
			- \frac{1}{2} & 1 & - \frac{1}{2} \\
			1 & - \frac{1}{2} & - \frac{1}{2}
		\end{pmatrix}$
		& $\begin{pmatrix}
			- \frac{1}{2} & - \frac{\sqrt{3}}{2} \\
			- \frac{\sqrt{3}}{2} &   \frac{1}{2}
		\end{pmatrix}$
		& $\begin{pmatrix}
			- \frac{1}{2} &  \frac{\sqrt{3}}{2} \\
			- \frac{\sqrt{3}}{2} & - \frac{1}{2}
		\end{pmatrix}$
		& $\begin{pmatrix}
			1 & 0 \\
			0 & -1
		\end{pmatrix}$
	\end{tabular}
	\caption{Vacuum overlap matrices for each possible $\kappa$.} % Table caption
	\label{tab:kappa_matrices} 
\end{table}
Thus, for a given order of indices one can immediately deduce the corresponding vacuum overlap matrices. The extension to seniority-six is no more difficult, but much more tedious. There are 15 primitive vectors and 5 elements in an orthogonal basis. Thus, there are 15 matrices $\Upsilon^{(\omega)}$ each of size $5 \times 5$. 

\subsubsection{3 shared blocked levels}
Suppose that the two seniority-four RG states have blocked levels $\{a,b,c',d'\}$ and $\{e,b,c',d'\}$ such that $c' < d'$. If $a<b$, then interpret $c = c'$ and $d = d'$, and if $a>b$ interpret $c = d'$ and $d=c'$ in the expressions that follow. This choice guarantees that the two vacuums for the $\{a,b,c,d\}$ states are
\begin{align}
	\ket{\varphi^{(1)}_{abcd}} &= \ket{(ab)(cd)} \\
	\ket{\varphi^{(2)}_{abcd}} &= \frac{1}{\sqrt{3}} \left( \ket{(ac)(bd)} - \ket{(ad)(bc)} \right).
\end{align}
To evaluate overlaps, the states $\ket{(eb)(cd)}$,$\ket{(ec)(bd)}$ and $\ket{(ed)(bc)}$ must be sorted with the parameter $\omega$ so that the corresponding vacuum overlap matrices are obtained. The matrix elements in the blocked levels are then computed from the $J$-sums \eqref{eq:jsum_f0_bc} and \eqref{eq:jsum_b0_cb}, and the results are presented in Table \ref{tab:fs_3shared}. 
\begin{table}[ht!] % [h] for here, you can change this option according to your needs
	\centering % Center the table
	\begin{tabular}{c|c|c|c|c} % Specify the number of columns and alignment
		& $\braket{\varphi^{(1)}_{ebcd} || \varphi^{(1)}_{abcd}}$ 
		& $\braket{\varphi^{(2)}_{ebcd} || \varphi^{(1)}_{abcd}}$ 
		& $\braket{\varphi^{(1)}_{ebcd} || \varphi^{(2)}_{abcd}}$ 
		& $\braket{\varphi^{(2)}_{ebcd} || \varphi^{(2)}_{abcd}}$   \\ % Table header row
		\hline % Draw another horizontal line
		$\gamma_{ea}$   & $\Upsilon^{(1)}_{11} \eta(4) \mathcal{F}^{(0)}_{ea}$ 
						& $\Upsilon^{(1)}_{21} \eta(4) \mathcal{F}^{(0)}_{ea}$  
						& $\Upsilon^{(1)}_{12} \eta(4) \mathcal{F}^{(0)}_{ea}$
						& $\Upsilon^{(1)}_{22} \eta(4) \mathcal{F}^{(0)}_{ea}$\\
		$\Gamma_{baeb}$ & $\gamma_{ea}$ & $\gamma_{ea}$ & $-\gamma_{ea}$ & $-\gamma_{ea}$ \\
		$\Gamma_{caec}$ & $  \Upsilon^{(3)}_{11}  \eta(4) \mathcal{F}^{(0)}_{ea}$ 
						& $  \Upsilon^{(3)}_{21}  \eta(4) \mathcal{F}^{(0)}_{ea}$
						& $- \Upsilon^{(3)}_{12}  \eta(4) \mathcal{F}^{(0)}_{ea}$
						& $- \Upsilon^{(3)}_{22}  \eta(4) \mathcal{F}^{(0)}_{ea}$\\
		$\Gamma_{daed}$ & $\Upsilon^{(2)}_{11} \eta(4) \mathcal{F}^{(0)}_{ea}$ 
						& $\Upsilon^{(2)}_{21} \eta(4) \mathcal{F}^{(0)}_{ea}$
						& $\Upsilon^{(2)}_{12} \eta(4) \mathcal{F}^{(0)}_{ea}$
						& $\Upsilon^{(2)}_{22} \eta(4) \mathcal{F}^{(0)}_{ea}$ \\
		\hline
		$\gamma_{ae}$   & $           \Upsilon^{(1)}_{11} \eta(4) \mathcal{B}^{(0)}_{ae}$ 
		                & $           \Upsilon^{(1)}_{21} \eta(4) \mathcal{B}^{(0)}_{ae}$
						& $           \Upsilon^{(1)}_{12} \eta(4) \mathcal{B}^{(0)}_{ae}$
						& $           \Upsilon^{(1)}_{22} \eta(4) \mathcal{B}^{(0)}_{ae}$\\
		$\Gamma_{beab}$ & $-2\gamma_{ae}$ & $-2\gamma_{ae}$ & 0 & 0 \\
		$\Gamma_{ceac}$ & $           \Upsilon^{(2)}_{11} \eta(4) \mathcal{B}^{(0)}_{ae}$ 
						& $           \Upsilon^{(2)}_{21} \eta(4) \mathcal{B}^{(0)}_{ae}$
						& $ -\sqrt{3} \Upsilon^{(2)}_{11} \eta(4) \mathcal{B}^{(0)}_{ae}$
						& $ -\sqrt{3} \Upsilon^{(2)}_{21} \eta(4) \mathcal{B}^{(0)}_{ae}$ \\
		$\Gamma_{dead}$ & $           \Upsilon^{(3)}_{11} \eta(4) \mathcal{B}^{(0)}_{ae}$
						& $           \Upsilon^{(3)}_{21} \eta(4) \mathcal{B}^{(0)}_{ae}$
						& $  \sqrt{3} \Upsilon^{(3)}_{11} \eta(4) \mathcal{B}^{(0)}_{ae}$
						& $  \sqrt{3} \Upsilon^{(3)}_{21} \eta(4) \mathcal{B}^{(0)}_{ae}$
	\end{tabular}
	\caption{TDM elements in blocked levels for two seniority-four states sharing 3 blocked levels. The first block of four rows are the 1-TDM and exchange 2-TDM elements for forward-scattering while the second block of four rows are for backward-scattering. The $J$-sums \eqref{eq:jsum_f0_bc} and \eqref{eq:jsum_b0_cb} are computed from the common matrix $J(ecbd)$ \eqref{eq:matsuno_4}.} % Table caption
	\label{tab:fs_3shared} % Label for referencing the table
\end{table}
For all four combinations of vacuums, the direct elements are
\begin{align}
	0 &= \Gamma_{aaea} = \Gamma_{eeea} \\
	\gamma_{ea} &= \Gamma_{bbea} = \Gamma_{ccea} = \Gamma_{ddea} \\
	\gamma_{ae} &= \Gamma_{aaae} = \Gamma_{eeae} = \Gamma_{bbae} = \Gamma_{ccae} = \Gamma_{ddae}.
\end{align}
The elements involving unblocked levels are essentially the same as those for the coupling of seniority-two RG states with one common blocked level:
\begin{align}
	\Gamma^{Vebcd(\mu),Uabcd(\nu)}_{kkea} &= \;\;\;2 \Upsilon^{(1)}_{\mu \nu} \eta(4) \mathcal{F}^{(0)}_{kkea}(J(ebcd)) \\
	\Gamma^{Vebcd(\mu),Uabcd(\nu)}_{kkae} &=      -2 \Upsilon^{(1)}_{\mu \nu} \eta(4) \mathcal{B}^{(0)}_{kkae}(J(ebcd)) \\
	\Gamma^{Vebcd(\mu),Uabcd(\nu)}_{akek} &= \;\;\;\;\Upsilon^{(1)}_{\mu \nu} \eta(4) \mathcal{P}^{(0)}_{akek}(J(ebcd)) \\
	\Gamma^{Vebcd(\mu),Uabcd(\nu)}_{kake} &= \;\;\;\;\Upsilon^{(1)}_{\mu \nu} \eta(4) \mathcal{P}^{(0)}_{keka}(J(ebcd)),
\end{align}
using the $J$-sums \eqref{eq:jsum_f0_kkbc}, \eqref{eq:jsum_b0_kkcb}, \eqref{eq:jsum_p0_ckbk} and \eqref{eq:jsum_p0_kbkc} with the common matrix $J(ebcd)$ \eqref{eq:matsuno_4}. Again, to be coherent in $\mathcal{P}^{(0)}_{keka}$, $e$ is unblocked in the right state so that $U_e \neq 0$. In all cases, the exchange elements are
\begin{align}
	\Gamma^{Vebcd(\mu),Uabcd(\nu)}_{kaek} &= - \frac{1}{2} \Gamma^{Vebcd(\mu),Uabcd(\nu)}_{kkea} \\
	\Gamma^{Vebcd(\mu),Uabcd(\nu)}_{keak} &= - \frac{1}{2} \Gamma^{Vebcd(\mu),Uabcd(\nu)}_{kkae}.
\end{align}

\subsubsection{2 shared blocked levels}
With two shared blocked levels, there is no possibility for one-electron transitions, and the only non-zero two-electron transitions necessarily involve all the blocked levels which are not shared. As was the case for the coupling of seniority-two with seniority-four states, there are different results depending on which indices are shared. Suppose the two shared indices are $a$ and $b$, i.e. that we are considering transitions from states with blocked levels $a,b,c,d$ to states with blocked levels $c,d,e,f$. Again, the first task is to build the vacuum overlap matrices $\Upsilon^{(\omega)}$.

For shared blocked levels $a$ and $b$ (equivalently $c$ and $d$), the matrix elements are presented in Table \ref{tab:ab_44_2shared}.
\begin{table}[ht!] % [h] for here, you can change this option according to your needs
	\centering % Center the table
	\begin{tabular}{c|c|c|c|c} % Specify the number of columns and alignment
	& $\braket{\varphi^{(1)}_{cdef} || \varphi^{(1)}_{abcd}}$
	& $\braket{\varphi^{(2)}_{cdef} || \varphi^{(1)}_{abcd}}$
	& $\braket{\varphi^{(1)}_{cdef} || \varphi^{(2)}_{abcd}}$
	& $\braket{\varphi^{(2)}_{cdef} || \varphi^{(2)}_{abcd}}$ \\ % Table header row
	\hline % Draw another horizontal line
	$\Gamma_{eafb}$
	& $ \Upsilon^{(1)}_{11}  \mathcal{F}^{(0)}_{eafb}$
	& $ \Upsilon^{(1)}_{21}  \mathcal{F}^{(0)}_{eafb}$
	& $ \Upsilon^{(1)}_{12}  \mathcal{F}^{(0)}_{eafb}$
	& $ \Upsilon^{(1)}_{22}  \mathcal{F}^{(0)}_{eafb}$ \\
	$\Gamma_{ebfa}$
	& $ \Upsilon^{(1)}_{11}  \mathcal{F}^{(0)}_{eafb}$
	& $ \Upsilon^{(1)}_{21}  \mathcal{F}^{(0)}_{eafb}$
	& $-\Upsilon^{(1)}_{12}  \mathcal{F}^{(0)}_{eafb}$
	& $-\Upsilon^{(1)}_{22}  \mathcal{F}^{(0)}_{eafb}$ \\
	\hline
	$\Gamma_{abef}$ 
	& $2 \Upsilon^{(1)}_{11} \mathcal{X}^{(0)}_{abef}$
	& $2 \Upsilon^{(1)}_{21} \mathcal{X}^{(0)}_{abef}$
	& 0 & 0 \\
	$\Gamma_{afeb}$ 
	& $- \Upsilon^{(1)}_{11}   \mathcal{X}^{(0)}_{abef}$
	& $- \Upsilon^{(1)}_{21}   \mathcal{X}^{(0)}_{abef}$
	& $  \Upsilon^{(1)}_{12}   \mathcal{X}^{(0)}_{abef}$
	& $  \Upsilon^{(1)}_{22}   \mathcal{X}^{(0)}_{abef}$ \\
	$\Gamma_{abfe}$ 
	& $2 \Upsilon^{(1)}_{11}  \mathcal{X}^{(0)}_{abfe}$
	& $2 \Upsilon^{(1)}_{21}  \mathcal{X}^{(0)}_{abfe}$
	& 0 & 0 \\
	$\Gamma_{aefb}$
	& $- \Upsilon^{(1)}_{11}  \mathcal{X}^{(0)}_{abfe}$
	& $- \Upsilon^{(1)}_{21}  \mathcal{X}^{(0)}_{abfe}$
	& $  \Upsilon^{(1)}_{12}  \mathcal{X}^{(0)}_{abfe}$
	& $  \Upsilon^{(1)}_{22}  \mathcal{X}^{(0)}_{abfe}$ \\
	$\Gamma_{baef}$
	& $2 \Upsilon^{(1)}_{11}  \mathcal{X}^{(0)}_{baef}$
	& $2 \Upsilon^{(1)}_{21}  \mathcal{X}^{(0)}_{baef}$
	& 0 & 0 \\
	$\Gamma_{bfea}$
	& $- \Upsilon^{(1)}_{11}  \mathcal{X}^{(0)}_{baef}$
	& $- \Upsilon^{(1)}_{21}  \mathcal{X}^{(0)}_{baef}$
	& $  \Upsilon^{(1)}_{12}  \mathcal{X}^{(0)}_{baef}$
	& $  \Upsilon^{(1)}_{22}  \mathcal{X}^{(0)}_{baef}$ \\
	$\Gamma_{bafe}$
	& $2 \Upsilon^{(1)}_{11}  \mathcal{X}^{(0)}_{bafe}$
	& $2 \Upsilon^{(1)}_{21}  \mathcal{X}^{(0)}_{bafe}$
	& 0 & 0 \\
	$\Gamma_{befa}$
	& $- \Upsilon^{(1)}_{11}  \mathcal{X}^{(0)}_{bafe}$
	& $- \Upsilon^{(1)}_{21}  \mathcal{X}^{(0)}_{bafe}$
	& $  \Upsilon^{(1)}_{12}  \mathcal{X}^{(0)}_{bafe}$
	& $  \Upsilon^{(1)}_{22}  \mathcal{X}^{(0)}_{bafe}$ \\
	\hline
	$\Gamma_{aebf}$
	& $ \Upsilon^{(1)}_{11}  \mathcal{B}^{(0)}_{aebf}$
	& $ \Upsilon^{(1)}_{21}  \mathcal{B}^{(0)}_{aebf}$
	& $ \Upsilon^{(1)}_{12}  \mathcal{B}^{(0)}_{aebf}$
	& $ \Upsilon^{(1)}_{22}  \mathcal{B}^{(0)}_{aebf}$ \\
	$\Gamma_{afbe}$
	& $ \Upsilon^{(1)}_{11}  \mathcal{B}^{(0)}_{aebf}$
	& $ \Upsilon^{(1)}_{21}  \mathcal{B}^{(0)}_{aebf}$
	& $-\Upsilon^{(1)}_{12}  \mathcal{B}^{(0)}_{aebf}$
	& $-\Upsilon^{(1)}_{22}  \mathcal{B}^{(0)}_{aebf}$ 
	\end{tabular}
	\caption{2 shared blocked levels, $(ab)$ pattern. In all cases, the $J$-sums \eqref{eq:jsum_f0_acbd}, \eqref{eq:jsum_x0_abcd} and \eqref{eq:jsum_b0_cadb} are computed from the matrix $J(cdef)$ \eqref{eq:matsuno_4}, and each element is multiplied by $\eta(4)$.} % Table caption
	\label{tab:ab_44_2shared} % Label for referencing the table
\end{table}
These elements appear intuitive enough. There are direct and exhcange elements that behave in a predictable way, along with pair type elements behaving in a distinct manner.

When the shared levels are $a,c$ ($\rho = 1$) or $a,d$ ($\rho = -1$), the same $J$-sums are required, but the prefactors become much more complicated and are presented in Table \ref{tab:ac_44_2shared}. 
\begin{table}[ht!] % [h] for here, you can change this option according to your needs
	\centering % Center the table
	\begin{tabular}{c|c|c|c|c} % Specify the number of columns and alignment
		& $\braket{\varphi^{(1)}_{bdef} || \varphi^{(1)}_{abcd}}$
		& $\braket{\varphi^{(2)}_{bdef} || \varphi^{(1)}_{abcd}}$
		& $\braket{\varphi^{(1)}_{bdef} || \varphi^{(2)}_{abcd}}$
		& $\braket{\varphi^{(2)}_{bdef} || \varphi^{(2)}_{abcd}}$ \\ % Table header row
		\hline % Draw another horizontal line
		$\Gamma_{eafc}$
		& $ \Upsilon^{(2)}_{11} \mathcal{F}^{(0)}_{eafc}$
		& $ \Upsilon^{(2)}_{21} \mathcal{F}^{(0)}_{eafc}$
		& $\rho \Upsilon^{(2)}_{12} \mathcal{F}^{(0)}_{eafc}$
		& $\rho \Upsilon^{(2)}_{22} \mathcal{F}^{(0)}_{eafc}$ \\
		$\Gamma_{ecfa}$
		& $ \Upsilon^{(3)}_{11} \mathcal{F}^{(0)}_{eafc}$
		& $ \Upsilon^{(3)}_{21} \mathcal{F}^{(0)}_{eafc}$
		& $\rho \Upsilon^{(3)}_{12} \mathcal{F}^{(0)}_{eafc}$
		& $\rho \Upsilon^{(3)}_{22} \mathcal{F}^{(0)}_{eafc}$ \\
		\hline
		$\Gamma_{acef}$
		& $-\Upsilon^{(1)}_{11}  \mathcal{X}^{(0)}_{acef}$
		& $-\Upsilon^{(1)}_{21}  \mathcal{X}^{(0)}_{acef}$
		& $ \sqrt{3} \rho \Upsilon^{(1)}_{11}  \mathcal{X}^{(0)}_{acef} $
		& $ \sqrt{3} \rho \Upsilon^{(1)}_{21}  \mathcal{X}^{(0)}_{acef} $ \\
		$\Gamma_{afec}$
		& $-\Upsilon^{(3)}_{11} \mathcal{X}^{(0)}_{acef}$
		& $-\Upsilon^{(3)}_{21} \mathcal{X}^{(0)}_{acef}$
		& $\;\;-\rho \Upsilon^{(3)}_{12} \mathcal{X}^{(0)}_{acef}$
		& $\;\;-\rho \Upsilon^{(3)}_{22} \mathcal{X}^{(0)}_{acef}$\\
		$\Gamma_{acfe}$
		& $-\Upsilon^{(1)}_{11} \mathcal{X}^{(0)}_{acfe}$
		& $-\Upsilon^{(1)}_{21} \mathcal{X}^{(0)}_{acfe}$
		& $ \sqrt{3} \rho \Upsilon^{(1)}_{11}  \mathcal{X}^{(0)}_{acfe} $
		& $ \sqrt{3} \rho \Upsilon^{(1)}_{21}  \mathcal{X}^{(0)}_{acfe} $ \\
		$\Gamma_{aefc}$
		& $-\Upsilon^{(2)}_{11} \mathcal{X}^{(0)}_{acfe}$
		& $-\Upsilon^{(2)}_{21} \mathcal{X}^{(0)}_{acfe}$
		& $\;\;-\rho \Upsilon^{(2)}_{12} \mathcal{X}^{(0)}_{acfe}$
		& $\;\;-\rho \Upsilon^{(2)}_{22} \mathcal{X}^{(0)}_{acfe}$\\
		$\Gamma_{caef}$
		& $-\Upsilon^{(1)}_{11} \mathcal{X}^{(0)}_{caef}$
		& $-\Upsilon^{(1)}_{21} \mathcal{X}^{(0)}_{caef}$
		& $ \sqrt{3} \rho \Upsilon^{(1)}_{11} \mathcal{X}^{(0)}_{caef} $
		& $ \sqrt{3} \rho \Upsilon^{(1)}_{21} \mathcal{X}^{(0)}_{caef} $ \\
		$\Gamma_{cfea}$ 
		& $-\Upsilon^{(2)}_{11} \mathcal{X}^{(0)}_{caef}$
		& $-\Upsilon^{(2)}_{21} \mathcal{X}^{(0)}_{caef}$
		& $\;\;-\rho \Upsilon^{(2)}_{12} \mathcal{X}^{(0)}_{caef}$
		& $\;\;-\rho \Upsilon^{(2)}_{22} \mathcal{X}^{(0)}_{caef}$\\
		$\Gamma_{cafe}$
		& $-\Upsilon^{(1)}_{11} \mathcal{X}^{(0)}_{cafe}$
		& $-\Upsilon^{(1)}_{21} \mathcal{X}^{(0)}_{cafe}$
		& $ \sqrt{3} \rho \Upsilon^{(1)}_{11} \mathcal{X}^{(0)}_{cafe} $
		& $ \sqrt{3} \rho \Upsilon^{(1)}_{21} \mathcal{X}^{(0)}_{cafe} $ \\
		$\Gamma_{cefa}$
		& $-\Upsilon^{(3)}_{11} \mathcal{X}^{(0)}_{cafe}$
		& $-\Upsilon^{(3)}_{21} \mathcal{X}^{(0)}_{cafe}$
		& $\;\;-\rho \Upsilon^{(3)}_{12} \mathcal{X}^{(0)}_{cafe}$
		& $\;\;-\rho \Upsilon^{(3)}_{22} \mathcal{X}^{(0)}_{cafe}$\\
		\hline
		$\Gamma_{aecf}$
		& $     \Upsilon^{(2)}_{11} \mathcal{B}^{(0)}_{aecf}$
		& $     \Upsilon^{(2)}_{21} \mathcal{B}^{(0)}_{aecf}$
		& $\rho \Upsilon^{(2)}_{12} \mathcal{B}^{(0)}_{aecf}$
		& $\rho \Upsilon^{(2)}_{22} \mathcal{B}^{(0)}_{aecf}$ \\
		$\Gamma_{afce}$
		& $ \Upsilon^{(3)}_{11} \mathcal{B}^{(0)}_{aecf}$
		& $ \Upsilon^{(3)}_{21} \mathcal{B}^{(0)}_{aecf}$
		& $\rho \Upsilon^{(3)}_{12} \mathcal{B}^{(0)}_{aecf}$
		& $\rho \Upsilon^{(3)}_{22} \mathcal{B}^{(0)}_{aecf}$
	\end{tabular}
	\caption{2 shared blocked levels, $(ac)$ pattern ($\rho = 1$). In all cases, the $J$-sums \eqref{eq:jsum_f0_acbd}, \eqref{eq:jsum_x0_abcd} and \eqref{eq:jsum_b0_cadb} are computed from the matrix $J(bdef)$ \eqref{eq:matsuno_4}, and each factor is multiplied by $\eta(4)$. For $(ad)$ patterns, swap $c\leftrightarrow d$ and take $\rho = -1$.} % Table caption
	\label{tab:ac_44_2shared} % Label for referencing the table
\end{table}

This completes the list of possible matrix elements that would couple seniorities zero, two and four.

\section{Discussion} \label{sec:dis}
The presented expressions for matrix elements between RG states always reduce to the computation of cofactors of the effective overlap matrix. Treated directly, this is unacceptably expensive. In particular, the coupling between each seniority-zero RG state and each seniority-four RG states requires fourth cofactors, of which there are $\binom{N}{4}^2$. Each cofactor is a determinant to compute with $\mathcal{O}(N^3)$ operations. Thankfully, almost all of these computations are redundant. 

The norm of a seniority-zero RG state is (up to the factor $\eta(0)$) the determinant of the effective overlap matrix which becomes the Jacobian of the EBV equations $\bar{J}$. Generally $\bar{J}$ is non-singular and hence invertible. (There are exceptional cases when the single-particle energies are exactly degenerate, which will be considered in a separate manuscript.) A consequence of Cramer's rule is that the inverse of $\bar{J}$
\begin{align}
	\bar{J}^{-1} = \frac{\text{adj}(\bar{J})}{\det (\bar{J})}
\end{align}
is the adjugate matrix (matrix of first cofactors of $\bar{J}$) scaled by its determinant. Thus, inverting $\bar{J}$ numerically provides the $N^2$ first cofactors in one stroke
\begin{align}
	\frac{[\bar{J}]^{p,q}}{\det(\bar{J})} = \bar{J}^{-1}_{qp},
\end{align}
noticing that the indices are swapped relative to the inverse. Second cofactors may then be computed on-the-fly from a theorem of Jacobi: the scaled $k$th order cofactors of $\bar{J}$ are a $k\times k$ determinant of its scaled first cofactors.\cite{vein_book} In particular, the scaled second cofactors are
\begin{align} \label{eq:jacobi_2}
	\frac{[\bar{J}]^{pq,rs}}{\det (\bar{J})} = 
	\frac{[\bar{J}]^{p,r}}{\det(\bar{J})} \frac{[\bar{J}]^{q,s}}{\det(\bar{J})} -
	\frac{[\bar{J}]^{p,s}}{\det(\bar{J})} \frac{[\bar{J}]^{q,r}}{\det(\bar{J})}.
\end{align}
All of the pertinent cofactors for RDM elements are thus constructed simply by inverting $\bar{J}$.

For two distinct seniority-zero RG states, the effective overlap matrix $J$ is necessarily singular but the first cofactors are still computable just as easily. With the singular value decomposition (SVD)
\begin{align}
	J = \mathcal{U} \varSigma \mathcal{V}^{\dagger}
\end{align}
one can compute the ``inverse''
\begin{align}
	J^{-1} &= \mathcal{V} \varSigma^{-1} \mathcal{U}^{\dagger} \\
	&= \sum_{p} \mathcal{V}_p \frac{1}{\varsigma_p} \mathcal{U}^{\dagger}_p \label{eq:diverge_inverse}
\end{align}
in terms of the columns of the unitary matrices $\mathcal{U}$ and $\mathcal{V}$ and the singular values $\varsigma$. Strictly speaking $J$ is singular and thus $J^{-1}$ diverges, but the same approach yields the first cofactors. Each small singular value contributes a simple pole in equation \eqref{eq:diverge_inverse}, which when multiplied by $\det(J)$ will yield a finite expression. In particular,
\begin{align}
	\tilde{J} = \det(\mathcal{U}) \det(\mathcal{V}) \prod_r \varsigma_r J^{-1}.
\end{align}
If $J$ has one singular value equal to zero, then $\tilde{J}$ is the residue of the single simple pole in \eqref{eq:diverge_inverse} and the first cofactors are
\begin{align}
	[J]^{p,q} = \tilde{J}_{qp}.
\end{align}
If $J$ has two singular values equal to zero, the first cofactors will all vanish as the simple poles will be scaled by two factors of zero. Second cofactors are computed in the same way, by plugging \eqref{eq:diverge_inverse} into \eqref{eq:jacobi_2} and isolating the non-zero components (see refs. \citenum{johnson:2024b} and \citenum{chen:2023} for details). Third and fourth cofactors are computable in precisely the same manner. Thus, in all cases a single linear algebra operation, here an SVD rather than matrix inversion, yields all the required information to compute the cofactors. As understood in ref. \citenum{johnson:2024b}, the number of near-zero singular values depends on the RG states involved, and changes as a function of the pairing strength $g$. The number and type of excitation here is much larger (compared with only the seniority-zero channel) and thus the presentation of the analogues of Slater-Condon rules will be the second manuscript in this series.

The complete set of matrix elements can be computed perfectly in parallel. The individual RG states are defined by their EBV, which are computed from the parameters $\{\varepsilon\}$ and $g$ of a reduced BCS Hamiltonian. These parameters have been variationally optimized for a single RG state and thus do not change. The EBV for each required state can thus be computed at once in parallel. Next, the matrix elements for each pair of states can be computed in parallel as the only common required information, the EBV for each state, $\{\varepsilon\}$ and $g$, have been precomputed and do not change.

While the focus in this contribution is coupling between RG states of a spin-conserving operator, the required modifications for spin-changing operators are straightforward from the Wigner-Eckart theorem. Consider for example seniority-two RG states with the triplet pair creators
\begin{align}
	\mathcal{Q}^{(1)}_{pq} &= a^{\dagger}_{p\uparrow} a^{\dagger}_{q\uparrow} \\
	\mathcal{Q}^{(0)}_{pq} &= \frac{1}{\sqrt{2}} \left( a^{\dagger}_{p\uparrow} a^{\dagger}_{q\downarrow} + 
	a^{\dagger}_{p\downarrow} a^{\dagger}_{q\uparrow} \right) \\
	\mathcal{Q}^{(-1)}_{pq} &= a^{\dagger}_{p\downarrow} a^{\dagger}_{q\downarrow}.
\end{align}
These will couple to seniority-zero RG states through the triplet excitation operators
\begin{align}
	\mathcal{T}^{(1)}_{pq} &= - a^{\dagger}_{p\uparrow}a_{q\downarrow} \\
	\mathcal{T}^{(0)}_{pq} &= \frac{1}{\sqrt{2}} \left( a^{\dagger}_{p\uparrow}a_{q\uparrow} - a^{\dagger}_{p\downarrow}a_{q\downarrow} \right) \\
	\mathcal{T}^{(-1)}_{pq} &= a^{\dagger}_{p\downarrow}a_{q\uparrow}.
\end{align}
With $m$ and $m'$ arbitrary labels, the only non-zero one-body elements are
\begin{align}
	\braket{\{v\}^M | \mathcal{T}^{(m)}_{ab} \mathcal{Q}^{(m')}_{ab} | \{u\}^{M-1} } &= 
	- \sqrt{3} \braket{1m,1m'|00} g \eta(0) \mathcal{D}^{(2)}_{ab}(J)
\end{align}
in terms of the Clebsch-Gordan (CG) coefficients $\braket{1m,1m'|00}$. For two triplets coupling to a singlet, the only non-zero coefficients are
\begin{align}
	\braket{11,1-1|00} = \braket{1-1,11|00} = - \braket{10,10|00} = \frac{1}{\sqrt{3}}.
\end{align}
The result is that the matrix element factors into a CG coefficient and a reduced element, here the $J$-sum, which has already been computed. This is the Wigner-Eckart theorem. Couplings between higher spin states will follow in a similar manner. The development is tedious but no new $J$-sums will appear for two-body operators.

\begin{figure} 
	\begin{subfigure}{\textwidth}
		\includegraphics[width=0.485\textwidth]{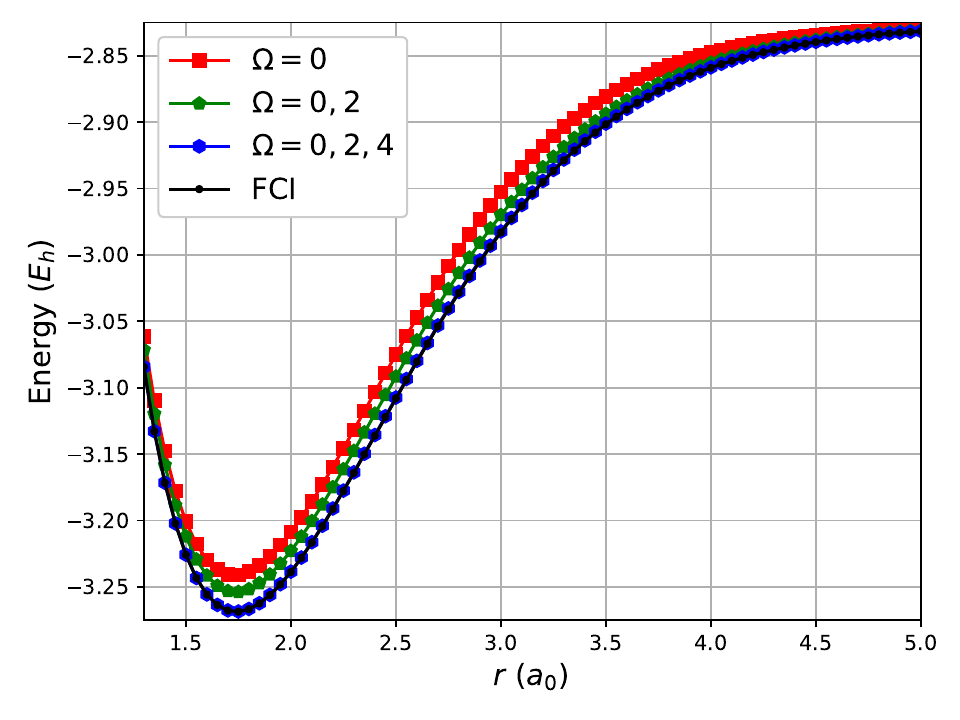}
	\end{subfigure}
	\caption[]{Seniority-based CI curves for the symmetric bond dissociation of linear H$_6$. Slater determinant and RG state seniority-based CI results are indiscernible. Results are computed in the optimal orbitals for seniority-zero Slater determinant CI obtained in ref. \citenum{johnson:2020} in the STO-6G basis.}
	\label{fig:s024_CI_h6}
\end{figure}

Finally, a numerical demonstration that these matrix elements are indeed correct is warranted. The dissociation of linear H$_6$ is small enough that all the necessary cofactors can be computed by brute-force. With orbitals optimized for the seniority-zero Slater determinant CI from ref. \citenum{johnson:2020}, seniority-based Slater determinant CI curves were computed with PyCI\cite{richer:2024} and are shown in Figure \ref{fig:s024_CI_h6}. Complete seniority-based RG state CI curves were also computed. These results should match numerically as on paper they are different representations of exactly the same space. Indeed they are found to match to at least 12 decimal places everywhere.
A few points for the symmetric dissociation of linear H$_8$ were also computed to verify the agreement between Slater determinant and RG state CI, but this represents the practical limit of what may be achieved by computing the cofactors by brute force. 

\begin{figure} 
	\begin{subfigure}{\textwidth}
		\includegraphics[width=0.485\textwidth]{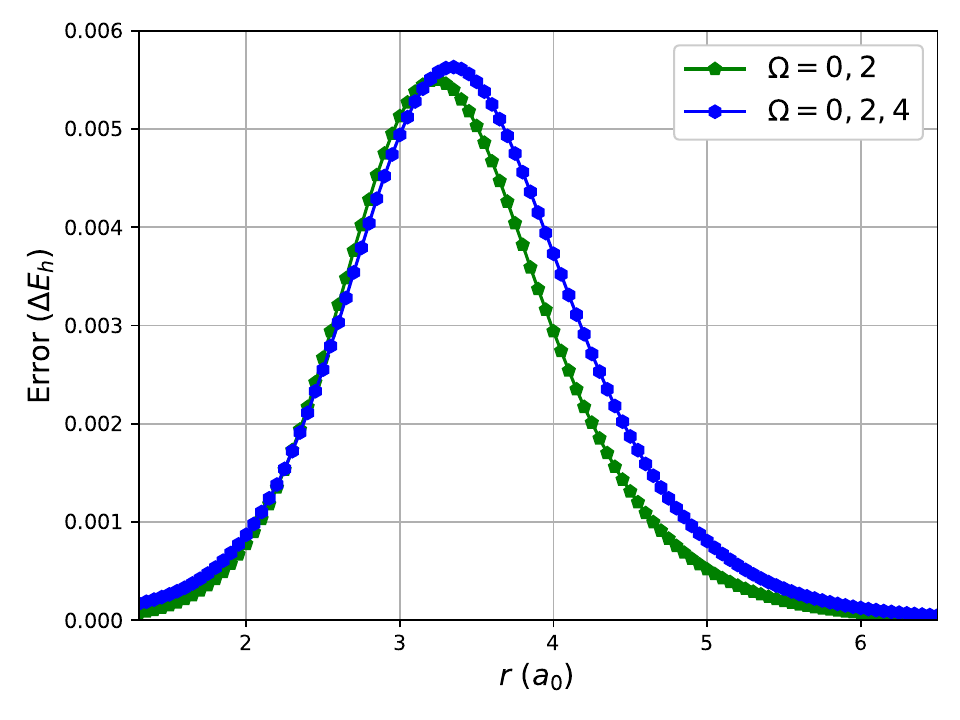}
	\end{subfigure}
	\caption[]{Errors of RGCISD in $\Omega=0,2$ and $\Omega=0,2,4$ channels for the symmetric bond dissociation of linear H$_6$. Results are computed in the optimal orbitals for seniority-zero Slater determinant CI obtained in ref. \citenum{johnson:2020} in the STO-6G basis.}
	\label{fig:s024_CISD_h6}
\end{figure}

The endgame is to replace seniority-based CI with a short excitation-based CI of RG states. Seniority-zero RG states can be classified by excitation level by comparing their corresponding bitstrings.\cite{johnson:2024b} A pair-single excitation has a bitstring that differs from the reference by one 1 and one 0, while a pair-double has a bitstring that differs from the reference by two 1s and two 0s. The same is generally true for RG states of non-zero seniorities. Start with a particular seniority-zero RG state as a reference. For linear H$_6$, the reference is 101010.\cite{fecteau:2022} Seniority-two \emph{singles} are represented by bitstrings where one 1 is replaced by an x and one 0 is replace by an x. For USDs, this is precisely a one-electron excitation. Seniority-two doubles occur in two manners: two 1s become x and one 0 becomes 1 or one 1 becomes 0 and two 0s become x. There are $M(N-M)$ seniority-two singles and $\binom{M}{2}(N-M) + M\binom{N-M}{2}$ seniority-two doubles. In seniority-four, there are no singles and only one type of double: two 1s become x and two 0s become x. A CI of one RG state plus its singles and doubles was employed for linear H$_6$. The results are not visually discernible from the previous curves, but the respective errors in each seniority channel are presented in Figure \ref{fig:s024_CISD_h6}.
The errors are computed relative to the complete CI in the given seniorities. In both cases, the excitation based RGCI is a good approximation for a small fraction of the cost. This treatment is naive and meant as a proof of principle. The identification of Slater-Condon rules will allow for the removal of deadweight and the inclusion of the necessary triples or quadruples.

\section{Conclusion}
In this contribution, matrix elements are computed between RG states of seniorities zero, two and four as a general two-body operator can only change the seniority by four. In each case, all that is required are cofactors of the effective overlap matrix $J$, which may be computed very efficiently by a singular value decomposition.\cite{johnson:2024b} Numerical computation of cofactors is left to the next manuscript in the series as this generally requires more information about particular sets of states to discern general rules. Matrix elements factor into common reduced quatities, $J$-sums, and specific coupling constants. Many elements appear for RG states that would vanish for CSFs. Couplings between RG states of higher seniorities introduces no additional types of elements. The only complication will be the tedious bookkeeping to correctly account for an orthonormal basis. Small proof of principle calculations demonstrate that the matrix elements have been correctly computed and that an excitation-based CI of RG states is a promising alternative to seniority-based CI. 

\section{Acknowledgments}
P. A. J. was supported by NSERC and the Digital Research Alliance of Canada.

\section{Data Availability}
The data that support the findings of this study are available from the corresponding author upon reasonable request.

\appendix

\section{Spin-coupled states} \label{sec:spin_coupling}
This section is included for convenience, principally as the content is not difficult but I could not find a reference which presents the complete solution (ref. \citenum{pauncz_book} gets most, but not all of it). The usual approach is the use of Rumer diagrams which yields a linearly independent but not orthogonal basis. Coupling an even number of fermions to a singlet amounts to enumerating the Catalan numbers in two ways, an exercise in combinatorics\cite{stanley_ec1,stanley_ec2} (see exercise 6.19 in ref. \citenum{stanley_ec2} for \emph{many} examples). 

Note: the letters $J$ and $M$ will refer in this appendix only to angular momentum quantum numbers as it is a standard choice. While this clashes with $J$ the effective overlap matrix and $M$ the number of pairs in the remainder of the manuscript, no confusion should arise.

\subsection{Singlets}
The correct way to construct an orthonormal basis of fermionic states with good spin labels is to build them iteratively with CG coupling: a state with spin labels $JM$ is explicitly constructed
\begin{align}
	\ket{JM} = \sum^{j_1}_{m_1=-j_1} \sum^{j_2}_{m_2=-j_2} \ket{j_1 m_1 j_2 m_2} \braket{j_1 m_1 j_2 m_2 | JM}
\end{align}
as a linear combination of uncoupled states $\ket{j_1 m_1 j_2 m_2}$. The CG coefficients $\braket{j_1 m_1 j_2 m_2 | JM}$ are generally obtained from large tables. Many conventions are used, and all are valid provided that they are used coherently. 

Starting from the doublet of states for the first fermion
\begin{align}
	\Ket{\frac{1}{2},\frac{1}{2}} &= a^{\dagger}_{1\uparrow}\ket{\theta} \\
	\Ket{\frac{1}{2},-\frac{1}{2}} &= a^{\dagger}_{1\downarrow}\ket{\theta}
\end{align}
one obtains a set of 4 states for two fermions by coupling with the doublet of states for the second fermion
\begin{align}
	\Ket{1,1} &= a^{\dagger}_{1\uparrow} a^{\dagger}_{2\uparrow} \ket{\theta} \\
	\Ket{1,0} &= \frac{1}{\sqrt{2}} \left( a^{\dagger}_{1\uparrow} a^{\dagger}_{2\downarrow} + a^{\dagger}_{1\downarrow} a^{\dagger}_{2\uparrow} \right) \ket{\theta} \\
	\Ket{1,-1} &= a^{\dagger}_{1\downarrow} a^{\dagger}_{2\downarrow} \ket{\theta} \\
	\Ket{0,0} &= \frac{1}{\sqrt{2}} \left( a^{\dagger}_{1\uparrow} a^{\dagger}_{2\downarrow} - a^{\dagger}_{1\downarrow} a^{\dagger}_{2\uparrow} \right) \ket{\theta}.
\end{align}
The eight states for three fermions are built by coupling the four states for two electrons to the doublet of the third fermion, etc. This procedure quickly becomes very tedious.

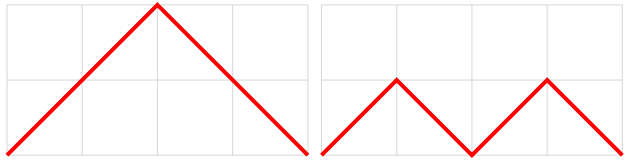
\begin{figure}
	\begin{tikzpicture}[scale=1.0]
		% Draw subtle grid
		\draw[gray!30] (0,0) grid (4,2);
		% Draw Dyck path
		\draw[line width=1.5pt,red] (0,0) -- (1,1) -- (2,2) -- (3,1) -- (4,0);
	\end{tikzpicture}
	\begin{tikzpicture}[scale=1.0]
		% Draw subtle grid
		\draw[gray!30] (0,0) grid (4,2);
		% Draw Dyck path
		\draw[line width=1.5pt,red] (0,0) -- (1,1) -- (2,0) -- (3,1) -- (4,0);
	\end{tikzpicture}
	\caption{Dyck paths for 4-electron singlets.}
	\label{fig:lowdin_path}
\end{figure}

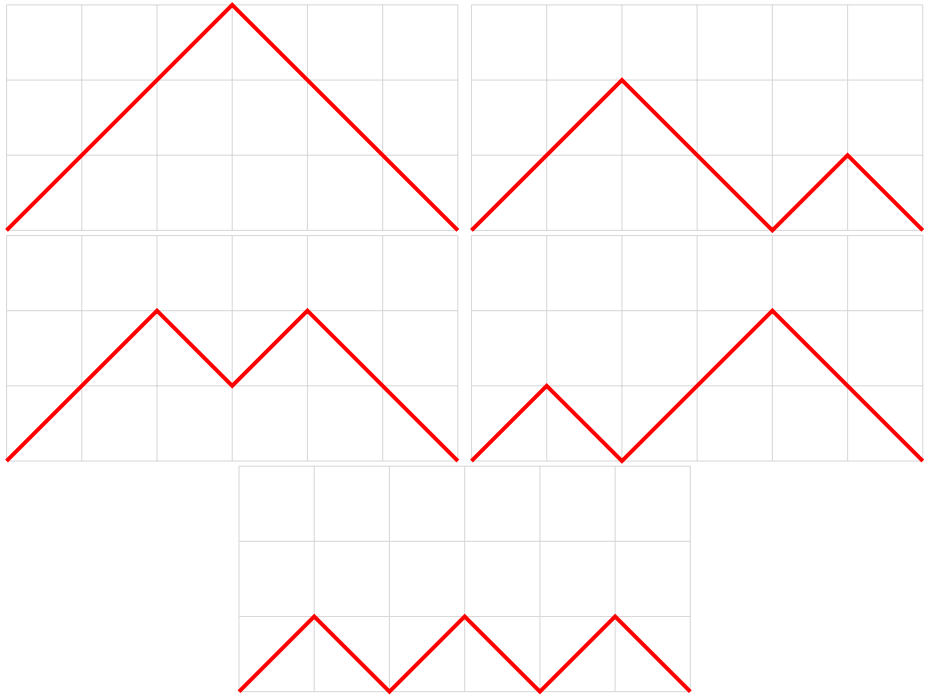
\begin{figure}
	\begin{center}
		\begin{tikzpicture}[scale=1.0]
			% Draw subtle grid
			\draw[gray!30] (0,0) grid (6,3);
			% Draw Dyck path
			\draw[line width=1.5pt,red] (0,0) -- (1,1) -- (2,2) -- (3,3) -- (4,2) -- (5,1) -- (6,0);
		\end{tikzpicture}
		\begin{tikzpicture}[scale=1.0]
			% Draw subtle grid
			\draw[gray!30] (0,0) grid (6,3);
			% Draw Dyck path
			\draw[line width=1.5pt,red] (0,0) -- (1,1) -- (2,2) -- (3,1) -- (4,0) -- (5,1) -- (6,0);
		\end{tikzpicture}\\
		\begin{tikzpicture}[scale=1.0]
			% Draw subtle grid
			\draw[gray!30] (0,0) grid (6,3);
			% Draw Dyck path
			\draw[line width=1.5pt,red] (0,0) -- (1,1) -- (2,2) -- (3,1) -- (4,2) -- (5,1) -- (6,0);
		\end{tikzpicture}
		\begin{tikzpicture}[scale=1.0]
			% Draw subtle grid
			\draw[gray!30] (0,0) grid (6,3);
			% Draw Dyck path
			\draw[line width=1.5pt,red] (0,0) -- (1,1) -- (2,0) -- (3,1) -- (4,2) -- (5,1) -- (6,0);
		\end{tikzpicture}\\
		\begin{tikzpicture}[scale=1.0]
			% Draw subtle grid
			\draw[gray!30] (0,0) grid (6,3);
			% Draw Dyck path
			\draw[line width=1.5pt,red] (0,0) -- (1,1) -- (2,0) -- (3,1) -- (4,0) -- (5,1) -- (6,0);
		\end{tikzpicture}
	\end{center}
	\caption{Dyck paths for 6-electron singlets.}
	\label{fig:lowdin_path_3}
\end{figure}

At each step in CG coupling, the last fermion is added to either increase or decrease $J$, which is strictly non-negative. This may be summarized as a diagram of upward and downward strokes that never cross the horizon. For singlets, the number of upward and downward strokes must always be the same. An example for four electrons coupled to a singlet is shown in Figure \ref{fig:lowdin_path}.
In combinatorics, such diagrams are known as Dyck paths, and enumerated by the Catalan numbers
\begin{align}
	\mathcal{C}_n = \frac{1}{2n+1}\binom{2n+1}{n}.
\end{align}
In particular, there are $\mathcal{C}_n$ Dyck paths with $n$ upward strokes and $n$ downward strokes. For $n=2$, the two possibilities are shown in Figure \ref{fig:lowdin_path}, while for $n=3$, the 5 possibilities are shown in Figure \ref{fig:lowdin_path_3}.
Dyck paths are useful as they are visually intuitive and easy to count. However, the orthogonal states not immediately obvious. The Catalan numbers also enumerate the set of standard Young tableaux (SYT) with two rows and $n$ columns (of shape $(n,n)$), hence there is a bijection between Dyck paths and SYT with shape $(n,n)$: each up-stroke in a Dyck path contributes a box in the first row while each down-stroke contributes a box in the second row. An SYT with $2n$ boxes contains the integers 1 to $2n$ such that the entries increase along each row and each column. In particular for $n=2$, the two distinct SYT are
\begin{align}
	\young(13,24) \quad \young(12,34).
\end{align}
The entries in the first (second) row correspond to the upward (downward) strokes in the Dyck path. 

The entries in the second row of the SYT label the states in the orthgonal basis. To get the individual primitive vectors (products of open-shell singlet creators) contributing to each state, begin with the SYT and permute the entries, keeping track of the sign, of the first row. If in the result the entries in each column increase, the tableau gives a contribution, e.g. 
\begin{align}
	\young(13,24) &\mapsto A^+_{12}A^+_{34} \ket{\theta} \\
	\young(12,34) - \young(21,34) &\mapsto \left( A^+_{13}A^+_{24} -A^+_{23} A^+_{14} \right) \ket{\theta}.
\end{align}
In the first line, the exchange of the entries in the first row is not allowed as the result would have a decrease in the first column. Thus, each column corresponds to an open-shell singlet creator. 

Going to $n=3$ states causes no further complications, but the notation may be economized. The five SYT are
\begin{align} \label{eq:yt_s6}
	\young(135,246) \quad \young(125,346) \quad \young(134,256) \quad \young(124,356) \quad \young(123,456)
\end{align}
but as the state labels are the entries of the second row, the SYT can be replaced with these labels. Further, these labels can be ordered. While sorting a list of integers in ascending order is unambiguous, there are different ways of sorting lists of sets of integers. It is often most convenient to sort lists of integers by comparing the \emph{last} element first, the so-called reverse lexicographic (revlex) ordering, which in the present case gives
\begin{align}
	(2,4,6) < (3,4,6) < (2,5,6) < (3,5,6) < (4,5,6).
\end{align}
By permuting the entries of the first row, keeping the contributions that increase in each column gives
\begin{align}
	(2,4,6) &\mapsto A^+_{12}A^+_{34}A^+_{56}\ket{\theta} \label{eq:s6_1} \\
	(3,4,6) &\mapsto \left(A^+_{13}A^+_{24} - A^+_{23}A^+_{14}\right)A^+_{56} \ket{\theta} \label{eq:s6_2} \\
	(2,5,6) &\mapsto A^+_{12}\left( A^+_{35}A^+_{46} - A^+_{45}A^+_{36} \right)\ket{\theta} \label{eq:s6_3} \\
	(3,5,6) &\mapsto A^+_{13}\left(A^+_{25}A^+_{46}-A^+_{45}A^+_{26} \right)\ket{\theta}
	- A^+_{23}\left(A^+_{15}A^+_{46} - A^+_{45}A^+_{16} \right)\ket{\theta} \label{eq:s6_4} \\
	(4,5,6) &\mapsto A^+_{14}\left(A^+_{25}A^+_{36} - A^+_{35}A^+_{26} \right)\ket{\theta}
	- A^+_{24}\left( A^+_{15}A^+_{36} - A^+_{35}A^+_{16} \right) \ket{\theta} \\
	&- A^+_{34}\left( A^+_{25}A^+_{16} - A^+_{15}A^+_{26} \right) \ket{\theta} \label{eq:s6_5}
\end{align}
an orthogonal basis for seniority-six singlets. 

A very tedious combinatorial development leads to the norm of each state, but the results are simply stated. When permuting the entries of the first row of \eqref{eq:yt_s6}, keep track of the number of allowed permutations with each number of transpositions. In particular, \eqref{eq:s6_1} involves only the identity permutation, \eqref{eq:s6_2} and \eqref{eq:s6_3} each involve the itentity and a permutation reducible to a single transposition, \eqref{eq:s6_4} includes the identity, two single transpositions and one double transposition, while \eqref{eq:s6_5} includes the identity, 3 single transpositions and 2 double transpositions. Denote the number of permutations with $t$ transpositions as $\pi_t$, and the number of primitive vectors contributing to the state $\ket{\varphi}$ as $\sharp(\varphi)$ to compute the norm
\begin{align}
	\vert \varphi \vert^2 = \sharp(\varphi) \sum^{n-1}_{t=0} \frac{\pi_t}{2^t}.
\end{align}
With these simple rules, an orthonormal basis for singlets of any seniority may be constructed.

\subsection{Higher spin states}
\begin{figure}
	\begin{center}
		\begin{tikzpicture}[scale=1.0]
			% (2,4) Lowdin path
			% Draw subtle grid
			\draw[gray!30] (0,0) grid (5,3);
			% Draw Dyck path
			\draw[line width=1.5pt,red] (0,0) -- (1,1) -- (2,0) -- (3,1) -- (4,0) -- (5,1);
		\end{tikzpicture}
		\begin{tikzpicture}[scale=1.0]
			% (3,4) Lowdin path
			% Draw subtle grid
			\draw[gray!30] (0,0) grid (5,3);
			% Draw Dyck path
			\draw[line width=1.5pt,red] (0,0) -- (1,1) -- (2,2) -- (3,1) -- (4,0) -- (5,1);
		\end{tikzpicture}\\
		\begin{tikzpicture}[scale=1.0]
			% (2,5)
			% Draw subtle grid
			\draw[gray!30] (0,0) grid (5,3);
			% Draw Dyck path
			\draw[line width=1.5pt,red] (0,0) -- (1,1) -- (2,0) -- (3,1) -- (4,2) -- (5,1);
		\end{tikzpicture}
		\begin{tikzpicture}[scale=1.0]
			% (3,5)
			% Draw subtle grid
			\draw[gray!30] (0,0) grid (5,3);
			% Draw Dyck path
			\draw[line width=1.5pt,red] (0,0) -- (1,1) -- (2,2) -- (3,1) -- (4,2) -- (5,1);
		\end{tikzpicture}\\
		\begin{tikzpicture}[scale=1.0]
			% (4,5)
			% Draw subtle grid
			\draw[gray!30] (0,0) grid (5,3);
			% Draw Dyck path
			\draw[line width=1.5pt,red] (0,0) -- (1,1) -- (2,2) -- (3,3) -- (4,2) -- (5,1);
		\end{tikzpicture}
	\end{center}
	\caption{L\"{o}wdin paths for 5-electron doublets.}
	\label{fig:lowdin_path_5}
\end{figure}
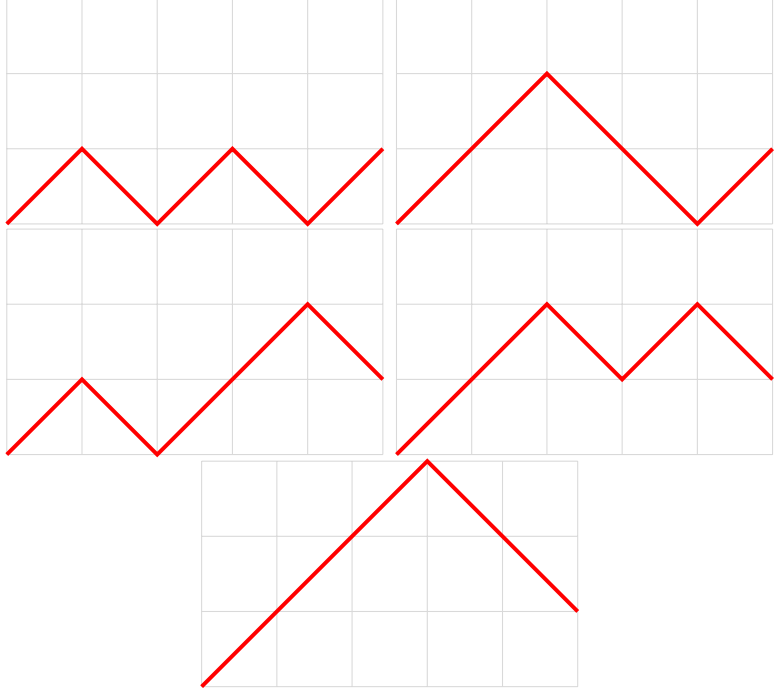

Orthogonal bases for other spin spates are constructed in the same manner, the main difference being that the sequence of upward and downward strokes does not end on the horizon. The name Dyck path is no longer used, but this type of diagram was suggested by L\"{o}wdin, so the name L\"{o}wdin path will be employed. An example of seniority-five doublets will be presented before the general expressions. There are five L\"{o}wdin paths shown in Figure \ref{fig:lowdin_path_5}
with five corresponding SYT
\begin{align}
	\young(135,24) \quad \young(125,34) \quad \young(134,25) \quad \young(124,35) \quad \young(123,45).
\end{align}
The entries of the second row of these SYT again label the positions of the downward strokes in the corresponding L\"{o}wdin paths, which may again be ordered in a revlex manner
\begin{align}
	(2,4) < (3,4) < (2,5) < (3,5) < (4,5).
\end{align}
The same rule for permuting the entries of the first row yields the orthogonal basis (with spin-projection $\sigma$)
\begin{align}
	(2,4) &\mapsto A^+_{12}A^+_{34}a^{\dagger}_{5\sigma}\ket{\theta} \\
	(3,4) &\mapsto \left( A^+_{13}A^+_{24} - A^+_{23}A^+_{14}\right) a^{\dagger}_{5\sigma} \ket{\theta} \\
	(2,5) &\mapsto A^+_{12} \left( A^+_{35}a^{\dagger}_{4\sigma} - A^+_{45}a^{\dagger}_{3\sigma} \right) \ket{\theta} \\
	(3,5) &\mapsto \left(A^+_{13}A^+_{25} - A^+_{23}A^+_{15}\right)a^{\dagger}_{4\sigma}\ket{\theta} 
				 - A^+_{45} \left(A^+_{13} a^{\dagger}_{2\sigma} - A^+_{23}a^{\dagger}_{1\sigma}\right) \ket{\theta} \\
	(4,5) &\mapsto \left( A^+_{14}A^+_{25} - A^+_{24}A^+_{15} \right) a^{\dagger}_{3\sigma} \ket{\theta} 
	             - \left( A^+_{14}A^+_{35} - A^+_{34}A^+_{15} \right) a^{\dagger}_{2\sigma} \ket{\theta} \nonumber \\
	            &- \left( A^+_{34}A^+_{25} - A^+_{24}A^+_{35} \right) a^{\dagger}_{1\sigma} \ket{\theta}.	             
\end{align}

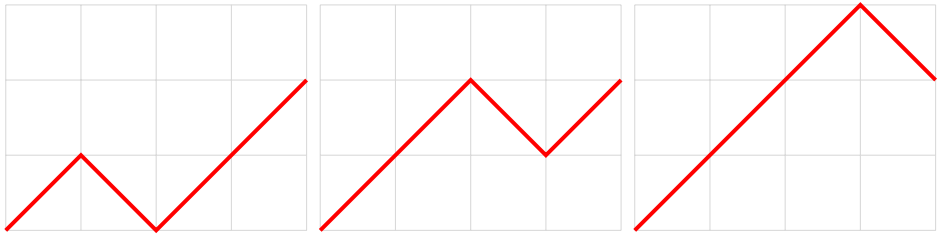
\begin{figure}
	\begin{center}
		\begin{tikzpicture}[scale=1.0]
			% (2,4) Lowdin path
			% Draw subtle grid
			\draw[gray!30] (0,0) grid (4,3);
			% Draw Dyck path
			\draw[line width=1.5pt,red] (0,0) -- (1,1) -- (2,0) -- (3,1) -- (4,2);
		\end{tikzpicture}
		\begin{tikzpicture}[scale=1.0]
			% (3,4) Lowdin path
			% Draw subtle grid
			\draw[gray!30] (0,0) grid (4,3);
			% Draw Dyck path
			\draw[line width=1.5pt,red] (0,0) -- (1,1) -- (2,2) -- (3,1) -- (4,2);
		\end{tikzpicture}
		\begin{tikzpicture}[scale=1.0]
			% (2,5)
			% Draw subtle grid
			\draw[gray!30] (0,0) grid (4,3);
			% Draw Dyck path
			\draw[line width=1.5pt,red] (0,0) -- (1,1) -- (2,2) -- (3,3) -- (4,2);
		\end{tikzpicture}
	\end{center}
	\caption{L\"{o}wdin paths for 4-electron triplets.}
	\label{fig:lowdin_path_4t}
\end{figure}

As a final example, there are three seniority-four triplets with L\"{o}wdin paths shown in Figure \ref{fig:lowdin_path_4t}, 
with corresponding SYT
\begin{align}
	\young(134,2) \quad \young(124,3) \quad \young(123,4).
\end{align}
With the vector of three triplet creators $\vec{\mathcal{Q}}^+_{pq}$, the orthonormal basis is
\begin{align}
	\ket{\varphi^{(1)}} &= A^+_{12} \vec{\mathcal{Q}}^+_{34} \ket{\theta} \\
	\ket{\varphi^{(2)}} &= \frac{1}{\sqrt{3}} \left( A^+_{13} \vec{\mathcal{Q}}^+_{24} - A^+_{23} \vec{\mathcal{Q}}^+_{14} \right) \ket{\theta} \\
	\ket{\varphi^{(3)}} &= \frac{1}{\sqrt{6}} \left(
		A^+_{14} \vec{\mathcal{Q}}^+_{23} - A^+_{24} \vec{\mathcal{Q}}^+_{13} + A^+_{34} \vec{\mathcal{Q}}^+_{12}
	\right) \ket{\theta}.
\end{align}
The first two basis vectors are obtained from the set of established rules, while the third requires a final detail. Permuting the entries in first row as before yields the three choices
\begin{align}
	\young(123,4) - \young(213,4) - \young(321,4)
\end{align}
where the last would contribute $-A^+_{34}\vec{\mathcal{Q}}^+_{21}$. For the purposes of computing the norm, the last two permutations are each single transpositions. Since $\vec{\mathcal{Q}}^+_{pq} = - \vec{\mathcal{Q}}^+_{qp}$, the last two entries of the first row can be permuted, keeping track of the sign. These permutations \emph{do not} contribute to the norm. For an arbitrary spin $S$ the last $2S$ entries of the first row should be arranged, keeping track of the sign, in ascending order. Up to an overall sign for $\ket{\varphi^{(2)}}$, these are the same results obtained from iterative Clebsch-Gordan coupling.

The general construction is the same: $n$-electron states with a spin of $2S$ are constructed as all the SYT with a first row of $\frac{1}{2}(n+2S)$ and a second row of $\frac{1}{2}(n-2S)$ entries. The primitive vectors are products of open-shell singlet creators $A^+_{pq}$ and a spin multiplet. Permuting the entries of the first row, such that in each column the entries increase, leads to a choice for an orthogonal basis, and the norm is computed from the same formula. The number of SYT (and hence the dimension of the orthogonal basis) is no longer a Catalan number, but is computable from the hook length formula. In a given Young tableau, the hook length of each box is the number of boxes to the right and below, including itself, e.g. for $n=8$, $S=1$ 
\begin{align}
	\young(65421,321)\;.
\end{align}
The number of SYT of a given shape, $\vert\lambda\vert$, is $n!$ divided by the product of all the hook lengths. In general, the entries of the first row descend from $\frac{1}{2}(n+2S)+1$ to 1, omitting the element $2S+1$, while the factors in the second row descend from $\frac{1}{2}(n-2S)$ to 1. The hook length formula gives
\begin{align}
	\vert\lambda\vert = n! \frac{2S+1}{\left(\frac{1}{2}(n+2S)+1\right)!} \frac{1}{\left(\frac{1}{2}(n-2S)\right)!}
	= \frac{2S+1}{\frac{1}{2}n + S + 1} \binom{n}{\frac{1}{2}n-S},
\end{align}
the known result for the number of CSFs with $n$ electrons coupled to a spin $S$.\cite{helgaker_book}

\section{Rapidities} \label{sec:rap_cf}
Matrix elements for RG states are straightforward to compute, on paper, from rapidities. Norms and RDM elements in particular have clear interpretations and are stable to compute, while TDM elements are in general much more expensive. In principle, this cost could be reduced, but the bigger problem is that these elements behave \emph{very poorly} numerically. For that reason, matrix elements from rapidities are not to be considered anything other than consistency checks. Even there, they are not to be trusted for TDM elements. The problem with rapidities is the TDM expressions involve denominators of products of differences of rapidities from the two states. In principle these rapidities are distinct, but when the single-particle energies $\{\varepsilon\}$ are even somewhat close to degenerate rapidities for distinct states will tend to be close to one another. The eigenvectors of the reduced BCS Hamiltonian and their correlation functions are well-defined in all cases so these poles could be managed by taking appropriate limits of the expressions. With EBV expressions, these problems are entirely avoided. 

For $\ket{\{v\}}$ on-shell and $\ket{\{u\}}$ arbitrary, Slavnov's theorem\cite{slavnov:1989,korepin_book,zhou:2002,belliard:2019} ensures that the scalar product is
\begin{align}
	\braket{ \{v\} | \{u\} } = \frac{ \det L(\{v\},\{u\})}{\mathfrak{d}(\{v\},\{u\})}
\end{align}
with $\mathfrak{d}(\{v\},\{u\})$ the Cauchy determinant
\begin{align}
	\mathfrak{d}(\{v\},\{u\}) &= \begin{vmatrix}
		\frac{1}{v_1 - u_1} & \dots & \frac{1}{v_1 - u_M} \\
		\vdots & \ddots & \vdots \\
		\frac{1}{v_M - u_1} & \dots & \frac{1}{v_M - u_M}
	\end{vmatrix} 
	= \frac{\prod_{\alpha < \beta} (v_{\alpha} - v_{\beta}) (u_{\beta} - u_{\alpha}) } { \prod_{\alpha\beta} (v_{\alpha} - u_{\beta}) }
\end{align}
and $L$ the matrix with elements
\begin{align} \label{eq:app_L_mat}
	L(\{v\},\{u\})_{\alpha\beta} = \frac{1}{(v_{\alpha}-u_{\beta})^2} \left( \frac{2}{g} + \sum_i \frac{1}{u_{\beta}-\varepsilon_i} - \sum_{\mu (\neq \alpha)} \frac{2}{u_{\beta}-v_{\mu}} \right).
\end{align}
For the norm, this expression reduces to the determinant of the \emph{Gaudin} matrix
\begin{align}
	\braket{\{v\} | \{v\}} = \det G(\{v\})
\end{align}
with elements
\begin{align}
	G(\{v\})_{\alpha \beta} &= \begin{cases}
		\sum_i \frac{1}{(v_{\alpha}-\varepsilon_i)^2} - \sum_{\mu (\neq \alpha)} \frac{2}{(v_{\mu} - v_{\alpha})^2}, \quad & \alpha = \beta \\
		\frac{2}{(v_{\alpha}-v_{\beta})^2}, \quad & \alpha \neq \beta.
	\end{cases}
\end{align}
Notice that $G$ is the Jacobian of Richardson's equations, just as $\bar{J}$ is the Jacobian of the EBV equations. Generally, Gaudin's hypothesis is that the norm of a Bethe ansatz state is the Jacobian of the corresponding on-shell equations. 

Instead of cofactors of $J$, the primitive summands to compute are determinants of $G$ with column replacements. Determinants with $p$ column replacements, scaled by the original determinant, are identical to $p \times p$ determinants of single-column updates. These updates are obtained in one stroke with Cramer's rule. In particular, one must solve the linear equations
\begin{align}
	G \frac{\partial \textbf{v}}{\partial \varepsilon_k} = 
	\begin{pmatrix}
		\frac{1}{(v_1 - \varepsilon_k)^2} \\
		\frac{1}{(v_2 - \varepsilon_k)^2} \\
		\vdots \\
		\frac{1}{(v_M - \varepsilon_k)^2}
	\end{pmatrix}
\end{align}
for each $k$. As the matrix $G$ is common, further optimization is possible through pLU or Cholesky decomposition. The 1-RDM elements are
\begin{align}
	\gamma_{kk} = 2 \sum^M_{\alpha=1} \frac{\partial v_{\alpha}}{\partial \varepsilon_k},
\end{align}
the direct 2-RDM elements are
\begin{align}
	\Gamma_{kkll} = 4 \sum_{\alpha < \beta} \frac{(v_{\alpha}-\varepsilon_k)(v_{\beta}-\varepsilon_l)+(v_{\alpha}-\varepsilon_l)(v_{\beta}-\varepsilon_k)}{(\varepsilon_k-\varepsilon_l)(v_{\beta}-v_{\alpha})} 
	\left( \frac{\partial v_{\alpha}}{\partial \varepsilon_k} \frac{\partial v_{\beta}}{\partial \varepsilon_l}
	- \frac{\partial v_{\alpha}}{\partial \varepsilon_l} \frac{\partial v_{\beta}}{\partial \varepsilon_k} \right),
\end{align}
while the pair-transfer elements are
\begin{align}
	\Gamma_{klkl} &= 2\sum_{\alpha} \frac{v_{\alpha}-\varepsilon_k}{v_{\alpha}-\varepsilon_l} \frac{\partial v_{\alpha}}{\partial \varepsilon_k}
	-4\sum_{\alpha < \beta} \frac{(v_{\alpha}-\varepsilon_k)(v_{\beta}-\varepsilon_k)}{(\varepsilon_k-\varepsilon_l)(v_{\beta}-v_{\alpha})} 
	\left( \frac{\partial v_{\alpha}}{\partial \varepsilon_k} \frac{\partial v_{\beta}}{\partial \varepsilon_l}
	- \frac{\partial v_{\alpha}}{\partial \varepsilon_l} \frac{\partial v_{\beta}}{\partial \varepsilon_k} \right).
\end{align}

For two states with the same seniorities, TDM elements require the evaluation of the form-factors 
\begin{align} \label{eq:app_ff1}
	\braket{\{v\}^M | S^+_p | \{u\}^M_{\alpha}} &= \lim_{u_{\alpha} \rightarrow \varepsilon_p} (u_{\alpha} - \varepsilon_p) \braket{\{v\}^M | \{u\}^M} \\
	&= \frac{\prod^M_{\mu=1} (v_{\mu} - \varepsilon_p)}{\prod^M_{\lambda=1 (\neq \alpha)} (u_{\lambda} - \varepsilon_p)}
	\frac{\prod^M_{\mu = 1} \prod^M_{\lambda=1 (\neq \alpha)} (v_{\mu} - u_{\lambda})}
	{\prod_{\mu < \mu'}(v_{\mu'} - v_{\mu}) \prod_{\lambda<\lambda'(\neq \alpha)}	(u_{\lambda} - u_{\lambda'})} \det L^p_{\alpha}	
\end{align}
and
\begin{align} \label{eq:app_ff2}
	\braket{\{v\}^M | S^+_p S^+_q | \{u\}^M_{\alpha,\beta} } &= \lim_{u_{\alpha} \rightarrow \varepsilon_p} \lim_{u_{\beta} \rightarrow \varepsilon_q}
	(u_{\alpha} - \varepsilon_p) (u_{\beta} - \varepsilon_q) \braket{\{v\}^M | \{u\}^M} \\
	&= \frac{1}{\varepsilon_p - \varepsilon_q} 
	\frac{\prod^M_{\mu=1} (v_{\mu} - \varepsilon_p)(v_{\mu} - \varepsilon_q)}
	{\prod^M_{\lambda=1 (\neq \alpha,\beta)} (u_{\lambda} - \varepsilon_p)(u_{\lambda} - \varepsilon_q)} \times \nonumber \\
	&\quad\quad\frac{\prod^M_{\mu = 1} \prod^M_{\lambda=1 (\neq \alpha,\beta)} (v_{\mu} - u_{\lambda})}
	{\prod_{\mu < \mu'}(v_{\mu'} - v_{\mu}) \prod_{\lambda<\lambda'(\neq \alpha,\beta)}	(u_{\lambda} - u_{\lambda'})} \det L^{pq}_{\alpha\beta}.	
\end{align}
Here, the matrix $L^p_{\alpha}$ is the matrix \eqref{eq:app_L_mat} with the $\alpha$th column replaced with the vector
\begin{align}
	\textbf{q}_p = \begin{pmatrix}
		\frac{1}{(v_1 - \varepsilon_p)^2} \\
		\frac{1}{(v_2 - \varepsilon_p)^2} \\
		\vdots \\
		\frac{1}{(v_M - \varepsilon_p)^2}
	\end{pmatrix}
\end{align}
while $L^{pq}_{\alpha\beta}$ is \eqref{eq:app_L_mat} with two updates. Decomposing the two-column updates into primitives from one-column updates may be possible with an appropriate modification of Cramer's rule.\cite{shafarevich_book} For states with seniorities differing by two, the same procedure applies, but one must first introduce a fictitious rapidity $w$, so that the required form factors may be computed as
\begin{align}
	\braket{\{v\}^M | S^+_p | \{u\}^{M-1}} &= \lim_{w\rightarrow \varepsilon_p} (w - \varepsilon_p) \braket{\{v\}^M | \{u\}^{M-1},w} \\
	\braket{\{v\}^M | S^+_p S^+_q | \{u\}^{M-1}_{\alpha}} &= 
	\lim_{w\rightarrow \varepsilon_p} \lim_{u_{\alpha}\rightarrow \varepsilon_q} (w - \varepsilon_p) (u_{\alpha} - \varepsilon_q) 
	\braket{\{v\}^M | \{u\}^{M-1},w} \\
	\braket{\{v\}^M | S^+_p S^+_q S^+_r | \{u\}^{M-1}_{\alpha,\beta}} &= 
	\lim_{w\rightarrow \varepsilon_p} \lim_{u_{\alpha}\rightarrow \varepsilon_p} \lim_{u_{\beta}\rightarrow \varepsilon_p} 
	(w - \varepsilon_p) (u_{\alpha} - \varepsilon_q) (u_{\beta} - \varepsilon_r) \braket{\{v\}^M | \{u\}^{M-1},w},
\end{align}
while for states with seniorities differing by four, two fictitious rapidites are introduced. In all cases the construction is essentially the same.

Numerical computation of the TDM elements proceeds by first computing the form-factors \eqref{eq:app_ff1} and \eqref{eq:app_ff2}, then taking the corresponding sums. As stated above, these elements become unstable when rapidities between different states are close to one another. Given that the only purpose is numerical verification, these summations can instead be computed by brute force from Slater determinant expansions. Individual RG states may be expanded in Slater determinants\cite{fecteau:2021}
\begin{align}
	\ket{\{u\}^M} = \sum_{\{i\}} C^{\{u\}}_{\{i\}} \ket{\{i\}},
\end{align}
where $\ket{\{i\}}$ is a Slater determinant with double-occupations indexed by the set $\{i\}$. The expansion coefficients are permanents
\begin{align}
	C^{\{u\}}_{\{i\}} =  \underset{\alpha,i} {\text{per}} \left( \frac{1}{u_{\alpha} - \varepsilon_i } \right) 
	= \begin{vmatrix}
		\frac{1}{u_1 - \varepsilon_{i_1}} & \dots & \frac{1}{u_1 - \varepsilon_{i_M}} \\
		\vdots & \ddots & \vdots \\
		\frac{1}{u_M - \varepsilon_{i_1}} & \dots & \frac{1}{u_M - \varepsilon_{i_M}}
	\end{vmatrix}^+
\end{align}
of Cauchy matrices. Permanents are generally intractable to compute, though Cauchy matrices may be computed through Borchardt's theorem,\cite{borchardt:1857} or EBV determinants.\cite{faribault:2012,gaudin_book,claeys:2017b} For the present purposes the goal is to numerically validate the EBV expressions, so the permanents are computed directly. Unsurprisingly, Borchardt's theorem also struggles with stability issues. In this basis the scalar products are
\begin{align}
	\braket{\{v\}^M | \{u\}^M} = \sum_{\{i\}} C^{\{i\}}_{\{v\}} C^{\{u\}}_{\{i\}}.
\end{align}
First form-factors are
\begin{align}
	\braket{\{v\}^M | S^+_p | \{u\}_{\alpha}} = \sum_{\{i\} : p \in \{i\}} C^{\{i\}}_{\{v\}} C^{\{u\}_{\alpha}}_{\{i\}_p}
\end{align}
where the summation is restricted to sets $\{i\}$ which include $p$. The coefficient $C^{\{u\}_{\alpha}}_{\{i\}_p}$ is again the permanent of a Cauchy matrix whose $\alpha,p$ element is equal to 1, while the remainder of the $\alpha$th row and $p$th column are equal to zero. Second form-factors are evaluated similarly
\begin{align}
	\braket{\{v\}^M | S^+_p S^+_q | \{u\}_{\alpha,\beta}} = \sum_{\{i\} : p,q \in \{i\}} C^{\{i\}}_{\{v\}} C^{\{u\}_{\alpha,\beta}}_{\{i\}_{p,q}}
\end{align}
with the summation restricted to sets $\{i\}$ including both $p$ and $q$, and corresponding coefficient matrices with $\alpha,p$ and $\beta,q$ elements equal to 1, withe zeroes elsewhere on the corresponding rows and columns. Extensions to states with seniorities differing by two or four is straightforward. The EBV expressions for \emph{each} DM element reported in the main text have been numerically verified against brute force expansion in terms of Slater determinants.

\bibliography{non_zero_sen}
\bibliographystyle{unsrt}

\end{document}